\documentclass[namedreferences]{solarphysics}
\usepackage[hyperref,optionalrh]{spr-sola-addons} 
\usepackage{breakurl}        

\usepackage{txfonts}
\usepackage{graphicx}           
\usepackage{ipa}                
\usepackage[usenames,dvipsnames]{color}  
\usepackage{natbib}             
\usepackage{multirow}
\usepackage{enumerate} 

\newcommand{\mcorr}[1]{#1}
\newcommand{\corr}[1]{#1}

\newcommand{\corrbis}[1]{#1}

\newcommand{\corrter}[1]{#1}

\newcommand{\corrqua}[1]{#1}

\newlength{\imsize}
\newlength{\labshift}
\newcommand{\labfigure}[5]{              
  \setlength{\labshift}{#3}
  \begin{minipage}[t]{#1}
    \includegraphics[width= #1 ]{#2}
    \hspace{- #1 } 
    \hspace{\labshift} 
    \raisebox{#4}{\textbf{#5}}
  \end{minipage}
 }

\newcommand{\eq}[1]{Equation~\ref{eq:#1}}
\newcommand{\Eq}[1]{Equation~\ref{eq:#1}}
\newcommand{\eqs}[2]{Equations~\ref{eq:#1} and \ref{eq:#2}}
\newcommand{\Eqs}[2]{Equations~\ref{eq:#1} and \ref{eq:#2}}

\newcommand{\sect}[1]{Section~\ref{s:#1}}
\newcommand{\sects}[2]{Sections.~\ref{s:#1}\,--\,\ref{s:#2}}
\newcommand{\tab}[1]{Table~\ref{t:#1}}
\newcommand{\fig}[1]{Figure~\ref{f:#1}}
\newcommand{\figs}[2]{Figures~\ref{f:#1}\,--\,\ref{f:#2}}
\newcommand{\Fig}[1]{Figure~\ref{f:#1}}
\renewcommand{\vec}[1]{ {\mathbfit #1} }
\newcommand{\BI}{\begin{itemize}}
\newcommand{\EI}{\end{itemize}}
\newcommand{\BE}{\begin{equation}}
\newcommand{\EE}{\end{equation}}
\newcommand{\BA}{\begin{eqnarray}}
\newcommand{\EA}{\end{eqnarray}}

\newcommand{\vB}{\vec{B}}

\newcommand{\Jz}{J_{\rm z}}

\newcommand{\vn}{\vec{n}}
\newcommand{\vl}{\vec{l}}
\newcommand{\vt}{\vec{t}}
\newcommand{\vw}{\vec{w}}

\newcommand{\hatn}{\hat{\vn}}
\newcommand{\hatl}{\hat{\vl}}
\newcommand{\hatt}{\hat{\vt}}
\newcommand{\hatw}{\hat{\vw}}

\newcommand{\nA}{\hatn_{\rm A}}
\newcommand{\lA}{\hatl_{\rm A}}
\newcommand{\wA}{\hatw_{\rm A}}
\newcommand{\nB}{\hatn_{\rm B}}
\newcommand{\lB}{\hatl_{\rm B}}
\newcommand{\wB}{\hatw_{\rm B}}
\newcommand{\SA}{S_{\rm A}}
\newcommand{\SB}{S_{\rm B}}
\newcommand{\sgn}{\mcorr{\zeta}}
\newcommand{\sA}  {\sgn^{\rm A}}
\newcommand{\vBtr}{\vB_{\rm tr}}
\newcommand{\vBlos}{\vB_{\rm los}}
\newcommand{\Btr}{B_{\rm tr}}
\newcommand{\Blos}{B_{\rm los}}
\newcommand{\BlA} {\Blos^{\rm A}}
\newcommand{\vBtrA}{\vBtr^{\rm A}}
\newcommand{\BtrA}{\Btr^{\rm A}}
\newcommand{\BwA} {B_{\rm w}^{\rm A}}

\newcommand{\thA} {\mcorr{\alpha^{\rm A}}}
\newcommand{\sB}  {\sgn^{\rm B}}
\newcommand{\BlB} {B_{\rm los}^{\rm B}}
\newcommand{\BtrB}{\Btr^{\rm B}}
\newcommand{\BwB} {B_{\rm w}^{\rm B}}

\newcommand{\thB} {\mcorr{\alpha^{\rm B}}}
\newcommand{\Bl} {B_{\rm los}}
\newcommand{\Bw} {B_{\rm w}}
\newcommand{\Bn} {B_{\rm n}}

\newcommand{\TD}{TD}
\newcommand{\Feng}{PENCIL-AR}
\newcommand{\Muram}{MURaM-QS}
\newcommand{\SDO}{SDO}
\newcommand{\HMI}{\SDO{}/HMI}
\newcommand{\SolO}{\textit{Solar Orbiter}}
\newcommand{\PHI}{{SO/PHI}}
\newcommand{\HRT}{\PHI{}-HRT}
\newcommand{\FDT}{\PHI{}-FDT}

\newcommand{\sophism}{SOPHISM}
\newcommand{\direct}{\textit{direct}}   
\newcommand{\reverse}{\textit{reverse}}  
\newcommand{\SDM}{SDM}

\newcommand{\Dnat}{\Delta_{\rm native}}
\newcommand{\DHMI}{\Delta_{\rm \HMI{}}}
\newcommand{\DHRT}{\Delta_{\rm \PHI{}}}
\newcommand{\rD}{r_\Delta}
\newcommand{\epsW}{\epsilon_W}
\newcommand{\epsBw}{\mcorr{\delta\Bw}}

\newcommand{\Tau}{\mcorr{\mathcal{T}}}                                       
\newcommand{\Marea}{\mcorr{\mathcal{M}_{\rm area}}}       
\newcommand{\MJz}{\mcorr{\mathcal{M}_{J_z}}}               
\newcommand{\MBt}{\mcorr{\mathcal{M}_{B_\perp > \Tau}}}               
\newcommand{\MBtone}{\mcorr{\mathcal{M}_{B_\perp > 100 \rm G}}}               
\newcommand{\MBtfive}{\mcorr{\mathcal{M}_{B_\perp > 500 \rm G}}}               

\newcommand{\eg}{e.g.}
\newcommand{\ie}{i.e.}
\newcommand{\locpath}{./}

\newcommand{\aap}{    {\it Astron. Astrophys.}}
\newcommand{\apj}{    {\it Astrophys. J.}}
\newcommand{\apjs}{   {\it Astrophys. J. Suppl.}}
\newcommand{\memsai}{  \it Memorie della Societi\'a Astronomica Italiana}
\newcommand{\solphys}{{\it Solar Phys.}}
\newcommand{\ssr}{    {\it Space Sci. Rev.}}

\begin{document}
\begin{article}
\begin{opening}
\title{Disambiguation of Vector Magnetograms by Stereoscopic Observations from the 
\textit{Solar Orbiter} (SO)/\textit{Polarimetric and Helioseismic Imager} (PHI)
and the 
\textit{Solar Dynamic Observatory} (SDO)/\textit{Helioseismic and Magnetic Imager} (HMI) 
}
\author[addressref={aff1,aff2},corref,email={valori@mps.mpg.de}]{\fnm{Gherardo}~\lnm{Valori}\orcid{0000-0001-7809-0067}} 
\author[addressref={aff1}]{\fnm{Philipp}~\lnm{L\"oschl}\orcid{0000-0002-0038-7968}}
\author[addressref={aff2}]{\fnm{David}~\lnm{Stansby}\orcid{0000-0002-1365-1908}}
\author[addressref={aff3,aff4}]{\fnm{Etienne}~\lnm{Pariat}\orcid{0000-0002-2900-0608}}
\author[addressref={aff1}]{\fnm{Johann}~\lnm{Hirzberger}} 
\author[addressref={aff5}]{\fnm{Feng}~\lnm{Chen}\orcid{0000-0002-1963-5319}}
%
\address[id=aff1]{Max-Planck-Institut f\"ur Sonnensystemforschung, Justus-von-Liebig-Weg 3, 37077, G\"ottingen, Germany}
\address[id=aff2]{University College London, Mullard Space Science Laboratory, Holmbury St. Mary, Dorking, Surrey, RH5 6NT, U.K.}
\address[id=aff3]{Laboratoire de Physique des Plasmas, \'Ecole Polytechnique, CNRS, Sorbonne Universit\'e, Observatoire de Paris, Univ. Paris-Sud, Palaiseau, France}
\address[id=aff4]{LESIA, Observatoire de Paris, Universit\'e PSL, CNRS, Sorbonne Universit\'e, Universit\'e de Paris, 5 place Jules Janssen, 92195 Meudon, France}
\address[id=aff5]{School of Astronomy and Space Science, Nanjing University, Nanjing 210023, China}
\runningauthor{G. Valori et al.}
\runningtitle{Stereoscopic Disambiguation}
\begin{abstract}
Spectropolarimetric reconstructions of the photospheric vector magnetic field are intrinsically limited by the $180^\circ$ ambiguity
in the orientation of the transverse component.
%
The successful launch and operation of \SolO{} has made the 
removal of the 180$^\circ$ ambiguity possible using solely observations
obtained from two different vantage points.  
While the exploitation of such a possibility is straightforward in principle, it is less so in practice and it is therefore important to assess the accuracy and limitations, as a function of both the 
spacecrafts' orbits and measurement principles. 
In this work we present a stereoscopic disambiguation method (\SDM{}) and discuss a thorough testing of its accuracy 
in applications to modelled active regions and quiet-Sun observations.  
%
\corr{In a first series of tests, we employ magnetograms extracted from three different numerical simulations as test fields, and model observations of the magnetograms from different angles and distances.
In these more idealized tests, the}
\SDM{} is proven to 
to reach a 100\,\% disambiguation accuracy when applied to 
moderately-to-well resolved fields. 
In such favorable conditions, the accuracy is almost independent of the relative position of the spacecrafts, with the obvious exceptions of configurations where the 
spacecrafts are within few degrees from co-alignment or quadrature. 
Even in the case of disambiguation of quiet-Sun magnetograms with significant under-resolved spatial scales, the \SDM{} provides an accuracy 
between 82\,\% and 98\,\% depending on the field strength. 
The accuracy of the \SDM{} is found to be mostly sensitive to the variable spatial resolution of \SolO{} in its highly elliptic orbit,
as well as to the intrinsic spatial scale of the observed field. 
Additionally, we provide an example of the expected accuracy as a function of time that can be used to optimally place remote-sensing observing windows during \SolO{} observation planning.
Finally, 
\corr{as a more realistic test, we consider magnetograms that are obtained using a radiative transfer inversion code and the SO/PHI Software siMulator (SOPHISM) applied to a 3D-simulation of a pore, and we present}
a preliminary discussion of the effect of the viewing angle on the observed field.
\corr{In this more realistic test of application of the \SDM{}, the method is able to successfully remove the ambiguity in strong-field areas.} 
\end{abstract}
\end{opening}
%
\section{Introduction}\label{s:intro}
The  magnetic field on the Sun is measured by remote sensing based on interpretation of spectropolarimetric observations of the emitted light \cite[see, \eg{}][]{Lites2000}.
In a traditional inversion technique applied to such measurements, the observed Stokes spectra are parametrically matched by synthetic spectra determined by the emission model (\eg{} the Zeeman effect), the propagation model (which requires a model of the solar atmosphere) and the detector system; see, \eg{} \cite{ToroIniesta2007}. 

However, even leaving aside any  measurement errors affecting the measured spectra and any systematic modeling errors affecting the spectropolarimetric inversions, the vector field at a given location still cannot be fully known.
While the line-of-sight (LoS, hereafter) component (to the observer) can be fully reconstructed, the transverse component can be inferred in magnitude and direction only, but not in orientation.
This results in an intrinsic 180$^\circ$ ambiguity in the orientation of the transverse component. 
\corrter{To fully determine the vector magnetic field, this ambiguity must be removed \citep{Harvey1969}.}

In the semi-classical picture, the ambiguity is due to the invariance of the Stokes vector to a 180$^\circ$ rotation of the reference system about the LoS-axis. 
Therefore, the 180$^\circ$ ambiguity is an intrinsic limitation of remote sensing that cannot be eliminated by improving spectropolarimetric measurements.

In actual observations, the magnetic field [$\vB$] in the detector image plane can be decomposed into  the sum of a LoS-component [$\vBlos$] and a transverse component [$\vBtr$] perpendicular to the LoS, as
\BE
  \vB = \vBlos + \sgn |\vBtr|\hatt, \qquad \mathrm{with \ } \sgn=\pm 1,
  \label{eq:B_im}
\EE
where the LoS is pointing from the Sun toward the observer and $\hatt$ is the unit vector along the transverse component in the image plane.
The removal of the ambiguity corresponds to determining the sign [$\sgn$] in each pixel.
In this sense, the ambiguity is a parity problem \citep{Semel1998} of the transverse component in each pixel of the image plane of the detector. 

Several empirical methods are available that propose solutions to remove the ambiguity.
A review is given by \cite{Metcalf2006}  with more recent testing by \cite{Leka2009}.
In particular, Table~1 of \cite{Leka2009} shows that all methods are based on choosing the sign [$\sgn$] in each pixel such that a specific quantity is minimized. 
The quantity to minimize can be, for instance, the angle between the observed transverse field and \corrter{a reference field}
(acute-angle method, \corrqua{with the reference field chosen to be, \eg{} the potential field; see discussion in \cite{Metcalf2006})}; or a weighed combination of current density and the divergence of the field (minimum-energy method; see again  \cite{Metcalf2006}), with the \corr{vertical derivatives of the field components approximated using 
\corrter{an extrapolation in height}}. 
Such 
\corrter{minimization methods}
require assumptions
\corrter{and, while} such methods have been tested and used extensively, they all propose hypothetical solutions: none can guarantee to correctly remove the ambiguity because they depend on models and assumptions, rather than solely on direct observations.

Remote observations of the same area on the Sun from different vantage points opens a novel possibility for removing the ambiguity of the transverse component of the magnetic field. 
From a purely geometrical point of view, if two observations of the same area on the Sun are available, then the  180$^\circ$ ambiguity can, in principle, be removed \citep[see, \eg][]{Solanki2015,Rouillard2020}.
The basic idea is that the unambiguous LoS-component measured by one vantage point will generally have a non-zero projection on the ambiguous transverse component measured by the second instrument.
In this way, the ``true'' orientation of the measured transverse component can be fixed as the sign of $\sgn$ that reproduces the alignment with the LoS of the other instrument.   
The stereoscopic disambiguation of the transverse-field orientation can be then derived directly from observations, without any additional hypothesis or modeling. 

\corr{Differently from traditional single-view disambiguation methods, the equations for the stereoscopic disambiguation are exact for continuous magnetic fields.
However, there are several factors that can limit the accuracy of the disambiguation in real applications.} 
For the sake of concreteness, we will consider stereoscopic disambiguation applied to vector magnetograms from the \textit{Helioseismic and Magnetic Imager} (HMI) instrument onboard the \textit{Solar Dynamic Observatory} \citep[][hereafter \SDO{}]{Scherrer2012a}, and from the \textit{High Resolution Telescope} (HRT) and \textit{Full Disk Telescope} (FDT) integrated on the \textit{Polarimetric and Helioseismic Imager} (PHI) onboard {\SolO} \citep[][hereafter SO]{Solanki2020}.
We point out, however, that the method and results illustrated in this article can be applied using magnetograms from other observatories, such as the forthcoming space-weather \textit{Lagrange} mission \citep[see, \eg{}][]{Hapgood2017}.

\corr{First, the} stereoscopic removal of the ambiguity requires the direct comparison of vector magnetograms taken from different viewpoints.
This implies that the field observations of one instrument need to be interpolated onto (the image plane and resolution of) the other.
Naturally, the finite instrumental resolution will set a limit to the accuracy of the disambiguation method. 
\corrter{For the sake of convenience we assume critical sampling which is almost perfectly correct for both \HRT{} and \FDT{} onboard Solar Orbiter and nearly correct for \HMI{}.}
Additionally, for the case of {\SolO}, the two available telescopes \HRT{} and \FDT{} have markedly different \corrter{plate scales} 
\citep{Solanki2020}. 

\corr{Second,} {\SolO} significantly changes distance from the Sun along its orbit, which implies that the effective resolution on the Sun of {\PHI} magnetograms, and with them the accuracy of the of their stereoscopic disambiguation, will change along the orbit.

\corr{Third, measurement errors of different nature are indeed present.
These can vary from systematic instrumental errors, such as the accuracy of pointing and the intrinsic sensitivity or spectral resolution of the instrument, to local inaccuracy in the calibration and inversion of spectropolarimetric measurements, due 
to approximations in the employed model atmosphere \citep[see, \eg][]{LandiDeglInnocenti2013, Borrero2014}. 
Such errors and properties are generally different for the magnetic-field maps built from different observatories \citep{Borrero2011,Schou2012,Albert2020}. 
}

\corr{Finally, aside} from these geometrical considerations, a more fundamental issue must be raised,  namely that two instruments pointing at exactly the same geometrical location on the Sun from different viewpoints do not measure exactly the same field. 
The reason is that, due to the difference in the optical path, the emission recorded by the two instruments originates from different depths in the photosphere. 
Hence, even if pointing at the same location on the solar surface, the measured emission would not come from the same parcel of plasma.  



\corr{
A realistic modeling of all of the sources of inaccuracies mentioned above is extremely challenging.
As the first one in a planned series of studies, the main goal of this article is to introduce the stereoscopic disambiguation method ({\SDM} hereafter), described in \sect{method}, and to 
verify the practical feasability of removing the 180$^\circ$ ambiguity that is intrinsic to spectropolarimetric inversions from remote-sensing reconstructions of the photospheric magnetic field from two observatories.
For this we consider two type of tests. 
The first one is a purely geometrical test, where a magnetogram is seen from different viewing angles and distances by the two detectors; see \sect{tests_geo}.
For this test, we use a set of three different numerical simulations, described in \sect{models}. 
The accuracy of the {\SDM} depending on the geometry of the observations is analysed in \sect{results_geo}.
In the second type of test we consider the synthetic emission from a three-dimensional numerical simulation of the photosphere (see \sect{tests_inv}) to treat the additional complication that, when observed from two different viewing angles, the geometrical height of the same optical depth [$\tau$] surface is different.
In this way, the accuracy of the \SDM{} is  tested against the influence of radiation detected by the two observatories that is not actually coming from the same parcel of plasma (\sect{results_inv}) -- an assumption that is implicitly made in the geometrical tests in \sect{tests_geo}.
}
The discussion and conclusions are finally given in \sect{conclusions}. 

\section{Stereoscopic Disambiguation Method (SDM)} \label{s:method}
%
\begin{figure}
 \setlength{\imsize}{0.8\textwidth}
 \centering
 \includegraphics[width=\imsize]{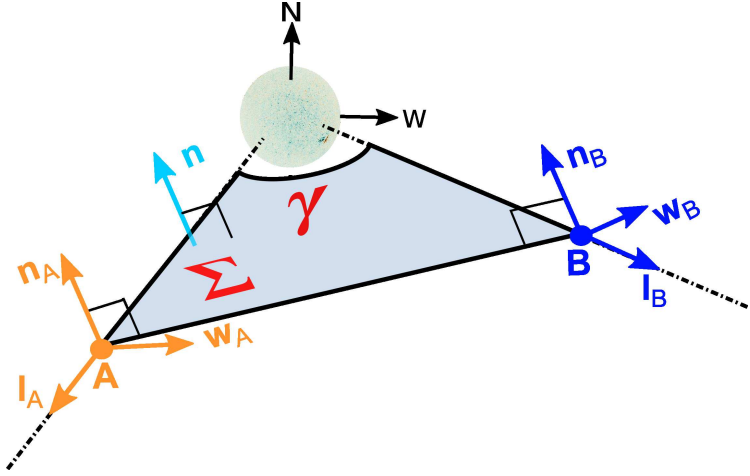}
\caption{Schematic representation of the reference systems for the application of the {\SDM}.
         The plane $\Sigma$ of normal $\hatn$ passes through the positions of the detectors A and B and the center of the Sun.
         For the detector A (respectively, B) the Cartesian reference system  $\SA=(\lA,\wA,\nA)$ (respectively,  $\SB=(\lB,\wB,\nB)$) is such that $\lA$ (respectively, $\lB$) is oriented along its LoS and $\nA$  (respectively, $\nB$) is parallel to $\hatn$.
         On the plane $\Sigma$, $\gamma$ is the separation angle between detectors A and B, defined as the angle between the LoS-directions, $\lA$ and $\lB$.
        }\label{f:multiview} 
\end{figure}


Let us consider the plane $\Sigma$ defined by the position of the two detectors, A and B, and the (center of the) Sun, and let $\hatn$ be the unit vector normal to $\Sigma$, as in \fig{multiview}.
For each detector A and B, let us define a base for the reference system defined by the unit vector $\hatn_X=\hatn$ normal to $\Sigma$ (positive toward solar North, with X in $[A,B]$), the unit vector $\hatl_X$ of the LoS-direction  (positive in the detector direction), and the (solar-West oriented) unit vector $\hatw_X=\hatn_X \times \hatl_X$, as $\SA=(\lA,\wA,\nA)$ and $\SB=(\lB,\wB,\nB)$.
By construction, $\nA=\nB=\hatn$, and the image plane of detector A (respectively, B) is parallel to the plane $(\wA,\nA)$ (respectively, $(\wB,\nB)$). 
The origin of each coordinate system is placed at the central pixel of the corresponding detector's image plane. 
\corr{With this choice, the plane identified by $(\wA,\nA)$ is basically a rotation of the detector plane of the telescope A (and similarly for B).}

Let us now consider the polar decomposition of the transverse component in \eq{B_im} as
\BE
\vBtrA= \BtrA\hatt^{\rm A}=\BtrA(\cos\thA \wA + \sin\thA \nA) 
\label{eq:polar}
\EE
where $\thA$ is the counter-clockwise polar angle from $\wA$ in the $(\wA,\nA)$ plane, and $\BtrA \ge 0$ is the amplitude of the transverse component.
\Eq{B_im} then becomes 
\BE
\vB=\BlA\lA+\sA\BtrA\left(\cos\thA\wA+\sin\thA\nA\right)
\label{eq:Adec}
\EE
where, due to the ambiguity in the observations, the values of $\thA$ are restricted to  $\thA  \in  [0, \pi]$.
For any $\thA$, the two possible transverse vectors correspond to $\sA=\pm 1$.
An analogous vector decomposition is done for $\vB$ in $\SB$.

Let $\gamma$ be the separation angle between the lines of sight $\lA$ and $\lB$ of the two detectors on $\Sigma$,  counted as the counter-clockwise rotation angle around $\nA=\nB$ with $\gamma \in [-\pi/2,\pi/2]$ and $\gamma=0$ for $\lA=\lB$, as in \fig{multiview}.  
By construction, the  vector $\vB$ in $\SA$ can be transformed into $\SB$ components by the simple rotation $\vB^{\rm B}=R(\gamma)\vB^{\rm A}$ of $\gamma$ around $\hatn$, with
\BE
 R_\gamma=\left (
  \begin{array}{ccc}
    \cos \gamma & \sin\gamma & 0 \\
   -\sin \gamma &  \cos\gamma & 0 \\
        0         &    0          & 1
 \end{array}
 \right ).
\EE
where the row and column order follows $(\lA,\wA,\nA)$.
This corresponds to a rotation of $\gamma$ around $\nA=\nB$ of the image plane, from ($\wA,\nA)$ to  ($\wB,\nB)$.  
Hence, the rotation $\gamma$ implies a foreshortening in the direction $\wB$ only. 
In components, we have 
\BA
 \BlA \cos\gamma +\sA\BtrA\cos\thA\sin\gamma &=&\BlB     \label{eq:dec_l}   \\
-\BlA \sin\gamma +\sA\BtrA\cos\thA\cos\gamma &=&\sB\BtrB\cos\thB  \label{eq:dec_w}   \\
 \sA\BtrA\sin\thA&=&\sB\BtrB\sin\thB                     \label{eq:dec_n}
\EA
\Eq{dec_n} can be rewritten as 
\BE
\frac{\sA}{\sB}=\frac{\BtrB}{\BtrA}\frac{\sin\thB}{\sin\thA}.
\label{eq:sigma_frac}
\EE
Because of the restrictions on the polar angles $\thA$ and $\thB$,  both sine functions are always positive, likewise $\BtrA$ and $\BtrB$ are by definition.
Hence, the right hand side of \eq{sigma_frac} is always positive and, since both $\sA$ and $\sB$ can only take the values $\pm 1$, it follows that 
\BE
\sA=\sB\equiv\sgn=\pm 1
\label{eq:sigma}
\EE
as a direct consequence of having chosen $\nA=\nB$ for the reference systems $\SA$ \corr{and} $\SB$.
In view of \eq{sigma}, we can then re-write \eqs{dec_l}{dec_w} as
\BA
\sgn&=&\frac{\BlB-\BlA\cos\gamma}{\BtrA\cos\thA\sin\gamma} \label{eq:s_l_full} \\
\sgn&=&\frac{\BlA\sin\gamma}{\BtrA\cos\thA\cos\gamma - \BtrB\cos\thB} \label{eq:s_w_full}
\EA
or, equivalently,
\BA
\sgn&=&\frac{\BlB-\BlA\cos\gamma}{\BwA\sin\gamma} \label{eq:s_l} \\
\sgn&=&\frac{\BlA\sin\gamma}{\BwA\cos\gamma-\BwB} \label{eq:s_w}
\EA
where we used the notation $\BwA=\BtrA\cos\thA$ (respectively, $\BwB=\BtrB\cos\thB$).
In this notation, all vector-field components appearing in \eqs{s_l}{s_w} are signed quantities, and $\gamma \in [-\pi/2,\pi/2]$.
Since $\sgn$ is defined such as to take only the values $\pm1$, in the numerical implementation of \eqs{s_l}{s_w} the sign of the right-hand side may be taken in order to avoid possible numerical oscillations.

Because of \eq{sigma}, either of Equations~\ref{eq:s_l} or \ref{eq:s_w} formally solves the ambiguity for the transverse components on both detectors simultaneously, assuming that they both provide vector information.
However, \eq{s_l}, as opposed to  \eq{s_w}, only involves $\BwA$, \ie{} the transverse component of detector A, hence \eq{s_l} can be used to solve the ambiguity in the transverse component of detector A also when detector B provides only LoS-information. 
This widens the applicability of \eq{s_l} to several space- and ground-based observatories, and it is particularly relevant for future space-weather monitoring spacecrafts that may not have full spectropolarimetric diagnostics. 

\Eqs{s_l}{s_w} are geometrically equivalent. 
However, while \eq{s_w} involves the transverse components on both detectors, \eq{s_l} involves only $\BwA$, \ie{} the transverse component on detector A.
Since the transverse component is intrinsically more noisy than the LoS one in observations, then \eq{s_l} is expected to be, broadly speaking, more accurate than \eq{s_w}.

The obvious limiting value where \eqs{s_l}{s_w} fail is $\gamma=0$, where the two systems are co-aligned and no stereoscopic view is available.
Similarly, close to quadrature ($\gamma=\pm \pi/2$), the extreme foreshortening of the field of view is expected to yield large errors, even though \eqs{s_l}{s_w} remain formally valid.
Other undefined situations may occur in individual pixels.
In  particular, \eq{s_l} cannot be used in pixels where $\BwA=0$. 
Similarly, $\sgn$ cannot be determined by \eq{s_w} if $\BlA=0$ (or, less obviously, if $\thA=0^\circ,180^\circ$). 
However, such conditions will not occur together, which opens the possibility of applying either \eq{s_w} or \eq{s_l} depending on which one is expected to yield the most accurate result in any given pixel (see \corr{\sect{appl_geo_ew} and} \sect{sdm_eq}). 

\corr{In practical applications, the field of view will be different for the two detectors, requiring restriction of the disambiguation area to the smaller of the two.
Also, in a real situation, the observed field in the image plane of each detector needs to be rotated into the frame defined by \eq{polar}. 
}
Such a transformation can be obtained based on the positions of the spacecrafts and the characteristics of the detectors employed, and it is not further discussed here.  

\corr{For the tests presented in this article, the simulations employed are Cartesian, and the magnetogram extracted from the simulation represents the solar magnetic field on the plane of the sky (\ie{} as viewed on the image plane), with uniform and homogeneous resolution.
In other words, in the following tests in practice we consider the above required transformations to be already performed.
In particular, we consider the test field to be already rotated such that $(\wA,\nA)$ are the Cartesian axis of the magnetogram, and that the detector A is placed vertically above the center of the test magnetogram at a fixed distance ({\ie} at constant resolution).
The detector B is placed at an angle $\gamma$ around $\nA$ and variable distance ({\ie} variable resolution). 
For simplicity, the fields of view of both detectors are assumed to cover the same area on the Sun, meaning that, at any angle $\gamma$, the field of view of the two detectors is identical and always covering the entire area of the magnetogram. 
\corrbis{Similarly, in each pixel we consider the corresponding field value to represent the pixel-average of the magnetic field.} 
}

%
\section{Geometrical Tests}\label{s:tests_geo}
We use three different numerical models of solar-like magnetic fields, including solutions of the nonlinear force-free equations and numerical MHD simulations.
From our point of view, these simulations model increasingly challenging test fields, from smooth and well-resolved fields, \corr{to proof of the feasibility of the method,} to more fine-scaled and unresolved \corr{fields, to estimate the performance in slightly more realistic scenarios}. 

However, we refer to the articles cited below for discussions on model limitations and their physical representation of the solar magnetic field.
\subsection{Numerical Models Used to Construct the Test Magnetograms}\label{s:models}

The synthetic magnetic-field data used in this test are presented in  \figs{mgm_td}{mgm_muram} and are obtained from a semi-analytical magnetic-field configuration of a coronal flux rope (\sect{td}) presenting a smooth distribution of the field, and from two synthetic models presenting more complex field distributions that represent active-region (\sect{feng}) and predominantly quiet-Sun (\sect{muram}) configurations.

\subsubsection{{\TD}: Smooth Test Case}\label{s:td}
\begin{figure}
 \setlength{\imsize}{0.25\columnwidth}
 \centering
 \labfigure{\imsize}{\locpath/Blos_Im_plane_HMI_td_sat_L_40_sat_B_0}{0.01\imsize}{1.3\imsize}{\textcolor{white}{a) $\Bl$}}
 \labfigure{\imsize}{\locpath/Bw_Im_plane_HMI_td_sat_L_40_sat_B_0}  {0.01\imsize}{1.3\imsize}{\textcolor{white}{b) $\Bw$}}
 \labfigure{\imsize}{\locpath/Bn_Im_plane_HMI_td_sat_L_40_sat_B_0}  {0.01\imsize}{1.3\imsize}{\textcolor{white}{c) $\Bn$}} 
 \vspace{5pt} \\
 \labfigure{\imsize}{\locpath/Blos_Im_plane_HRT_td_sat_L_40_sat_B_0}{0.03\imsize}{1.65\imsize}{\textcolor{white}{d) $\Bl$}}
 \labfigure{\imsize}{\locpath/Bw_Im_plane_HRT_td_sat_L_40_sat_B_0}  {0.03\imsize}{1.65\imsize}{\textcolor{white}{e) $\Bw$}}
 \labfigure{\imsize}{\locpath/Bn_Im_plane_HRT_td_sat_L_40_sat_B_0}  {0.03\imsize}{1.65\imsize}{\textcolor{white}{f) $\Bn$}}
 \vspace{5pt}\\ \includegraphics[width=0.45\columnwidth]{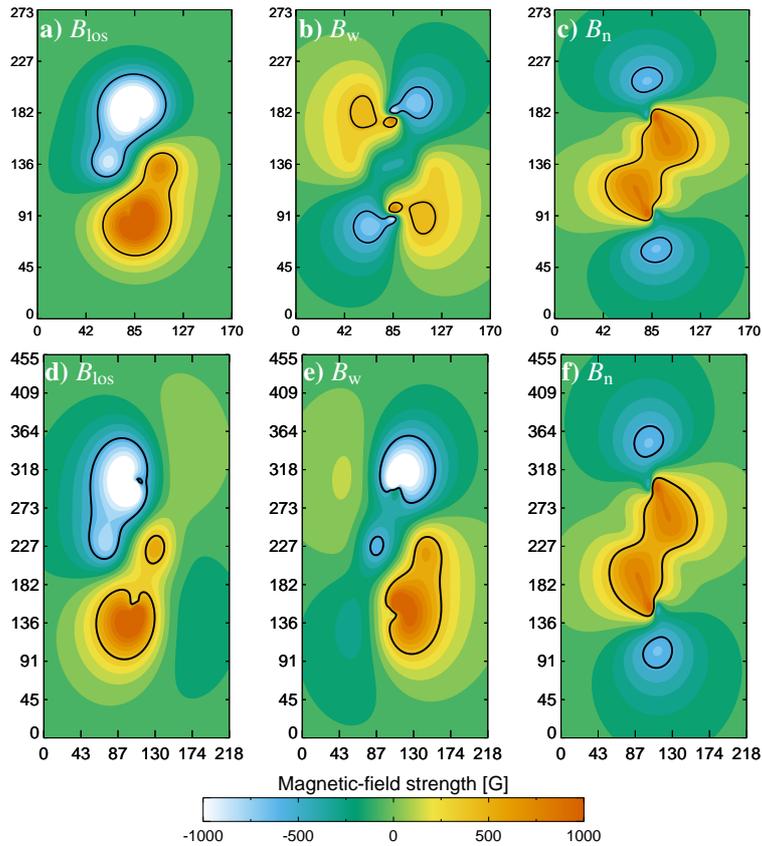}
 \caption{\corr{
{\TD} test model: 
Field distribution for the three components $\Bl$ (\textit{left column}), $\Bw$ (\textit{middle column}), and  $\Bn$ (\textit{right column}) in the image plane of {\HMI} (\textit{top row}) and {\HRT} (\textit{bottom row}), for the reference case (i.e. separated by \corrter{an} angle $\gamma = 40^\circ$, cf. \tab{reference}); the continuous isocontour represents the $\pm 500$\,G value.
The axes represent the number of pixels in the respective image planes that are used to cover the field of view by each detector in the reference observing configuration; see \sect{results_geo} for additional details. 
} }
\label{f:mgm_td}
\end{figure}
The first numerical test is constructed using the semi-analytical solution of the 3D force-free equation by \citet[][hereafter \TD{}]{titov1999}, consisting of a flux rope embedded in a potential field.
The parameters  of the {\TD} model used here are the same as those in the $N$=1 case in \cite{valori2016}.
The original number of pixels (428$\times$684) is quadrupled by interpolation with respect to \cite{valori2016}, in order to provide a very smooth and extremely well-resolved field at all viewing angles, which can be used as a proof of principle for the \SDM{}; see \fig{mgm_td}.

\subsubsection{{\Feng}: Active-Region Test Case}\label{s:feng}
\begin{figure}
 \setlength{\imsize}{0.31\columnwidth}
 \centering
 \labfigure{\imsize}{\locpath/Blos_Im_plane_HMI_feng_sat_L_40_sat_B_0}{-0.03\imsize}{0.44\imsize}{\textcolor{white}{a) $\Bl$}}
 \labfigure{\imsize}{\locpath/Bw_Im_plane_HMI_feng_sat_L_40_sat_B_0}  {-0.03\imsize}{0.44\imsize}{\textcolor{white}{b) $\Bw$}}
 \labfigure{\imsize}{\locpath/Bn_Im_plane_HMI_feng_sat_L_40_sat_B_0}  {-0.03\imsize}{0.44\imsize}{\textcolor{white}{c) $\Bn$}} 
 \vspace{5pt} \\
 \labfigure{\imsize}{\locpath/Blos_Im_plane_HRT_feng_sat_L_40_sat_B_0}{-0.02\imsize}{0.55\imsize}{\textcolor{white}{d) $\Bl$}}
 \labfigure{\imsize}{\locpath/Bw_Im_plane_HRT_feng_sat_L_40_sat_B_0}  {-0.02\imsize}{0.55\imsize}{\textcolor{white}{e) $\Bw$}}
 \labfigure{\imsize}{\locpath/Bn_Im_plane_HRT_feng_sat_L_40_sat_B_0}  {-0.02\imsize}{0.55\imsize}{\textcolor{white}{f) $\Bn$}}
 \vspace{5pt}\\ \includegraphics[width=0.45\columnwidth]{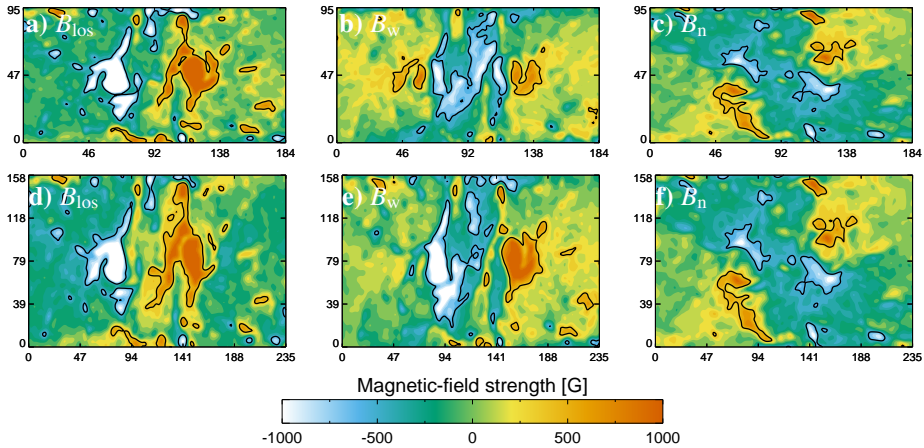}
 \caption{\corr{
{\Feng} test model: 
Field distribution for the three components $\Bl$ (\textit{left column}), $\Bw$ (\textit{middle column}), and  $\Bn$ (\textit{right column}) in the image plane of {\HMI} (\textit{top row}) and {\HRT} (\textit{bottom row}), for the reference case (i.e. separated by \corrter{an} angle $\gamma = 40^\circ$, cf. \tab{reference}); the continuous isocontour represents the $\pm 500$\,G value.
The axes represent the number of pixels in the respective image planes that are used to cover the field of view by each detector in the reference observing configuration; see \sect{results_geo} for additional details. 
} }
\label{f:mgm_feng}
\end{figure}

The second test model employs the numerical coronal evolution obtained by driving the PENCIL code as in \cite{Chen2014}, with a photospheric driver from the MURaM simulation described by \cite{Cheung2010}. 
The simulation was \corr{interpolated} to twice the original pixel size and then slightly cropped to reduce the influence of the boundaries.
The  extracted magnetogram, taken a few pixels above the PENCIL photospheric boundary, has an extent of 463$\times$239 pixels, with a uniform pixel size of 192\,km. 
The magnetogram shows some fine scales, but is still smooth and well resolved (see \fig{mgm_feng}).
This test, hereafter \Feng{}, is supposed to represent a mixture of field on small and large scales typical of an active region. 
 
\subsubsection{{\Muram}: Pore and  Quiet-Sun Test Case}\label{s:muram}
\begin{figure}
 \setlength{\imsize}{0.31\columnwidth}
 \centering
 \labfigure{\imsize}{\locpath/Blos_Im_plane_HMI_muram_sat_L_40_sat_B_0}{+0.01\imsize}{0.8\imsize}{\textcolor{white}{a) $\Bl$}}
 \labfigure{\imsize}{\locpath/Bw_Im_plane_HMI_muram_sat_L_40_sat_B_0}  {+0.01\imsize}{0.8\imsize}{\textcolor{white}{b) $\Bw$}}
 \labfigure{\imsize}{\locpath/Bn_Im_plane_HMI_muram_sat_L_40_sat_B_0}  {+0.01\imsize}{0.8\imsize}{\textcolor{white}{c) $\Bn$}} 
 \vspace{5pt} \\
 \labfigure{\imsize}{\locpath/Blos_Im_plane_HRT_muram_sat_L_40_sat_B_0}{+0.01\imsize}{1.11\imsize}{\textcolor{white}{d) $\Bl$}}
 \labfigure{\imsize}{\locpath/Bw_Im_plane_HRT_muram_sat_L_40_sat_B_0}  {+0.01\imsize}{1.11\imsize}{\textcolor{white}{e) $\Bw$}}
 \labfigure{\imsize}{\locpath/Bn_Im_plane_HRT_muram_sat_L_40_sat_B_0}  {+0.01\imsize}{1.11\imsize}{\textcolor{white}{f) $\Bn$}}
 \vspace{5pt}\\ \includegraphics[width=0.45\columnwidth]{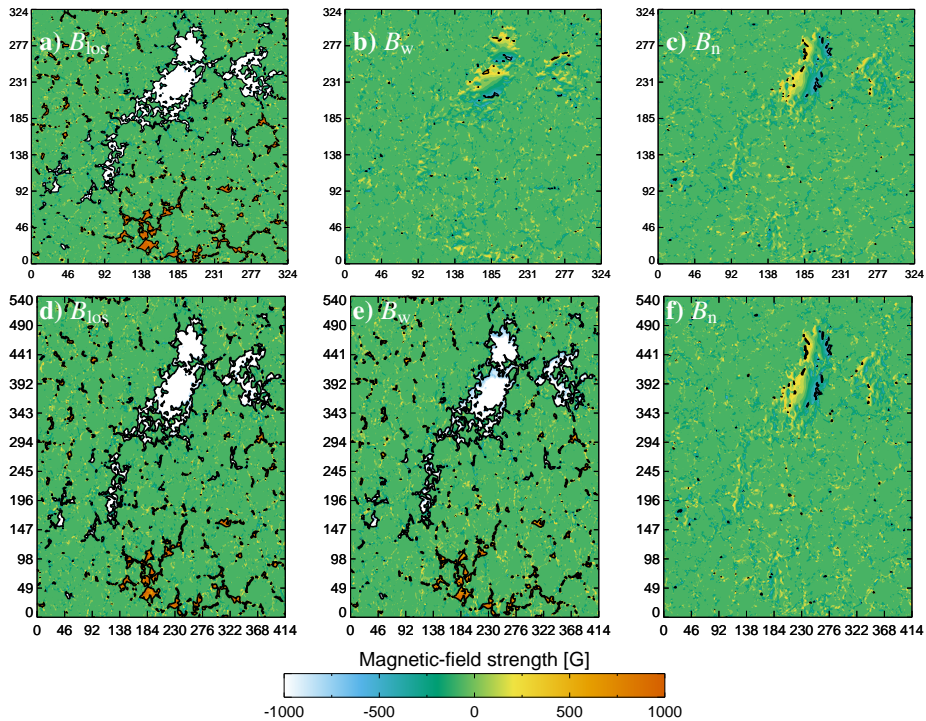}
\caption{\corr{
{\Muram} test model: 
Field distribution for the three components $\Bl$ (\textit{left column}), $\Bw$ (\textit{middle column}), and  $\Bn$ (\textit{right column}) in the image plane of {\HMI} (\textit{top row}) and {\HRT} (\textit{bottom row}), for the reference case (i.e. separated by \corrter{an} angle $\gamma = 40^\circ$, cf. \tab{reference}); the continuous isocontour represents the $\pm 500$\,G value.
The axes represent the number of pixels in the respective image planes that are used to cover the field of view by each detector in the reference observing configuration; see \sect{results_geo} for additional details. 
} }
\label{f:mgm_muram}
\end{figure}
Magneto-hydrodynamic (MHD) simulations of a typical small-scale photospheric structure, as observed with the Sunrise balloon-borne observatory, were carried out by \cite{Riethmueller2017} with the MURaM numerical simulation code \citep{Voegler2005}.  
The pixel sizes in the simulation are (41.67, 41.67, 15.89) km in $x$, $y$, and $z$, respectively.
These test data are considered as close as possible to real observational data provided by \HMI{} and \PHI{}. 
They contain a small unipolar flux concentration (pore) surrounded by small-scale magnetic structures and a large quiet-Sun field with strengths from zero up to the kilo-Gauss regime.
The slice of the {\Muram} numerical simulation used as test model in \sect{results_geo} is a plane of 812$\times$812 pixels taken at a depth of 700\,km, which was calculated to approximately represent the $\tau=1$ surface for the continuum at 500\,nm, for $\gamma=0$.

As a test for the \SDM{}, the \Muram{} case is extremely challenging as it contains length-scales of field variations down to the pixel scale; see \fig{mgm_muram}. 
As discussed in more details below, in the 
\corrter{disambiguation of \HMI{} magnetograms}
it is often the case that, in 
\corrter{quiet-Sun} areas, a random choice of the orientation is adopted because of the difficulties that 
\corrter{the annealing encounters in areas of low linear-polarization signal \citep[see, \eg][]{Hoeksema2014a, Liu2017a}.} 


\subsection {Construction of the Test Magnetograms}\label{s:mgm}
\begin{figure}
 \setlength{\imsize}{0.7\columnwidth}
    \centering
    \includegraphics[width=\imsize]{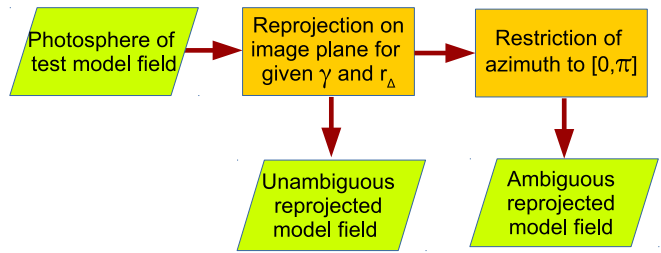}
    \caption{\corr{Flow-diagram of the construction of the test magnetograms for the geometrical tests; see \sect{mgm}. The procedure is applied on the image plane of each telescope for the given observing configuration specified by $\gamma$ and $\rD$. Input/output are visualized with \textit{green parallelograms}, operations with \textit{orange rectangles}.}}
    \label{f:flow_geo}
\end{figure}

For each model in \sect{models}, we extract the vector-field distribution at one (flat) layer at a given height, which we treat as the ``photosphere'' of the model to be used as a test (called a test model, for brevity).
In order to build a homogeneous set of test models out of the test cases in \sect{models}, the pixel size and the maximum field  strength are set to be the same for all models, as follows:

First, the vector field of each test model is rescaled such that the maximum value of the norm of the magnetic field is equal to 2000 G.

Second, the \corrter{spatial sampling} 
of the two detectors A and B of \fig{multiview} in \sect{method} and the pixel size of the test magnetograms must be specified.
From now on, we refer to a concrete case where detector A is {\HMI} and detector B is {\PHI}.
Therefore, we speak of  
``{\HMI} \corrter{spatial sampling}'' $\DHMI$ (respectively, ``{\PHI} \corrter{spatial sampling}'' $\DHRT$) 
as the helioprojective (linear) size in arcsec of the area on the Sun covered by one pixel of {\HMI} (respectively, {\PHI}) at the center of the solar disk.
Due to the choice of the reference system in \sect{method}, {\HMI} is by definition at $\gamma=0$ and at a fixed distance from the test magnetogram (\ie{} at 1\,AU distance from the Sun). 
Therefore,  {\HMI} can be assigned a fixed 
\corrter{spatial sampling} of $\DHMI=0.5$ arcsec and is never affected by foreshortening (see \sect{method}), whereas {\PHI} spatial 
\corrter{sampling} $\DHRT$ varies along its orbit (the platescale on the \HRT{} detector is 0.5 arcsec, that of \FDT{} is 3.75 arcsec).
In particular, for varying distance of {\SolO}, the ratio
\BE
 \rD=\frac{\DHRT}{\DHMI}
 \label{eq:rD}
\EE
is equal to the ratio of the {\PHI} 
\corrter{spatial sampling} $\DHRT$ to the {\HMI} 
\corrter{spatial sampling} $\DHMI$.
By changing $\rD$ we can explore the 
\corrter{spatial sampling} dependence of both detectors \HRT{} and \FDT{} onboard \SolO{}. 

Similarly, for a given distance, \ie{} for a given $\rD$, the variation of the separation angle $\gamma$ defines the foreshortened resolution (in the $\wB$ direction, see again \sect{method}) that is used by \PHI{} to image the field of view as  $\DHRT/\cos(\gamma)$. 
For testing purposes we vary $\rD$ and $\gamma$ independently, even though the two quantities are related by the real \SolO{} orbit (see discussion in \sect{conclusions} \corr{and \fig{eta_orbit} in particular}). 

Finally, there is a third resolution to be considered, that is the  resolution of the test model itself, the native resolution [$\Dnat$]. 
While for the \TD{} simulation the pixel size is not strictly given, the \Feng{} and \Muram{} simulations have horizontal pixel size approximately equal to 0.27 and 0.06\,arcsec, respectively.
In order to have comparable test cases and to widen the parameter space available for our testing, we chose to assign to all test cases the same pixel size ($\Dnat=0.2$\,arcsec).

Therefore, there are three basic quantities that define the parameters of each synthetic-observation configuration: the natural resolution of the test model [$\Dnat$], the 
\corrter{spatial sampling} ratio [$\rD$] parametrizing the distance of {\SolO}, and the separation angle between spacecrafts [$\gamma$].  
In summary, \corr{as visualize in \fig{flow_geo}}, we simulate an observation from a given detector with three steps:
\begin{enumerate}[i)]
 \item fix the observatory view by specifying $\gamma$ (finite for \PHI{}, zero for \HMI{}) and 
\corrter{spatial sampling} (variable for \PHI{}, fixed to 0.5\,arcsec for \HMI{});
 \item create the image-plane view of the test model by interpolating the model magnetogram to the effective 
\corrter{spatial sampling} of the considered detector, as seen from the given angle and distance.
  For {\HMI}, since it is always placed on the vertical of the test model, this entails only an interpolation of the test model to the effective {\HMI} 
\corrter{spatial sampling}. 
\corrbis{We use the term ``remapping'' for this procedure.}
  On the other hand, for {\PHI} the test model is first \corrbis{remapped} to the $\gamma$-foreshortened field of view at the given {\PHI} 
\corrter{spatial sampling}, and then the field components of the test model are re-projected on the \PHI{} reference system (\ie{} \corrbis{the field components are re-written in the $\SB$ reference system} of \sect{method}). 
  At this stage, the image-plane view of the test model field (the ``reprojected model field'') still has the correct orientation of the transverse component and is therefore considered to be the magnetogram to be compared with;
 \item restrict the orientation of the transverse components, \ie{} $\thA$ and $\thB$, to be within $[0,\pi]$, producing in this way the ambiguity to be removed in the reprojected model field of each instrument.  
\end{enumerate}  
The last step produces our basic ``observation-like''  {\HMI} and {\PHI}  magnetograms that reproduce effective 
\corrter{spatial sampling} and viewing angles, and have ambiguous transverse components complying with the definition in \eq{polar}.

\subsection{Application of the Disambiguation Equation}\label{s:appl_geo}
\begin{figure}
 \setlength{\imsize}{\columnwidth}
    \centering
    \includegraphics[width=\imsize]{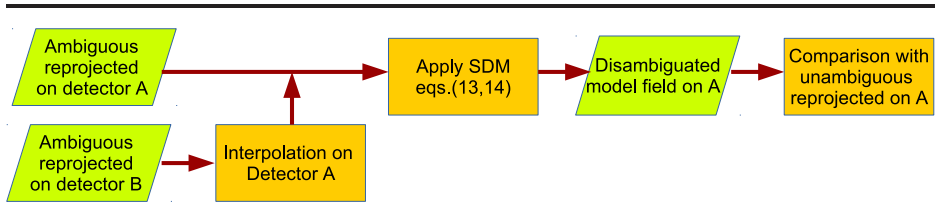}
    \caption{\corr{Flow-diagram of the \direct{} application of the \SDM{}; see \sect{appl_geo}. The \reverse{} application is obtained by formally exchanging A and B. Input/output are visualized with \textit{green parallelograms}, operations with \textit{orange rectangles}.}}
    \label{f:flow_direct}
\end{figure}
Next, \eqs{s_l}{s_w} are applied to the {\HMI} and {\PHI}  magnetograms as constructed in \sect{mgm}.
As discussed in \sect{method}, in principle either of Equations~\ref{eq:s_l} or \ref{eq:s_w} removes the ambiguity of the transverse component of both {\HMI} and {\PHI} at the same time. 
However, 
one needs to decide which detector grid is used for its application (see Sect. \ref{s:appl_geo_appproach}). In addition, one can choose to use \eq{s_l}, \eq{s_w} or a combination of the two (cf. Sect. \ref{s:appl_geo_ew}).

\subsubsection{Direct and reverse procedures}\label{s:appl_geo_appproach}

First, one can use the {\PHI} magnetogram to remove the ambiguity on the {\HMI} transverse component by spatially remapping (but not re-projecting) the {\PHI} grid onto the {\HMI} one. 
This requires interpolating the field components of the {\PHI} magnetogram onto the {\HMI} grid and applying \eqs{s_l}{s_w}. 
In this way, $\sgn$ is determined in each pixel of the {\HMI} grid, thereby removing the ambiguity.
The {\HMI} disambiguated magnetograms can be then directly compared with the correct reprojected model field obtained in step~2 of \sect{mgm} on the same ({\HMI}) grid to assess the correctness of the disambiguation.
The numerical implementation of \eqs{s_l}{s_w} that is used in the tests below employs this procedure.
In the following plots, we indicate the above procedure as the {\direct} case, plotted as a orange solid line in, {\eg} \fig{sep_res}. 
\corr{\Fig{flow_direct} shows a diagram of the  \direct{} application of the \SDM{}.}

At this point, one could remap the obtained value of $\sgn$ or the disambiguated $\Btr$ back onto the {\PHI} magnetogram and remove the ambiguity there too.
However, since this operation would imply an additional interpolation, we refrain from this.

On the other hand, one can simply formally exchange the positions of the two detectors and set the separation angle to $-\gamma$ in the disambiguation routine.
In this way, the disambiguation routine employs the field components of the {\HMI} magnetogram to remove the ambiguity of the {\PHI} transverse component, with the {\HMI} grid interpolated onto the {\PHI} one. 
Again, the {\PHI} disambiguated magnetogram is then compared with the correct reprojected model field obtained in step~2 of \sect{mgm} on the same ({\PHI}) grid.
We indicate this procedure as the {\reverse} case, plotted as a blue dashed  line in, {\eg} \fig{sep_res}. 
We present the results of the application of both direction of removal ({\ie} the {\direct} and {\reverse} cases) in the next sections.  

\subsubsection{Combined Application of SDM Equations: $\epsW$ Parameter}\label{s:appl_geo_ew}

As explained in \sect{method}, \eq{s_l} and \eq{s_w} are two geometrically equivalent ways of removing the ambiguity. 
In other words, the \SDM{} can be applied using only \eq{s_l}, only \eq{s_w}, or a combination of the two. 
One can use such equivalence to choose which equation to apply in each pixel depending on which one is expected to be more reliable for the given field values.
\corr{In particular, for generic values of $\gamma$, \eq{s_l} is not defined at locations where $\BwA\approx 0$ (see also \sect{method} for the component notation). 
In such locations, one can use \eq{s_w} instead.
Similarly, \eq{s_w} yields an undetermined $\sgn$ where $\BlA\approx 0$, \ie{} at the polarity inversion line as seen from detector A, and the employment of \eq{s_l} is to be favored instead of \eq{s_w} to remove the ambiguity at such locations.
There are of course other locations where, depending on the value of $\gamma$, either the numerator or the denominator of  \eqs{s_l}{s_w} vanishes, but these are expected to be isolated points.}  
Therefore, we consider a combined application where \eq{s_l} is applied everywhere except in pixels where the amplitude of the transverse component \corr{$\BwA$ is below a given value specified by the parameter  $\epsW$ and the relation} 
\BE
\mcorr{
|\BwA| < \epsW \max(|\BlA|) \ .
}
 \label{eq:ew}
\EE
\corr{The same above prescription for the combined application of the \SDM{} holds for both the \direct{} and \reverse{} application.}
As will be demonstrated by our analysis \corr{in \sect{sdm_eq}}, this combined use of \eqs{s_l}{s_w} generally yields the best results.

\section{Results of the Geometrical Tests}\label{s:results_geo}
In this section we report the results of testing {\SDM} as described in \sect{appl_geo} to the magnetograms obtained following \sect{mgm}.
Varying the test model, the 
\corrter{spatial sampling} and separation angles of orbital parameters, and different parameters of the {\SDM} testing procedure, 
we aim to thoroughly 
assess the feasibility of stereoscopic disambiguation.

\subsection{Evaluation Criteria}\label{s:success}
\corr{To evaluate the success of \SDM{} we adopt a selection of the metrics introduced by \cite{Metcalf2006} and \cite{Leka2009}.
In particular we consider the metrics:}
\corr{
\begin{itemize} 
\item \textit{Area}: 
For each individual tested configuration ({\ie} each data point in, \eg{} \fig{sep_res}), we quantify the rate of success of the disambiguation as the fraction
\BE
\Marea=\frac{\mathrm{Number\ of\ correctly\ disambiguated\ pixels}}{\mathrm{Total\ number\ of\ considered\ pixels}} 
\label{eq:eta}
\EE 
in each magnetogram.
Unless stated otherwise (as in, \eg{} \sect{thresh_results}), the number of pixels in the denominator of \eq{eta} is the total number of pixels in the relevant image plane (\ie{} in the magnetogram).
In \eq{eta}, the numerator is the number of pixels where the direction of the disambiguated transverse component (obtained applying the \SDM{} as described in  \sect{appl_geo} to the observation-like magnetogram obtained in Point iii of \sect{mgm}) matches the true direction as given by the reprojected test-model field (obtained in Point ii of \sect{mgm}), on the relevant image plane; see also \fig{flow_direct}.
Hence, the case of perfect disambiguation in every pixel corresponds to $\Marea=1$, whereas a systematically incorrect disambiguation in all pixels corresponds to $\Marea=0$. A random disambiguation would theoretically correspond to $\Marea=0.5$, hence a suitable disambiguation should obtain a score much higher than this.
\item
\corrbis{
\textit{Normalized vertical current density}: Difference in the vertical (to the test-model field) current density $\Jz$, computed as 
\BE
\MJz= 1-\frac{\sum |\Jz(\mathrm{Reprojected})-\Jz(\mathrm{Disambiguated})|}{2\sum |\Jz(\mathrm{Reprojected})|}
\label{eq:MJz}
\EE
where $\Jz$ is computed for the  \SDM{}-disambiguated field and for the re-projected model field (obtained in Point ii of \sect{mgm}), on the relevant image plane, \ie{} on the \HMI{} (respectively \PHI{}) image plane in the \direct{} (respectively \reverse{}) application of the \SDM{}. 
}
\item
\textit{Transverse Field}:
The fraction of transverse field above a specified threshold $\Tau$ that was resolved correctly,
\BE
\MBt=\frac{\sum B_{\rm{tr}>\Tau}(\mathrm{Correctly\ resolved\ pixels})}{ B_{\mathrm{tr}>\Tau}(\mathrm{All\ pixels})}
\label{eq:MBt}
\EE
on the relevant image plane.
\end{itemize}
The metrics $\Marea$ and $\MJz$ were judged by \cite{Leka2009} to provide enough information for noise-free cases such as those considered in this article.
However, in view of the different properties of the tests fields employed here, and of the \Muram{} case in particular, we find it useful to add the $\MBt$ metric introduced by \cite{Leka2009} for studying noise and unresolved scale effects.
Following \cite{Leka2009}, the $\MBt$ metric is computed for the threshold values $\Tau=100$ and $500$\,G.
}
\subsection{Reference Case}\label{s:reference}
\begin{table}
  \begin{tabular}{cccc}
   Parameter              & Symbol           & Reference      &  Section \\
   \hline
   SDM{} formula          & $\epsW$          & 0.01       & \ref{s:ref_results} \\
   Resolution ratio       & $\rD$            & 0.6        & \ref{s:orbit_results} \\
   Separation angle       & $\gamma$         & 40$^\circ$ & \ref{s:orbit_results} \\
   Threshold              & $\Tau$               & 0          & \ref{s:thresh_results} \\
   \hline 
  \end{tabular}
  \caption{Parameters of the tests (``Parameter'' column), with the corresponding symbol (``Symbol'' column)
  and values of the parameters for the reference case (``Reference'' column)
  \corr{for the geometrical tests in \sect{results_geo}}. 
  The ``Section'' column  refers to the section where the dependence of the \SDM{} accuracy on that particular parameter is discussed.}  
 \label{t:reference}
\end{table}
We consider a reference case summarized in \tab{reference}, where 
\begin{itemize}
 \item the model test resolution is $\Dnat=0.2$ arcsec; this formally results into a field of view of 85.6$\times$136.8, 92.6$\times$47.8, 162$\times$162 arcsec  for the {\TD}, {\Feng}, and {\Muram} test cases, respectively;
 \item the center-of-disk 
       \corrter{spatial sampling} of {\HMI} is $\DHMI=0.5$\,arcsec, with \SDO{} at 1\,AU distance from the Sun;  
 \item the \HRT{} telescope of \SolO{} has a center-of-disk 
        \corrter{spatial sampling} equal to $\DHRT=0.3$ arcsec, {\ie} $\rD=0.6$; this corresponds to a distance of {\SolO} from the Sun of 0.6\,AU;
 \item the separation angle between \SDO{} and \SolO{} is $\gamma=40^\circ$;
 \item for the given $\DHRT$ and $\gamma$, the effective 
        \corrter{spatial sampling}, \ie{} the $\wB$-foreshortened linear pixel size on the Sun, of \HRT{} is  $\DHRT/\cos(\gamma)=0.39$ arcsec;
 \item the \SDM{} is applied using both \eq{s_l} and \eq{s_w}, where the parameter for choosing the latter over the former is $\epsW=0.01$; see \sect{appl_geo} for details; 
\end{itemize}
Each of the plots in this section adopts the above values, except for one parameter at a time that is varied to study the dependence of the {\SDM} on that parameter, as indicated in the column ``Section'' of \tab{reference}.

The geometrical parameters of the reference case are chosen to be representative of a generic situation. 
In particular, specific combinations are avoided where the grids of the two instruments happen to overlap exactly ({\eg} for $\gamma=60^\circ$ and a 
\corrter{spatial sampling} of {\HRT} equal to exactly half the 
\corrter{spatial sampling} of {\HMI}).  
The corresponding 
\corr{field components} in the image plane of {\HMI} and {\HRT} for the parameters of the standard case are given in \corr{\figs{mgm_td}{mgm_muram}}.
We notice how the same field of view is rendered by a different number of pixels in the image plane of each detector, and that the model test field is interpolated in both cases to produce the detector image (\ie{} $\Dnat$ is smaller than both  $\DHMI$ and $\DHRT$).


%
\subsection{Test of the \SDM{}}\label{s:ref_results}
\begin{table}
 \begin{tabular}{lcccc}
  \multirow{2}{*}{Model}   &  $\Marea$           & $\MJz$               & $\MBtone$            & $\MBtfive$ \\
                           & {\it direct / reverse} & {\it direct / reverse}  &   {\it direct / reverse}  &   {\it direct / reverse}  \\
  \hline
     \TD{} &   1.0000  / 1.0000    & \corrbis{ 1.0000  / 1.000} &   1.0000 / 1.0000  &   1.0000 / 1.0000 \\  
   \Feng{} &   1.000   / 0.999     & \corrbis{ 0.994   / 0.988} &    1.000 / 0.999 &  0.999  / 0.999\\
  \Muram{} &     0.87 / 0.82       & \corrbis{ 0.82    / 0.79 } &  0.92 / 0.96 &  0.97 / 0.99\\
 \end{tabular}
 \caption{
   \corr{
    Success metrics area [$\Marea$], normalized current density [$\MJz$, and transverse field [$\MBt$], for $\Tau=100$ and $\Tau=500$\,G) defined in \sect{success}, referring to the application of \SDM{} to the reference case of \tab{reference}, for the three test models \TD{}, \Feng{}, and \Muram{} (see discussion in \sect{test_reference}).
    In each entry in the table the two values correspond to the \direct{} and \reverse{} cases.  
    The first column in this table [$\Marea$] is identical the first line in bold face in \tab{ref_results}.
   }
 }
 \label{t:leka}
\end{table}
\corr{
This section presents the basic tests of the accuracy of the \SDM{} using the geometrical configuration of the reference case of \sect{reference}.
A successful disambiguation is one that scores significantly better than a random choice of orientation in each pixel.
}
\subsubsection{Application of the \SDM{} to the Reference Case}\label{s:test_reference}
\corr{\tab{leka} presents the metrics defined in \sect{success} for application of the \SDM{} to the reference case in \sect{reference}. 
The two values in each entry are obtained from the \direct{} and \reverse{} application of SDM{}, respectively, as explained in \sect{appl_geo}.
} 

The entries for the \TD{} model \corr{are all unity, for all metrics. This} 
shows that, in case of a fully resolved and smooth case,  \SDM{} can remove the ambiguity in the transverse component of the reference case with 100\,\% accuracy, in both the \direct{} and \reverse{} cases. 
This proves the correctness of the \SDM{} equations and of their implementation.
Similarly, for the \Feng{} model, the \SDM{} rate of success is basically 100\,\% \corr{for all metrics}, despite the smaller fine scales that are present in the field with respect to the \TD{} model. 

In both the \TD{} and \Feng{} cases the rate of success is similar for the \direct{} and \reverse{} cases. 
In the \Muram{} case, on the other hand,  $\Marea$  is 87\,\% in the \direct{} case, where the higher-resolution \HRT{} magnetogram is used to remove the ambiguity on the lower-resolution \HMI{}.
Conversely, in the less favorable \reverse{} case, $\Marea$ is 82\,\%. 
\corr{The normalize vertical current density metric [$\MJz$] is similar for the \direct{} and \reverse{} cases (approximately 0.8), indicating that the errors in  disambiguation quantified by $\Marea$ induces only a moderate variation of  $\MJz$.
As a comparison, for the multipole test case in Table~2 of \cite{Metcalf2006}, a much smoother case than the \Muram{} discussed here, $\MJz$ varied from 1 to -0.42, depending on the disambiguation method.
Finally, the transverse-field metric, $\MBt$ for $\Tau=100$ and $\Tau=500$\,G, show that the \SDM{} errors are related to small-field values: with a threshold of $100$\,G, 92\,\% of the transverse field is correctly disambiguated in the \direct{} case, increasing to 97\,\% for a threshold of $500$\,G.
In the \reverse{} case, the fraction of the transverse field correctly represented is even larger, 96\,\% and 99\,\% for $\Tau=100$ and $500$\,G, respectively. 
\Fig{combi_other}b confirms that erroneously disambiguated pixels are all located in small-field areas in the \Muram{} case.
However, we refer to \sect{sdm_eq} for a more detailed discussion of the origin of \SDM{} errors.
}

As we have discussed, the \Muram{} case is extremely challenging, being a realistic simulation of a pore that is surrounded by much less active or even quiet regions.
Moreover, in the reference case the native resolution is smaller than both $\DHMI$ and the (foreshortened) $\DHRT$, which implies that small scales are also under-resolved. 
This is the first indication that the correct disambiguation of the transverse field is affected by the  interpolation on areas where small-scales are present. 
However, given the challenges posed by the \Muram{} test case, we regard a rate of success above 80\,\% over the whole magnetogram as a significant improvement with respect to the random choice ($\Marea=0.5$) that is usually employed in such cases; see also \sect{sdm_eq} and \sect{results_noise_ref}.

\subsubsection{Dependence on the \SDM{} Equations}\label{s:sdm_eq}
\begin{table}
  \begin{tabular}{clccc}
     Errors & \multicolumn{1}{c}{\multirow{2}{*}{Method}}       & \TD{}             & \Feng{}  &   \Muram{}   \\
     Added &         & {\it direct / reverse}             & {\it direct / reverse}   &   {\it direct / reverse}  \\
    \hline
 \multirow{6}{*}{No} &  \multicolumn{1}{c}{\bf Reference case:}    & \multirow{2}{*}{\bf 1.0000 / 1.0000}  &\multirow{2}{*}{\bf  1.000 / 0.999} &\multirow{2}{*}{\bf  0.87 / 0.82} \\
& \bf{\SDM{}, Eqs.~\ref{eq:s_l}, \ref{eq:s_w} with $\mathbf{\epsW=0.01}$}      &  & & \\
    \cline{2-5}
& \SDM{}, Eq.~\ref{eq:s_l} only                    & 0.9999 / 0.9999  &  0.997 / 0.994 &  0.85 / 0.80 \\
& \SDM{}, Eq.~\ref{eq:s_w} only                         & 0.9977 / 0.9964  &  0.984 / 0.976 &  0.84 / 0.81 \\
& Random                                      & 0.4986 / 0.4998  &  0.495 / 0.503 &  0.50 / 0.50 \\
    \hline
 \multirow{5}{*}{Yes} &  \SDM{}, Eqs.~\ref{eq:s_l}, \ref{eq:s_w} with $\epsW=0.01$     & 0.9943 / 0.9917  &  0.997 / 0.998 &  0.78 / 0.76 \\
&  \SDM{}, Eq.~\ref{eq:s_l}  only                       & 0.9864 / 0.9774  &  0.989 / 0.987 &  0.76 / 0.74 \\
&  \SDM{}, Eq.~\ref{eq:s_w}  only                       & 0.9380 / 0.9496  &  0.935 / 0.952 &  0.75 / 0.75 \\
& Random                                        & 0.5042 / 0.5004  &  0.502 / 0.498 &  0.50 / 0.50 \\
    \hline
  \end{tabular}
\caption{Area metric [$\Marea$] for different disambiguation methods, for the three test models \TD{}, \Feng{}, and \Muram{}.
  The reference case of \tab{reference} \corr{(in bold face) is same as the first column in \tab{leka}, and is repeated here for reference}.
  In each entry in the table the two values corresponds to the \direct{} and \reverse{} cases.  
  The column ``Method'' indicates which method was used for the disambiguation.
  In the bottom half of table  the same tests as in the top half are repeated with \corr{measurement errors added}; see \sect{results_noise_ref}.
     }
\label{t:ref_results}
\end{table}

We noticed in \sect{method} that \eq{s_l} and \eq{s_w} are geometrically equivalent, and in \sect{appl_geo} that a combination of both might be beneficial to the overall \SDM{} accuracy.  
Here we test such speculations.

\tab{ref_results} reports $\Marea$ in the reference case summarized in \tab{reference} for the three test models \TD{}, \Feng{}, and \Muram{}.
The different \SDM{} rows in \tab{ref_results} refer to the \SDM{} applied using only \eq{s_l}, only \eq{s_w}, or a combination of the two; see discussion in \sect{appl_geo}. 
The two values in each entry are obtained from the \direct{} and \reverse{} application of SDM{}, respectively, as explained in \sect{appl_geo}. 
In addition, the row ``Random'' corresponds to a random choice in each pixel of the orientation of the transverse component.
We consider first the top half of \tab{ref_results}.
The bottom part of \tab{ref_results} repeats the same tests as in the top half but with 
\corr{measurement errors}
added, and is discussed in \sect{results_noise_ref}.

The ``Random'' entry in the fourth line of \tab{ref_results} shows that a random choice of the orientation of the transverse component yields a \corr{area metric} $\Marea\approx 0.5$, in all cases, which reflects the parity nature of the disambiguation problem.

The dependence of $\Marea$ on which of Equations~\ref{eq:s_l} or \ref{eq:s_w} is used in the application of the \SDM{} can be assessed comparing the first three lines of \tab{ref_results}. 
Employing \Eq{s_l} almost always results in higher values of $\Marea$ than with \eq{s_w}, and the best results are obtained from a combination of the two (\ie the reference case, the first line in \tab{ref_results}). 

\begin{figure}
 \centering
 \setlength{\imsize}{0.48\columnwidth} 
\ \textbf{Without errors added} \vspace{1mm} \\
 \labfigure{0.8\imsize}{\locpath/td_sat_L_40_sat_B_0_error_map_HMI_method_los_Bx}{0.08\imsize}{1.08\imsize}{a) \eq{s_l}}
 \labfigure{0.8\imsize}{\locpath/td_sat_L_40_sat_B_0_error_map_HMI_method_tr_Bz} {0.08\imsize}{1.08\imsize}{b) \eq{s_w}} \\
\ \textbf{With errors added} \vspace{1mm} \\
 \labfigure{0.8\imsize}{\locpath/td_sat_L_40_sat_B_0_noise_error_map_HMI_method_los_Bx}{0.08\imsize}{1.08\imsize}{c) \eq{s_l}}
 \labfigure{0.8\imsize}{\locpath/td_sat_L_40_sat_B_0_noise_error_map_HMI_method_tr_Bz} {0.08\imsize}{1.08\imsize}{d) \eq{s_w}}
 \\
 \caption{
     \corrbis{Error map of wrongly disambiguated pixels (\textit{orange squares}) obtained in the application of \eq{s_l} (\textit{left column}) and \eq{s_w} (\textit{right column}). 
The background fields represent the component that is determining the error location in the two cases, \ie{} of $B_{\rm w}$  \corr{for \eq{s_l}} (\textit{left column}) and of $B_{\rm los}$ \corr{for \eq{s_w}} (\textit{right column}), shown as $\pm500$\,G-isolines, in the \direct{} application on the \HMI{} image plane, with the corresponding polarity inversion line shown by \textit{black-dashed} lines. 
     The  \textit{green} areas represent error-prone areas; \corr{see \sect{sdm_eq}  for details}.
     The \textit{top (respectively, bottom) row} corresponds to magnetogram without (respectively, with) \corr{measurement errors} added; see also \sect{results_noise_ref}.
   The axes represent the number of pixels on the \HMI{} image plane.
     }
  }
\label{f:error_maps}
\end{figure}

In this respect, it is interesting to study the locations where errors are found. 
For this we use plots of the simpler \TD{} model, but our conclusion holds for all models. 
The top two panels of \fig{error_maps} show the locations of the wrongly disambiguated pixels (orange squares) obtained in the application of \eq{s_l} (left column) and \eq{s_w} (right column). 
As expected from the discussion of \eqs{s_l}{s_w} in \sect{method} \corr{and  \ref{s:appl_geo_ew}}, the disambiguation formula are expected to fail in areas where $\sgn$ is ill-defined, \ie{} where \corr{$\BwA\approx 0$} at the denominator of \eq{s_l} and where \corr{$\BlA\approx 0$} at the numerator of \corrter{\eq{s_w}}
\corr{($\BlA$ and $\BwA$ are the LoS-component and the transverse component along $\hatw$, respectively, defined in \sect{method} and \fig{multiview}).}
This is confirmed by the panels in the top row of \fig{error_maps}, where errors in the disambiguation are found for small values of $B_{\rm w}$ for \eq{s_l} (\fig{error_maps}a) and of $B_{\rm los}$ for \eq{s_w} (\fig{error_maps}b), \eg{} around polarity inversion lines. 
Therefore, we define the error-prone areas for, \eg{} \eq{s_l} in the \direct{} case on the \HMI{} image plane, as the area where $\BtrA \le 5\times 10^{-3}\max{|\BlA|}$ or $(\pi/2 -0.02 \le \thA  \le \pi/2 + 0.02)$; see again \eqs{s_l}{s_w} for notation. 
An analogous definition of the error-prone areas employing the same heuristic values is done in the \reverse{} case and on the \PHI{} image plane (not shown).
This heuristic definition of the error-prone areas captures all wrongly disambiguated pixels, as \fig{error_maps}a,b shows.

Since such  areas can be identified prior to the disambiguation, this bears two important consequences. 
First, the two disambiguation equations have, in general, different  error-prone areas that are not overlapping, in this way confirming that \eq{s_l} and \eq{s_w}  can be used alternatively on every given pixel depending on which is the most accurate. 
The result of this combined application of \eqs{s_l}{s_w} as per \sect{appl_geo} is shown in the first line of the \tab{ref_results} for $\epsW=0.01$. 
In all test cases, the \corr{area metric} of the combined \SDM{} equations is higher than when only one of the two equations is applied.
\corr{In particular, the success rate is exactly $\Marea=1.0$, \ie{} no errors, for the \TD{} model in both \direct{} and \reverse{} cases, and for the \Feng{} \direct{} case. 
Only, a few pixels are erroneously disambiguated in the \Feng{} \reverse{} case ($\Marea=0.999$).   
In the \Muram{} case, the combined application of \eq{s_l} and \eq{s_w} also improves $\Marea$ but to a lesser extent. 
The spatial distribution of errors in the \Feng{} and \Muram{} \direct{} cases are shown in \fig{combi_other}, confirming the analysis of the error location detailed above for the \TD{} model.}


%
\begin{figure}
 \centering
 \setlength{\imsize}{0.70\columnwidth}
 \labfigure{\imsize}{\locpath/feng_sat_L_40_sat_B_0_error_map_HRT_method_combi_Bz}{-0.1\imsize}{0.70\imsize}{a)}
\\
 \labfigure{1.3\imsize}{\locpath/muram_sat_L_40_sat_B_0_error_map_HRT_method_combi_Bz}{-0.1\imsize}{1.60\imsize}{b)}
 \caption{
\corrbis{Error map of wrongly disambiguated pixels (\textit{orange squares} in \textbf{(a)}, \textit{orange small squares} in \textbf{(b)}) for the reference case (\ie{} the combined application of \eq{s_l} and \eq{s_w} with $\epsW=0.01$; see \tab{reference}), for the \textbf{(a)} \Feng{}; and \textbf{(b)} \Muram{} cases. 
The background fields represent the $B_{\rm los}$-component shown as $\pm500$\,G-isolines, in the \direct{} application on the \HMI{} image plane; see \fig{mgm_feng}d and \fig{mgm_muram}d).
The \TD{} case is not shown, since no errors are present in this case; see the first row in \tab{leka}.
Notice that wrongly disambiguated pixels in Panel \textbf{(b)} are only 13\,\% of the total; see \tab{leka}.  
The axes represent the number of pixels on the \HMI{} image plane.
} 
}
\label{f:combi_other}
\end{figure}

Then, the possibility of identifying error-prone areas allows the association of a confidence level to the \SDM{} result in each pixel. 
\corr{Thus, $\epsW$ can be used as a parameter to} 
be tuned considering measurements errors, in order to produce the most reliable disambiguation on the basis of specific uncertainties and noise levels. 

Finally, to complete the test on the combined use of \eqs{s_l}{s_w}, we consider the dependence of $\Marea$ on the parameter $\epsW$. 
\Fig{combi}a\,--\,c are obtained for the reference case of \tab{reference} by varying only $\epsW$, with solid orange and dashed blue lines representing the \direct{} and \reverse{}  direction of applications of \eqs{s_l}{s_w}, respectively.
The top row of \fig{combi}, shows that the dependence of the success rate $\Marea$ on $\epsW$ is negligible in the \TD{} and \Feng{} cases, and very weak (up to 1\,--\,2\,\%) in the \Muram{} case.
The insensitivity of $\Marea$ on $\epsW$ can be expected on the grounds of the limited extension of the error-prone areas shown in the top row of \fig{error_maps}, for both \direct{} and \reverse{} methods.

\begin{figure}
 \centering
 \setlength{\imsize}{0.30\columnwidth}
 \ \hspace{5mm} \textbf{\TD{}} \hspace{0.22\textwidth} \textbf{\Feng{}} \hspace{0.16\textwidth} \textbf{\Muram{}} \vspace{-4mm}\\
 \labfigure{\imsize}{\locpath/td_method_combi_eps_los}{-0.16\imsize}{0.80\imsize}{a)}
 \labfigure{\imsize}{\locpath/feng_method_combi_eps_los}{-0.16\imsize}{0.80\imsize}{b)}
 \labfigure{\imsize}{\locpath/muram_method_combi_eps_los}{-0.16\imsize}{0.80\imsize}{c)}\\
 \labfigure{\imsize}{\locpath/td_method_combi_noise_eps_los}{-0.16\imsize}{0.80\imsize}{d)}
 \labfigure{\imsize}{\locpath/feng_method_combi_noise_eps_los}{-0.16\imsize}{0.80\imsize}{e)}
 \labfigure{\imsize}{\locpath/muram_method_combi_noise_eps_los}{-0.16\imsize}{0.80\imsize}{f)}
\\
 \caption{Dependence of the success rate [$\Marea$] on the parameter $\epsW$ for the three test cases \TD{} (\textit{left column}), \Feng{} (\textit{middle column}), and \Muram{} (\textit{right column}), without (\textit{top row}) and with (\textit{bottom row})  \corr{measurement errors} added.
     The \textit{solid-orange} (respectively, \textit{blue-dashed}) line corresponds to the \corr{area metric} for the {\HMI} (respectively, {\PHI}) case where \eq{s_l} employs the LoS-component of  {\PHI} to remove the ambiguity on {\HMI} (respectively, employs the LoS-component of  {\HMI} to remove the ambiguity on {\PHI}), \ie{}  to the \direct{} (respectively, \reverse{}) direction of application of \eqs{s_l}{s_w}.  
     \corr{Note the difference in vertical scale between the \TD{}/\Feng{} cases (Panels \textbf{(a)}, \textbf{(b)}, \textbf{(d)}, \textbf{(e)}) and the \Muram{} cases (Panels \textbf{(c)}, \textbf{(f)}).}
     }
\label{f:combi}
\end{figure}

\subsubsection{Effect of \corr{Measurement Errors} on the \SDM{}}\label{s:results_noise_ref} 
\corr{
In this work we do not explicitly address the effect of noise or unresolved scales on the accuracy of the \SDM{}. Noise impacts will be the subject of a forthcoming, dedicated article.
However, given the unique possibility of \SDM{} to predict areas that are prone to errors, we discuss in 
}
this section how $\Marea$ in the reference case of \sect{ref_results} are changed by the presence of \corr{measurement errors.
We stress that we do not attempt to provide a physically motivated model of the origin, nature, and spatial distribution of such errors.
Instead, we consider}
\corr{an extremely simplified model for the measurement error in each pixel of the image plane given by} 
a random component of amplitude up to 70\,G, plus an  additional 30\,\% random \corr{relative} error on the transverse component. 
\corr{The error is added to each pixel of the (ambiguous) \HMI{} and \PHI{} magnetograms prior to application of the \SDM{}. 
The success metrics are again computed as explained in \sect{success}, namely comparing the disambiguated magnetogram with the corresponding (error-free) re-projected model field obtained as in Point ii of \sect{mgm}.   
}

\corr{
The bottom half of \tab{ref_results} reports the results of the same tests as in the top part but with  measurement errors  added.}
%
Measurement errors do not have any appreciable influence on the value of $\Marea$ obtained if a random orientation is chosen (``Random'' entry in row \corrter{eight}), which is to be expected given the random nature of the \corrter{error} model we adopted.
On the other hand, in all cases of separate application of \eq{s_l} and \eq{s_w} (\corrter{sixth and seventh} 
rows respectively), \corr{measurement errors cause}  a  noticeable decrease of the success rate with respect to the corresponding clean cases in the top half of \tab{ref_results}; see also \sect{sdm_eq}. 
This decrease in accuracy can be as small as 1\,\%, as, \eg{} for the application of \eq{s_l} in the \TD{} case, but also as large as 9\,\%, as for, \eg{} \eq{s_l} applied to the  \Muram{} case.
Also, consistent losses in accuracy due to \corr{measurement errors} are possible even for the smooth and resolved \TD{} test magnetogram (6\,\% for \eq{s_w}).
The distribution \corr{of wrongly disambiguated pixels} is clearly very well captured by the error-prone areas, as \fig{error_maps}c,d show.
However, and most importantly, the combined application of \eqs{s_l}{s_w}, again with $\epsW=0.01$, is able to recover an accuracy of practically 100\,\% in both the \TD{} and \Feng{} tests (cf. the \corrter{fifth} 
and first rows in  \tab{ref_results} and \sect{test_reference}).
The corresponding improvement in the \Muram{} case is more limited, of about 1\,--\,2\,\%.

\corr{Next, }let us consider how the dependence of $\Marea$ on the $\epsW$-parameter is changed by the presence of \corr{measurement errors} (see again \sect{sdm_eq}). 
\Fig{combi}d\,--\,f, compared with the clean corresponding cases in \fig{combi}a\,--\,c, shows that for all test fields such a dependence is stronger when \corr{measurement errors are} added than in the clean case.
The effect of \corr{measurement errors}  on such a dependence is comparatively larger for the \TD{} case than for the \Feng{} and \Muram{} cases (note the difference in the vertical scales between \corr{Panels \textbf{(a)}, \textbf{(b)}, \textbf{(d)}, \textbf{(e)} and Panels \textbf{(e)}, \textbf{(f)} in} \fig{combi}\corrbis{)}, 
bringing  the range of variation of $\Marea$ as a function of $\epsW$ to about 2\,--\,3\,\% for all cases. 
The value $\epsW=$ 0.01 in the reference case in \tab{reference} was chosen as a compromise between the maxima of all twelve curves in \fig{combi}.

In summary, the combined use of \eqs{s_l}{s_w} can recover the true field orientation even in the presence of significant \corr{measurement errors}, at least for field with moderately well-resolved scales such as in \TD{} and \Feng{} test models.

\subsection{Orbit Effects: 
\corrter{Spatial Sampling} and Separation Angle}\label{s:orbit_results}
\begin{figure}
 \centering
 \setlength{\imsize}{0.48\columnwidth}
 \textbf{\TD{}} \vspace{-8mm} \\
 \labfigure{\imsize}{\locpath/td_separation}{-0.07\imsize}{0.80\imsize}{a)}
 \labfigure{\imsize}{\locpath/td_resolution}{-0.07\imsize}{0.80\imsize}{b)}\\
 \textbf{\Feng{}} \vspace{-8mm} \\
 \labfigure{\imsize}{\locpath/feng_separation}{-0.05\imsize}{0.80\imsize}{c)}
 \labfigure{\imsize}{\locpath/feng_resolution}{-0.05\imsize}{0.80\imsize}{d)}\\
 \textbf{\Muram{}} \vspace{-8mm} \\ 
 \labfigure{\imsize}{\locpath/muram_separation}{-0.03\imsize}{0.80\imsize}{e)}
 \labfigure{\imsize}{\locpath/muram_resolution}{-0.03\imsize}{0.80\imsize}{f)}\\
 \caption{Rate of successful disambiguation [$\Marea$] as a function of the separation angle [$\gamma$] (\textit{left column}) and of the {\HRT} 
\corrter{spatial sampling} ratio [$\rD$] (\textit{right column}), for the {\TD} (Panels \textbf{(a)}, \textbf{(b)}), {\Feng} (Panels \textbf{(c)}, \textbf{(d)}), and {\Muram} (Panels \textbf{(e)}, \textbf{(f)}) test magnetograms, for the \direct{} (\textit{orange-solid line}) and the \reverse{} (\textit{blue-dashed line}) cases, respectively.  
     The \textit{light-green areas} in the \textit{right column} identify the range of 
      \corrter{spatial sampling} of the two telescopes (\HRT{} and \FDT{}) along the planned \SolO{} orbit.
     The \textit{black vertical-dotted lines} mark the native resolution [$\Dnat$] of the test model and 
     the \textit{red vertical-dotted lines} identify the point of equal 
     \corrter{spatial sampling} of {\HMI} and \HRT{}, $\DHMI=\DHRT$ nominally equal to 0.5\,arcsec. 
     \corr{Note the difference in vertical scale between the \TD{} case (Panels \textbf{(a)} and \textbf{(b)}) and the \Feng{} / \Muram{} cases (Panels \textbf{(c)}\,--\,\textbf{(f)}).}
     }
\label{f:sep_res} 
\end{figure}
In this section we study how the disambiguation of the reference case in \tab{reference} is affected by the changing separation angle and distance that is caused by spacecraft orbital motion.
In particular, \fig{sep_res} summarizes the \corr{area metric} [$\Marea$] of the {\SDM} as a function of the separation angle [$\gamma$] (left column) and of the 
\corrter{spatial sampling} ratio [$\rD$] (right column) in \eq{rD}. 
The  left and right columns of \fig{sep_res}, and following similar ones, are constructed using 36 sample points equally spaced in $\gamma$ and 11 equally spaced points in $\log (\rD)$.

In the {\TD} case, the \corr{area metric} of the {\SDM} is 100\,\% at all separation angles, see \fig{sep_res}a, in both directions of application of \eqs{s_l}{s_w} (corresponding to the {\direct} and {\reverse} cases, solid orange and dashed blue lines, respectively); see \sect{appl_geo}.
This plot, like the other two in the left column of \fig{sep_res}, is obtained using all the parameters from the standard case, {\eg} at 
\corrter{spatial sampling} ratio $\rD=0.6$, except for $\gamma$, which is varied.
In other words, in the idealized situation where the field is well resolved on both detectors, the application of the {\SDM} is able to remove the ambiguity at all values of the viewing angle [$\gamma$] without any significant inaccuracy. 
The only minimal variations in certainty correspond to the already identified situations of $\gamma=0$ and $\gamma=\pm 90^\circ$, in individual pixels, as expected from \sect{method}. 

Similarly, $\Marea$ as a function of the 
\corrter{spatial sampling} ratio [$\rD$] is above 99\,\% for all resolutions considered; see \fig{sep_res}b. 
In the panels on the right column of \fig{sep_res}, the two green areas show the 
\corrter{spatial sampling} intervals spanned by the higher-resolution telescope \HRT{} (left green area) and the lower-resolution telescope \FDT{} (right green area) during the orbit of {\SolO}. 
The reference case, used here for all parameters except for $\rD$, sets in particular $\gamma=40^\circ$ for all the plots in the right column of \fig{sep_res}.
The accuracy of {\SDM} is basically always 100\,\%  when {\HMI} is used to remove the ambiguity of either of the {\PHI} instruments transverse component  (\ie{} in the \reverse{} case, blue-dashed line).
In the {\direct} case, where the {\PHI} LoS component is interpolated onto the {\HMI} image plane (orange-solid line), $\Marea$ starts deviating from 100\,\% only in the \FDT{} range of resolutions, down to about 99\,\% for $\rD=10$.  
This departure is an additional indication  that the interplay between the instrument 
\corrter{spatial sampling} and the native resolution (\ie{} the intrinsic scales of the observed field) of the  test magnetogram plays a role in the accuracy of the {\SDM}. 

As a reference, the vertical black-dashed line in the right panels of \fig{sep_res} corresponds to the ratio of the resolution of the test magnetogram to the {\HMI} 
\corrter{spatial sampling} [$\Dnat/\DHMI$].
Its value is slightly inside the \HRT{} green area, meaning that, for the highest values of the {\HRT} 
\corrter{spatial sampling}, the test field is extrapolated rather than interpolated onto the {\HRT} grid. 
This is not optimal, but, since it affects the accuracy on the high-resolution side, it has no real effect on the results. 
On the other hand, a larger value of $\Dnat$ allows for wider extension of the right-hand side of the plot by still having enough grid points on the image plane of {\FDT} at the lowest resolutions, which is of more relevance to the tests we present here. 

For the {\Feng} test magnetogram, the variation of $\Marea$ with the separation angle $\gamma$ (\fig{sep_res}c) can be as large as 10\,\%.
However, such large errors are found only close to critical angles ($\gamma =0, \pm90^\circ$): at a wide range of separation angles, say for $|\gamma| \in [15^\circ,70^\circ]$, both the {\direct} and the {\reverse} case have accuracy in excess of 99\,\%.  

The dependence on 
\corrter{spatial sampling} in the {\Feng} case (\fig{sep_res}d) is much stronger, but affects the \FDT{} range of resolutions only. 
With respect to the \TD{} case (\fig{sep_res}b), the \Feng{} magnetogram presents a mixture of large and small scales, with the interpolation effect having a larger impact on accuracy. 
While $\Marea$ is above 99\,\% in the whole \HRT{} range of 
\corrter{spatial sampling} for both {\direct} and {\reverse} cases, its value for the {\direct} case drops rapidly for  $\rD >1$, reaching $\Marea=0.80$ for $\rD=10$. 
The \corr{area metric} in the {\direct} case is 99\,\%, unaffected by 
\corrter{spatial sampling}. 
Opposite to the {\reverse} case, in the {\direct} case the progressively less-resolved \PHI{} data are used to remove the ambiguity on (the image plane of) {\HMI}, hence requiring increasing under-sampling as $\rD$ grows. 
It is worth noticing, however, that in both directions $\Marea\approx 1$ within the \HRT{} resolution span.
  
In the {\Muram} case (\fig{sep_res}e,f) the small scales in the test magnetogram impact the accuracy $\Marea$ in the strongest way.
In this case the differences between the {\direct} and the {\reverse} cases are more clearly visible. 

In the {\direct}  case (\fig{sep_res}e)  $\Marea$ monotonically grows  between 54\,\% and 88\,\% as $|\gamma|$ goes from 0 to $90^\circ$. 
In the {\reverse} case, local maxima are present close to $\pm 50^\circ$, with an overall accuracy span between 60\,\% and 87\,\%.  

Similarly, \fig{sep_res}f shows how $\Marea$ as a function of the \PHI{} 
\corrter{spatial sampling} ratio [$\rD$] in the {\direct} case rapidly changes from 99\,\% at the highest 
\corrter{spatial sampling} ($\rD=0.1$) to almost 50\,\% at $\rD=10$. 
In the {\reverse} case, the \corr{area metric} is again independent of 
\corrter{spatial sampling}, and equal to 82\,\%. 
The slightly increased accuracy at $\rD=1$ and $\rD=10$ is likely only due to occasional overlap of pixels on the two grids that reduced the  interpolation error, thereby increasing the accuracy. 
Even in the  \HRT{} resolution interval, there is a significant change in $\Marea$  from 95\,\% to 77\,\%.

In summary, in the {\Muram} case, the interpolation of the small scales in the test magnetogram lowers the attainable accuracy of the disambiguation, with respect to the {\TD} and {\Feng} cases. 
However, we notice that when particular parameters are chosen such that interpolation is not required ({\eg} in the already mentioned example of $\DHRT=0.25$ arcsec and $\gamma=60^\circ$) the accuracy of the method in the {\Muram} case is also 100\,\%. 
As discussed in \sect{muram},  the \Muram{} test is to be considered as a very challenging example with a large fraction of the field below the 
\corrter{spatial sampling} of both detectors. 

In reality, $\gamma$ and the distance of {\SolO} from the Sun ({\ie} the 
\corrter{spatial sampling} ratio [$\rD$]) are not independent as we have treated them in this section, but linked by the actual orbit. 
We consider this practical application in \sect{conclusions}.

\subsection{Effect of Threshold on Significant Pixels}\label{s:thresh_results}
%
\begin{figure}
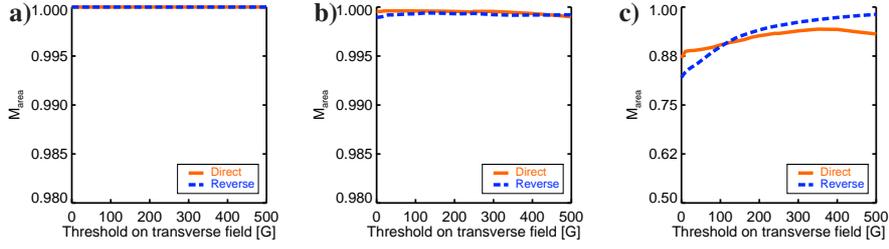

 \centering
 \setlength{\imsize}{0.30\columnwidth}
  \labfigure{\imsize}{\locpath/td_threshold}{-0.12\imsize}{0.80\imsize}{a)}
  \labfigure{\imsize}{\locpath/feng_threshold}{-0.12\imsize}{0.80\imsize}{b)}
  \labfigure{\imsize}{\locpath/muram_threshold}{-0.12\imsize}{0.80\imsize}{c)}
 \caption{
 \corr{Dependence of the success rate $\Marea$ as a function of the threshold $\Tau$ on the amplitude of the transverse component [G], for the \textbf{(a)} {\TD}, \textbf{(b)} {\Feng}, and \textbf{(c)} {\Muram} test magnetograms.
 See the caption to \fig{sep_res} for additional notation.
     Please, note the difference in vertical scale between the \TD{} and \Feng{} cases (Panels \textbf{(a)} and \textbf{(b)}) and \Muram{} cases (Panel \textbf{(c)}).
     }
 }
 \label{f:thr}
\end{figure}
The success rate [$\Marea$], \eq{eta}, is computed including all pixels in the image plane, regardless of the actual value of the field. 
However, since accuracy depends on interpolation, in this section we study how $\Marea$ changes if only pixels (on the image plane of the detector where the ambiguity is resolved) above a given threshold are included in the calculation of \eq{eta}.

In real applications, \corr{errors in the application of disambiguation methods have been related to high noise levels in the transverse-field component, or to the presence of under-resolved scales (see, \eg \corrbis{\cite{Leka2009,Hoeksema2014a}}).
As an order of magnitude estimation of the noise level in HMI magnetograms, \cite{Liu2017a} indicates the value of about 
\corrter{150Mxcm$^{-2}$ as an approximate threshold for the pixel-averaged signal} 
below which the transverse field outside active regions is assigned a random value of $\sgn$}.

\Fig{thr} shows the result for the reference case of \tab{reference} where such a threshold [$\Tau$] is applied, for value of the threshold varying from 0 to 500\,G in 23 steps of increasing amplitude from 1\,G to 50\,G (we recall that all test cases are normalized such as to have the maximum component at 2000\,G).
The accuracy in the {\TD} and {\Feng} cases is already very high even if all pixels are included in \eq{eta}, and there is little room for improvement by masking; see \fig{thr}a,b respectively.
In the {\Muram} case (\fig{thr}c), however, a strong dependence on the masking values is found. 
In particular, $\Marea$ is increased by about 10\,\% in the {\reverse} case by masking all pixels with transverse component below 150\,G, and reaches $\Marea\simeq98$\,\% if only pixels with the transverse component larger than 500\,G are considered. 
\begin{figure}
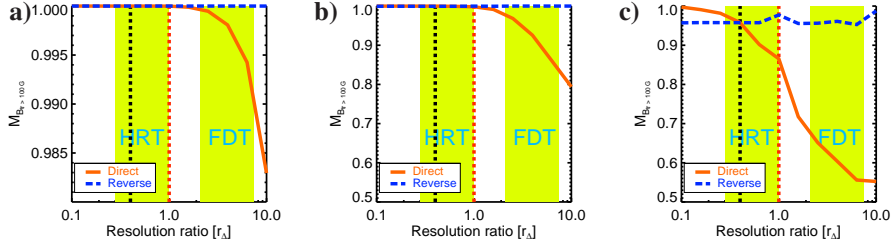

 \centering
 \setlength{\imsize}{0.30\columnwidth}
 \labfigure{\imsize}{\locpath/td_resolution_kdBt100}{-0.12\imsize}{0.80\imsize}{a)}
 \labfigure{\imsize}{\locpath/feng_resolution_kdBt100}{-0.12\imsize}{0.80\imsize}{b)}
 \labfigure{\imsize}{\locpath/muram_resolution_kdBt100}{-0.12\imsize}{0.80\imsize}{c)}
 \caption{\corr{Dependence of $\MBtone$ 
                as a function of the \PHI{} resolution ratio [$\rD$].
     Please, note the difference in vertical scale between the \TD{} case (Panel \textbf{(a)}) and the \Feng{} and \Muram{} cases (Panels \textbf{(b)} and \textbf{(c)}, respectively).
                }
     }
 \label{f:Mthr}
\end{figure}

\corr{
Because of the variation of 
\corrter{spatial sampling} experienced by \PHI{} along its orbit, we also study how the transverse field  is rendered for different re-binning factor $\rD$, as quantified by the transverse field metric $\MBtone$. 
\Fig{Mthr} shows the $\MBtone$ as a function of $\rD$, with all other geometrical parameters kept at the values of the reference case of \tab{reference}.
}

\corr{First, comparing Panels a, b, and c in \Fig{Mthr} with the corresponding panels  in \fig{sep_res} (Panels b, d, and f for the \TD{}, \Feng{}, and \Muram{} cases, respectively) we notice that the $\MBtone$ metric behaves very similarly to the $\Marea$ metric for all cases and $\rD$-values. 
In particular, for the \TD{} and \Feng{} cases $\MBtone$ is unity for the entire range of $\rD$ in the \reverse{} applications, and it departs from unity only for the higher $\rD$-values in the \FDT{} range of resolutions.  
In the \Muram{} case, the fraction of transverse field correctly represented by the disambiguated magnetogram drops significantly as a function of $\rD$ in the \direct{} case, while remaining at the almost constant value of 0.95 in the \reverse{} case. 
In other words, in the \Muram{} case, as the success of the disambiguation quantified as $\Marea$ decreases with $\rD$, so does the fraction of correctly represented transverse field $\MBtone$, and at the rate.  
}

\corr{The plots for the transverse field metric for $\Tau=500$\,G ($\MBtfive$, not shown here) is practically indistinguishable from \fig{Mthr}, except for the constant value of the \reverse{} application to the \Muram {} case attaining the slighter higher value 0.98, instead of 0.95 as for $\MBtone$ (see \fig{Mthr}c). 
Considering the \Muram{} case in the above test, and \fig{thr}c in particular, this implies that a higher threshold [$\Tau$] on the transverse field does increases the relative accuracy of the \SDM{}. 
However, such increased accuracy does not further improve how the transverse field is represented at different $\rD$ 
in the \Muram{} case, as quantified by $\MBt$.
}

Such a strong dependence of $\Marea$ on the masking of weak fields is an indirect confirmation that the small scales are the most important source of inaccuracies for the {\SDM}.
On the other hand, even for the quiet-Sun case in \Muram{}, the \corr{area metric} is above 80\,\%, which is significantly better than the random orientation.  
In this respect, \SDM{} is expected to significantly improve disambiguation with respect to state-of-the-art using only one viewpoint.

%
\section{Reconstructed Magnetogram Test}\label{s:tests_inv}
In this section we describe the test of {\SDM} with simulated observations from different viewpoints, rather than with a 2D-map from a numerical simulation that is simply reprojected at different angles, as in \sect{tests_geo}.
Such a test has a relevance that goes beyond testing the validity of the {\SDM} as such, as it addresses the question of how a field is rendered by the observation and inversion procedure from different viewpoints.
Without pretence to exhaust this topic, we include it here as an example to illustrate the complex challenges faced by any stereoscopic method.

\subsection{Test Construction}\label{s:inv_model}
\begin{table}
  \begin{tabular}{lcccc}
                Type          & Construction method      & Ambiguous &  \multicolumn{2}{c}{Reference sections} \\
                              &                          &           &  Definition & \SDM{} application\\
    \hline
    Reprojected               & geometrical reprojection & no        & \ref{s:mgm} &  \ref{s:results_geo} \\
    Synthetic       & SPINOR + SOPHISM         & yes       & \ref{s:synth_mgm}         & not used \\
    Reconstructed             & Response functions       & no        & \ref{s:rec_mgm}         & \ref{s:results_inv} \\
    \multirow{2}{*}{Reproj-reconstructed}
                              & geometrical reprojection & no        & \ref{s:geo_vs_inv}         & not used  \\[-2pt]
                              & of the $(\gamma=0)$-reconstructed &    &                  &             \\
  \end{tabular}
  \caption{Nomenclature of the magnetograms discussed in this article.
           ``Type'' indicates the name used for the magnetogram, the ``Construction method'' column contains keywords identifying how the magnetogram is built, ``Ambiguous'' column indicates if the obtained magnetogram is intrinsically affected by the ambiguity in the transverse component, and ``Reference sections'' indicates in which sections the given type of magnetogram is discussed (''Definition'') and used as a test for the disambiguation method (``\SDM{} application''). We recall that the ambiguity in the ''Reprojected'' and ''Reconstructed'' types is introduced prior to the application of the \SDM{}.}
           \label{t:mgm}
\end{table}

\begin{figure}
 \centering
 \setlength{\imsize}{0.25\columnwidth}
 \ \hspace{15mm} \textbf{Synthetic} \hspace{0.12\textwidth} \textbf{Reconstructed} \hspace{0.10\textwidth} \textbf{Difference} \vspace{1mm}\\
 \ \hspace{30mm} \includegraphics[width=\imsize]{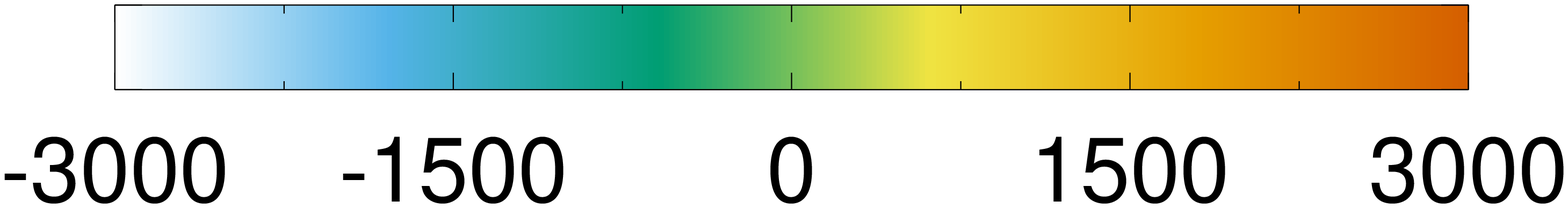} \hspace{0.13\textwidth}  \includegraphics[width=\imsize]{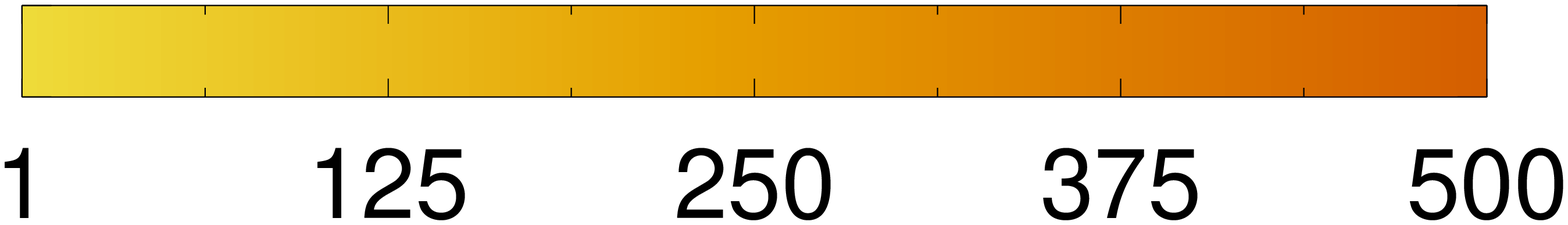} \vspace{-0mm}\\
 \raisebox{0.5\imsize}{$\gamma=\phantom{1}0^\circ$}
 \includegraphics[width=\imsize]{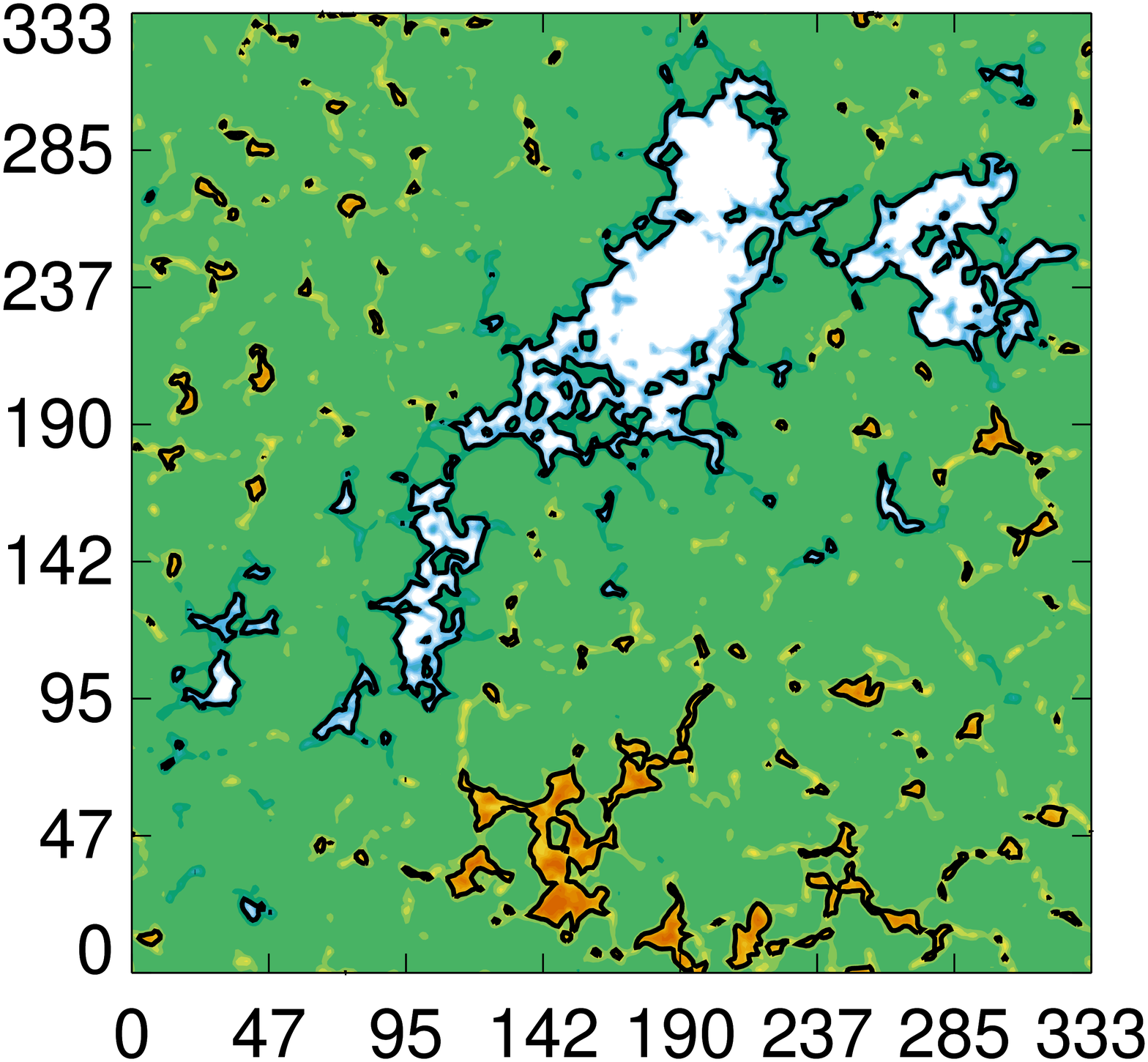}
 \includegraphics[width=\imsize]{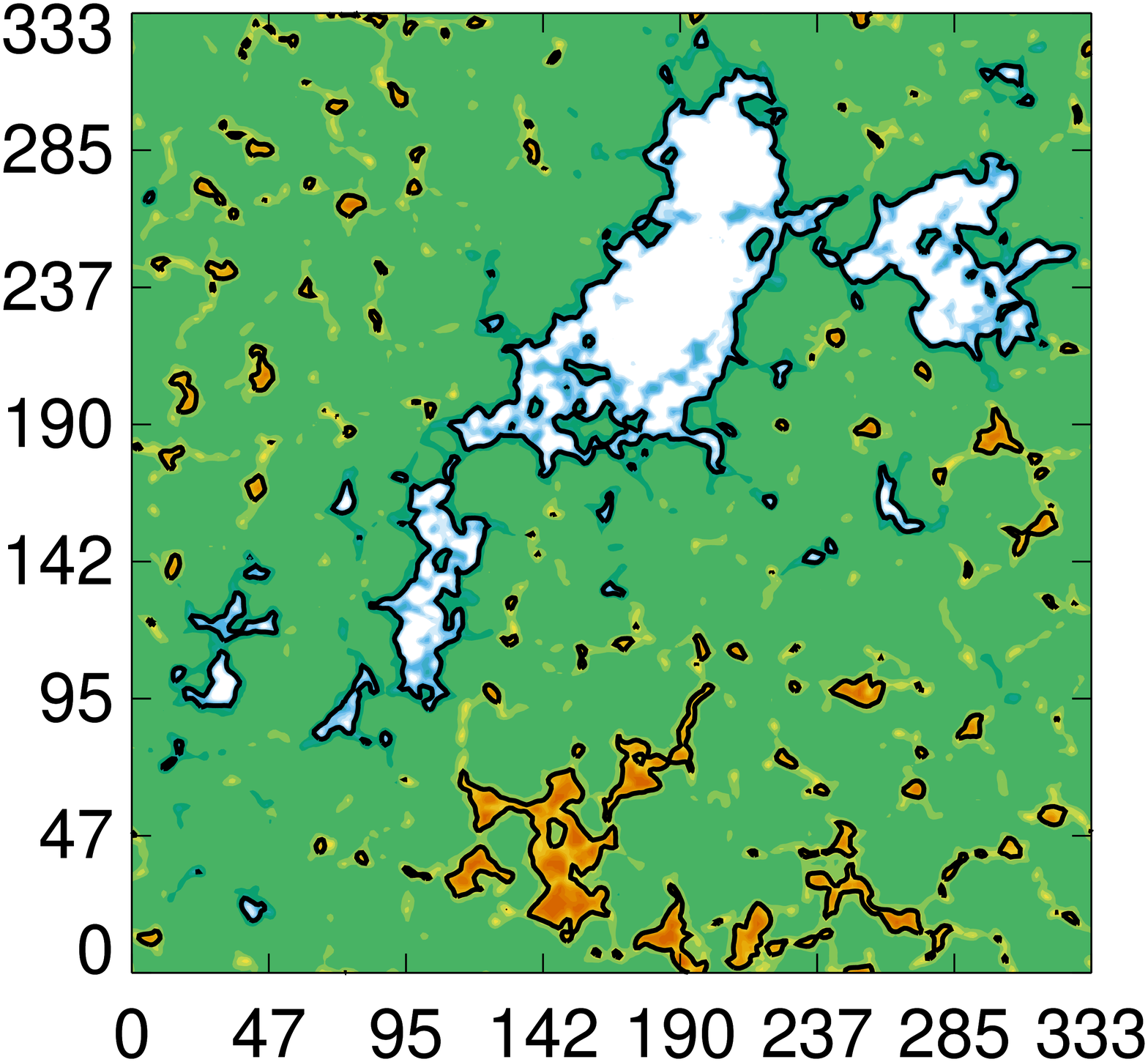}
 \includegraphics[width=\imsize]{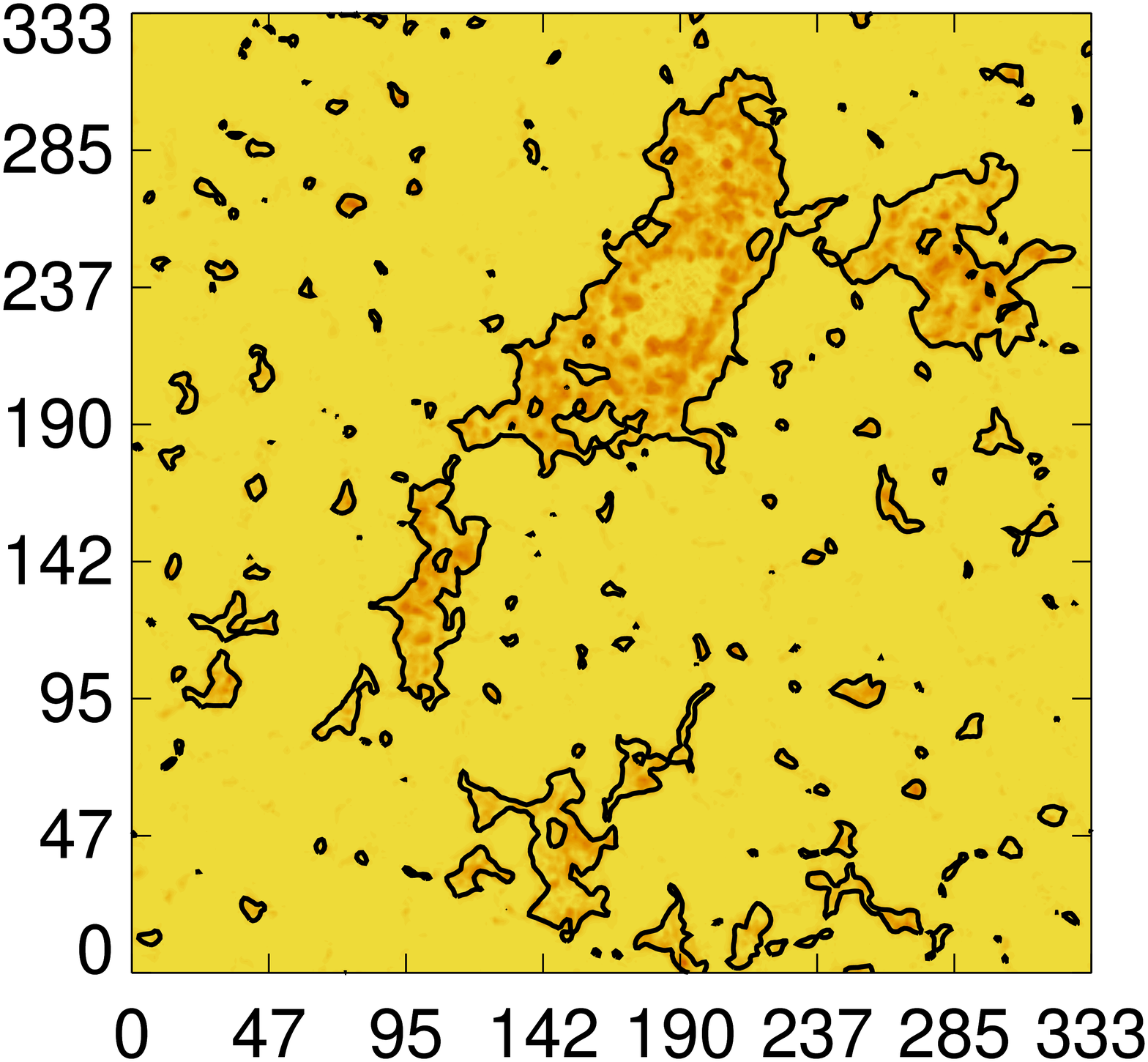}\vspace{-5mm}\\
 \raisebox{0.5\imsize}{$\gamma=10^\circ$}
 \includegraphics[width=\imsize]{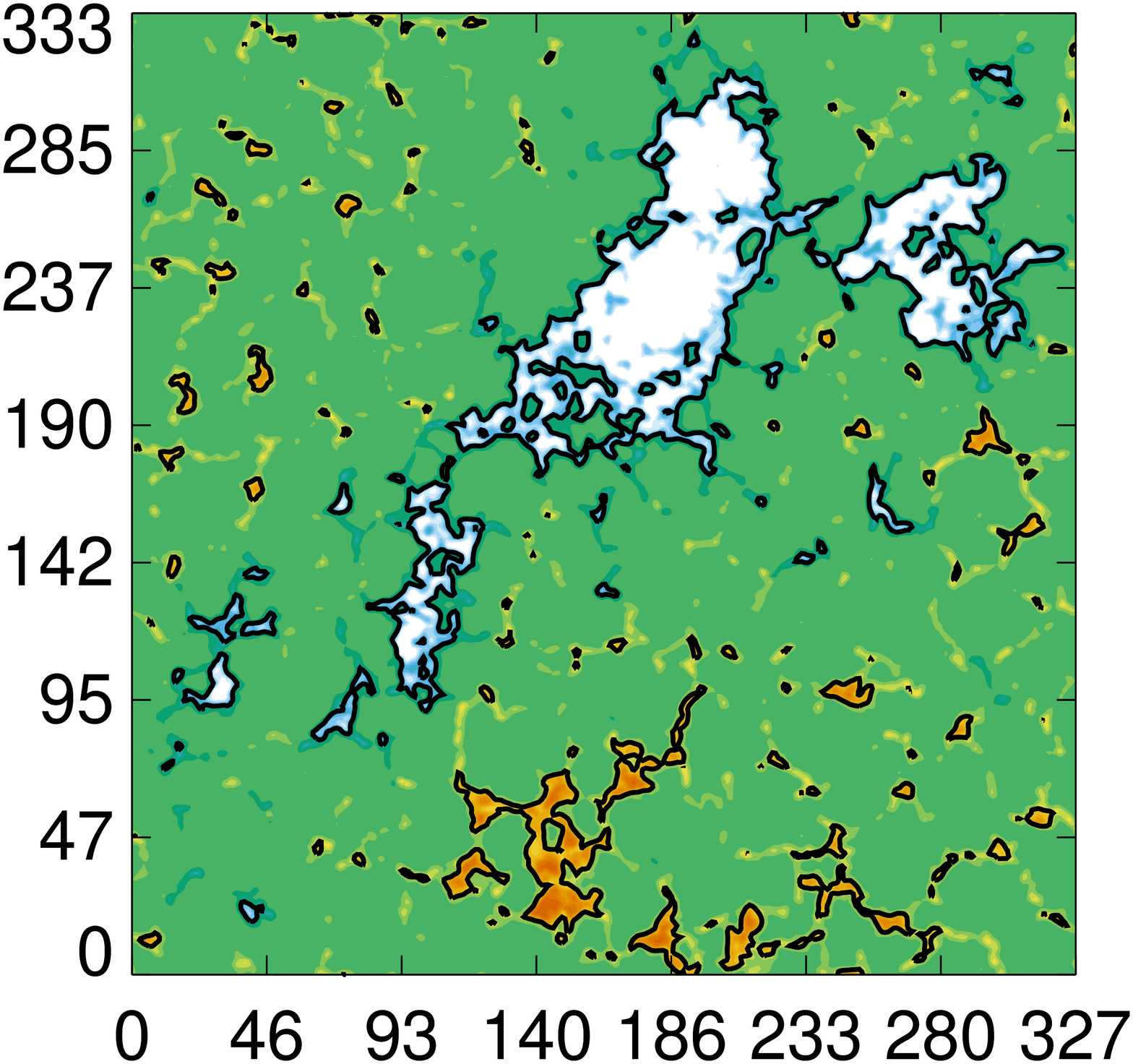}
 \includegraphics[width=\imsize]{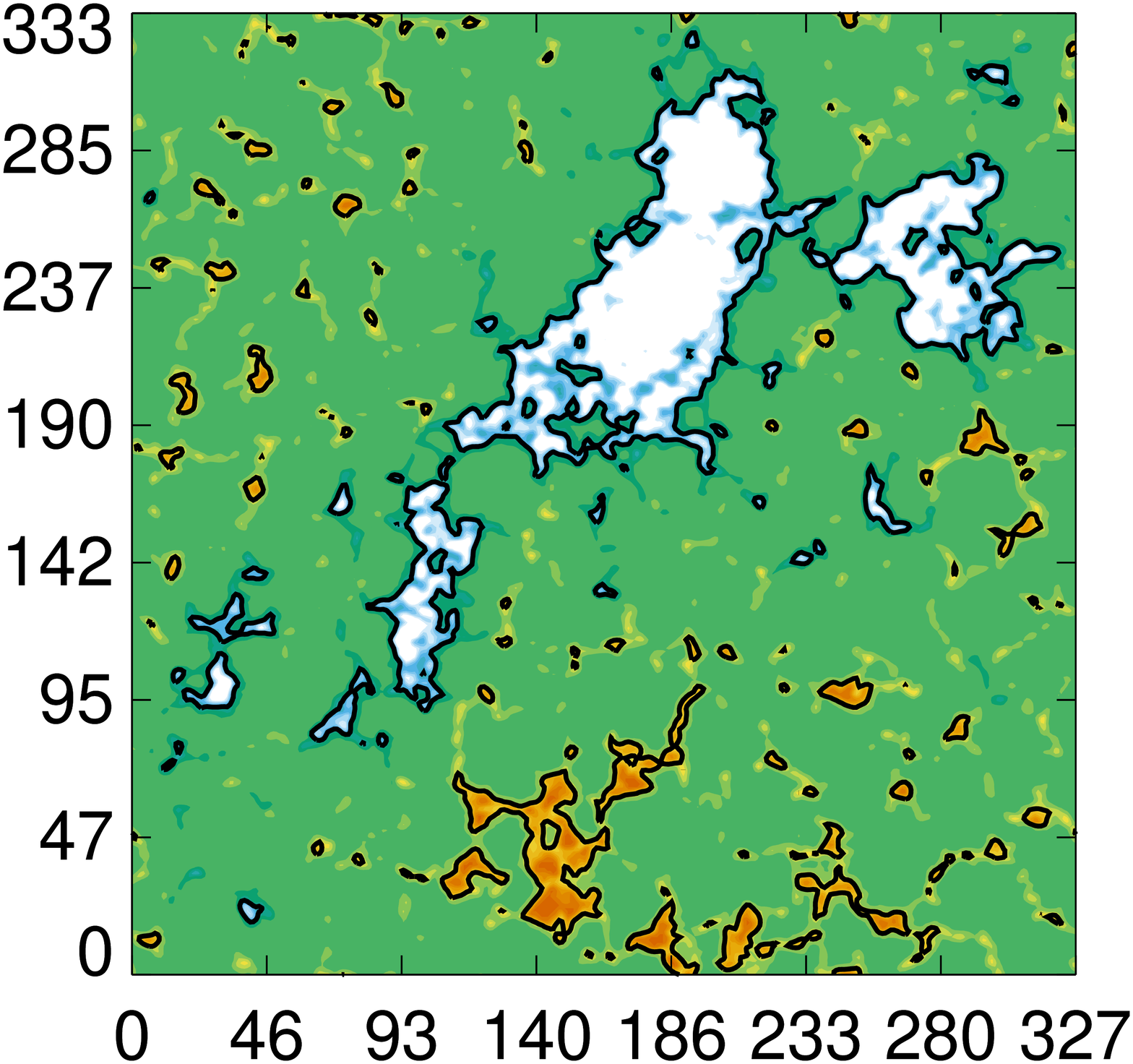}
 \includegraphics[width=\imsize]{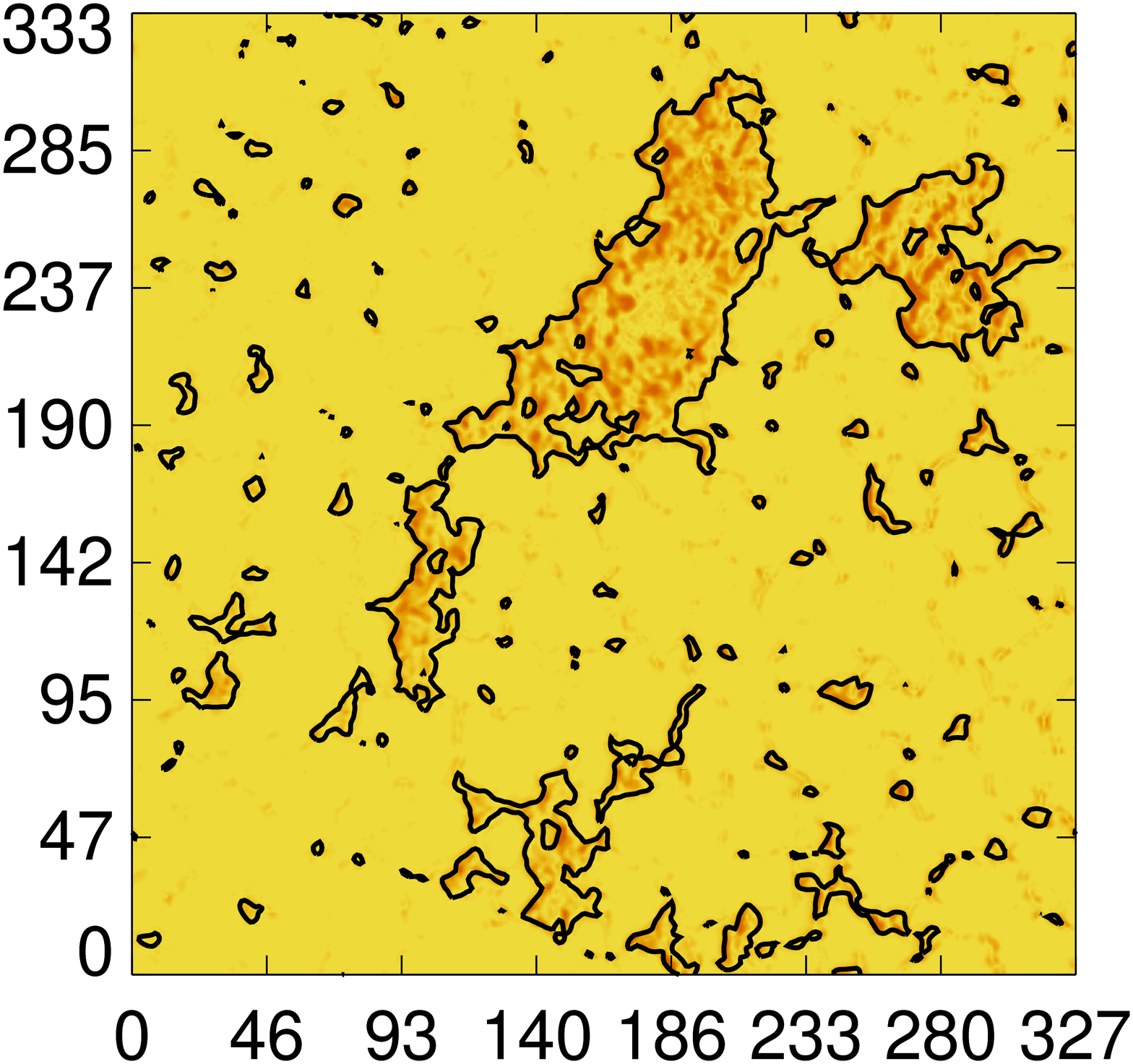}\vspace{-5mm}\\
 \raisebox{0.5\imsize}{$\gamma=20^\circ$}
 \includegraphics[width=\imsize]{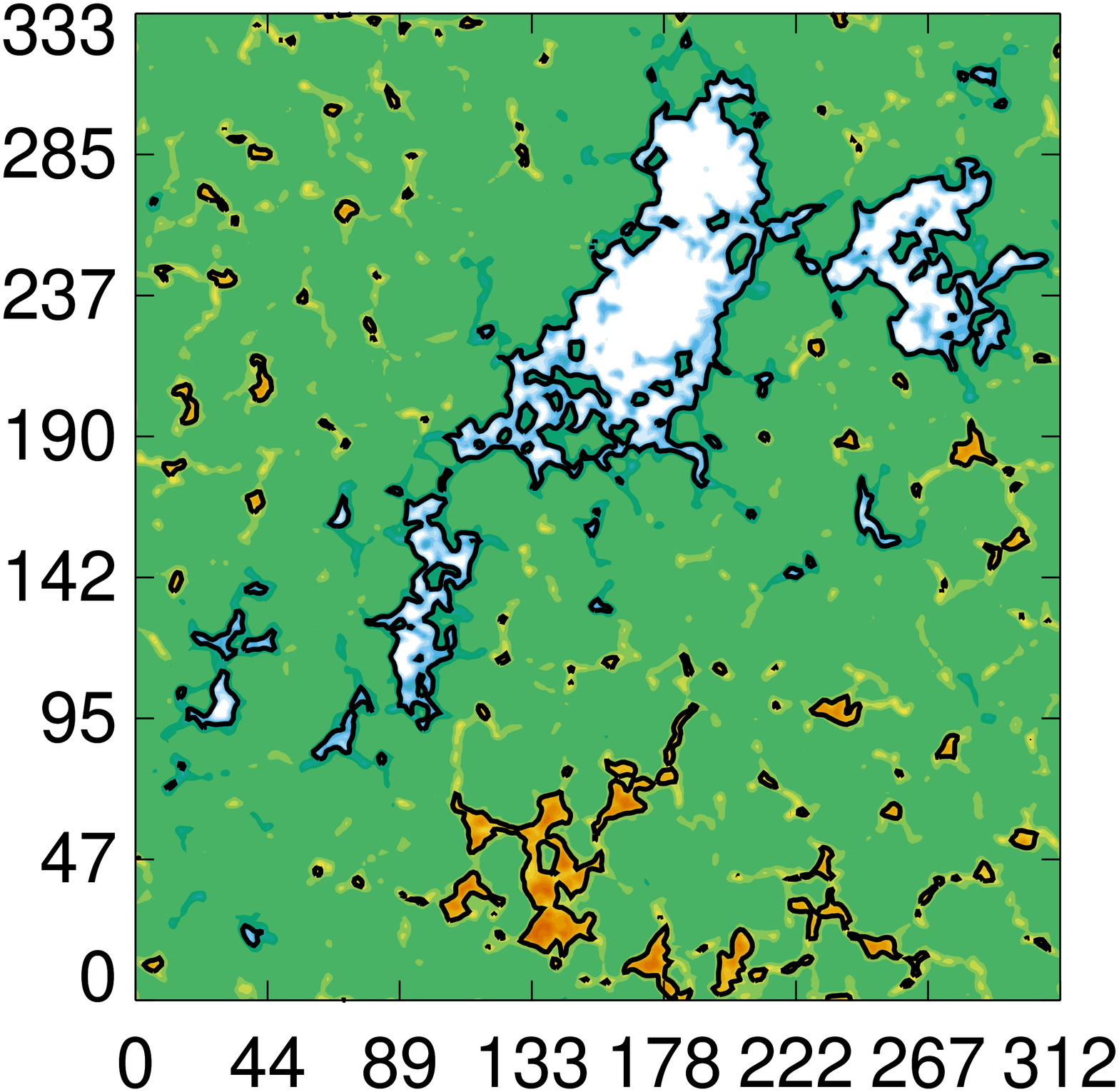}
 \includegraphics[width=\imsize]{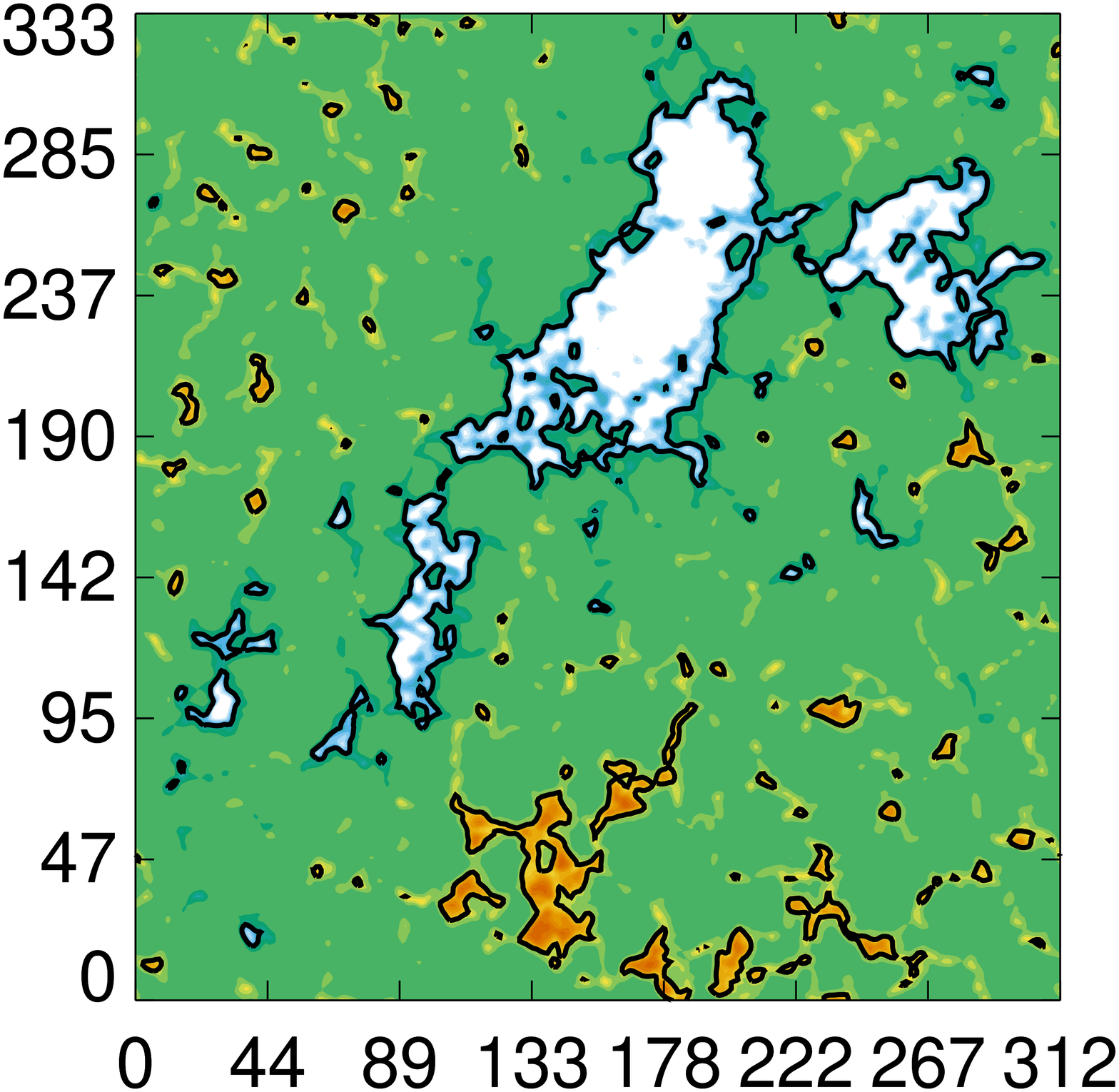}
 \includegraphics[width=\imsize]{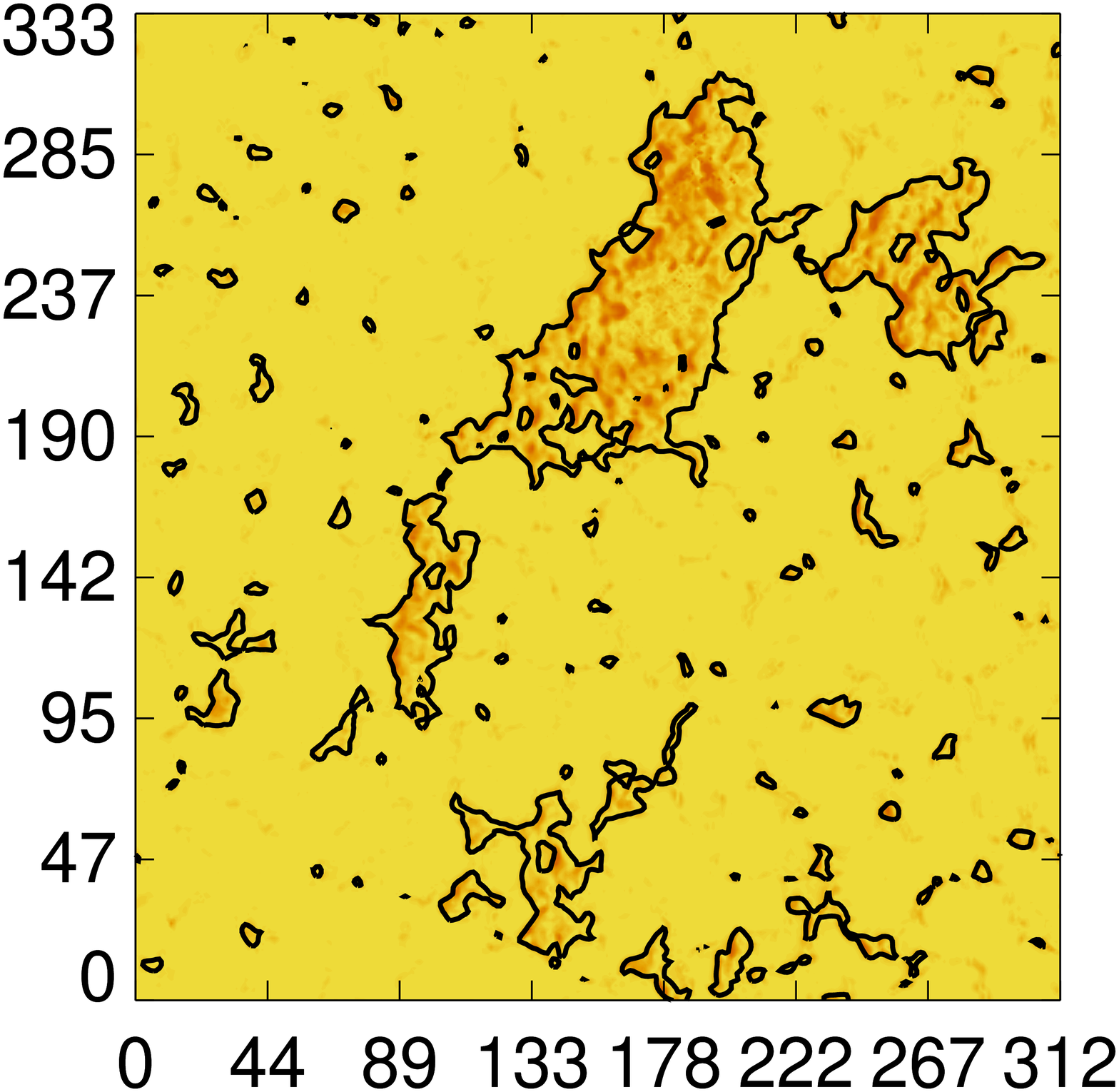}\vspace{-5mm}\\
 \raisebox{0.5\imsize}{$\gamma=30^\circ$}
 \includegraphics[width=\imsize]{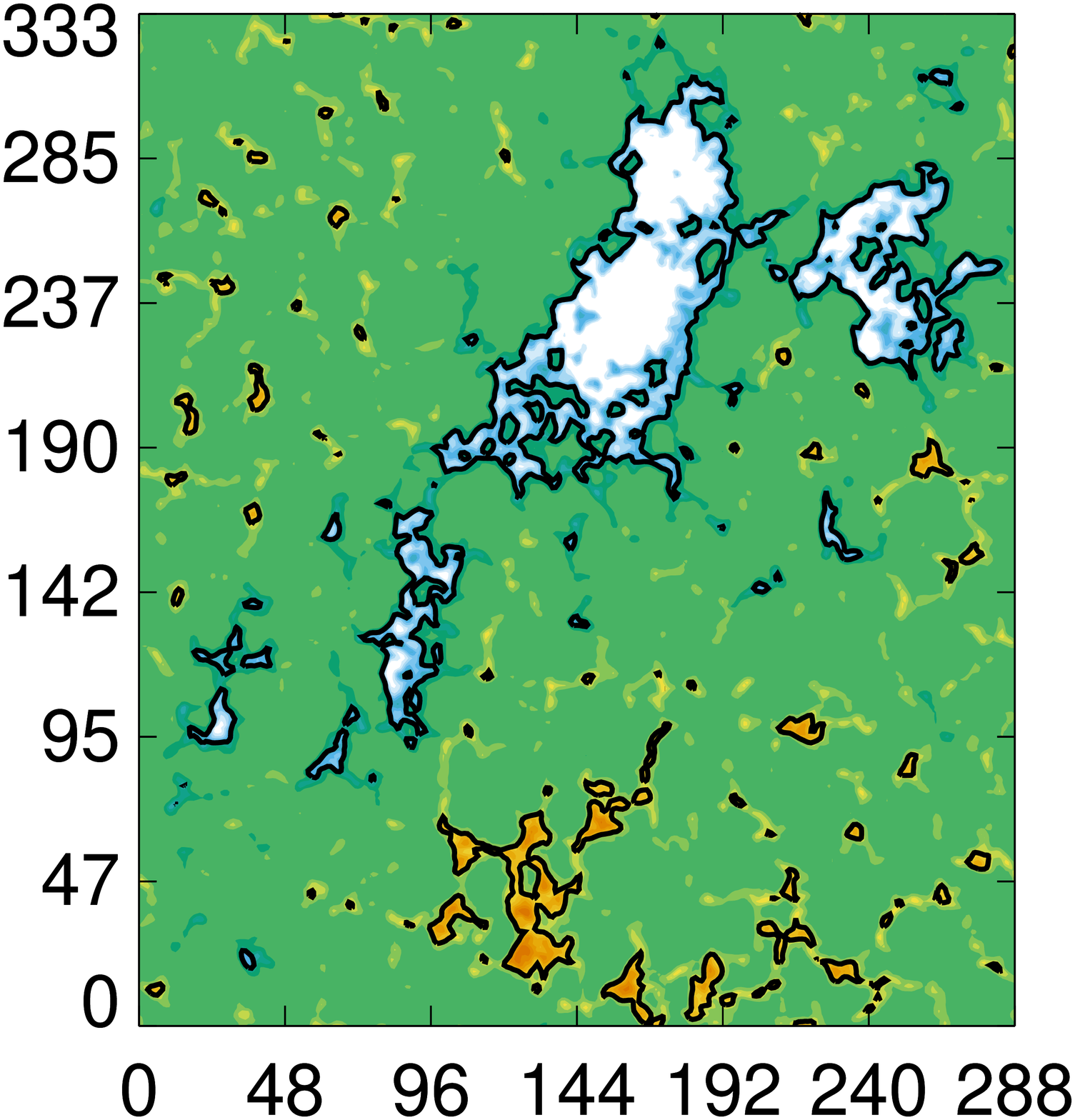}
 \includegraphics[width=\imsize]{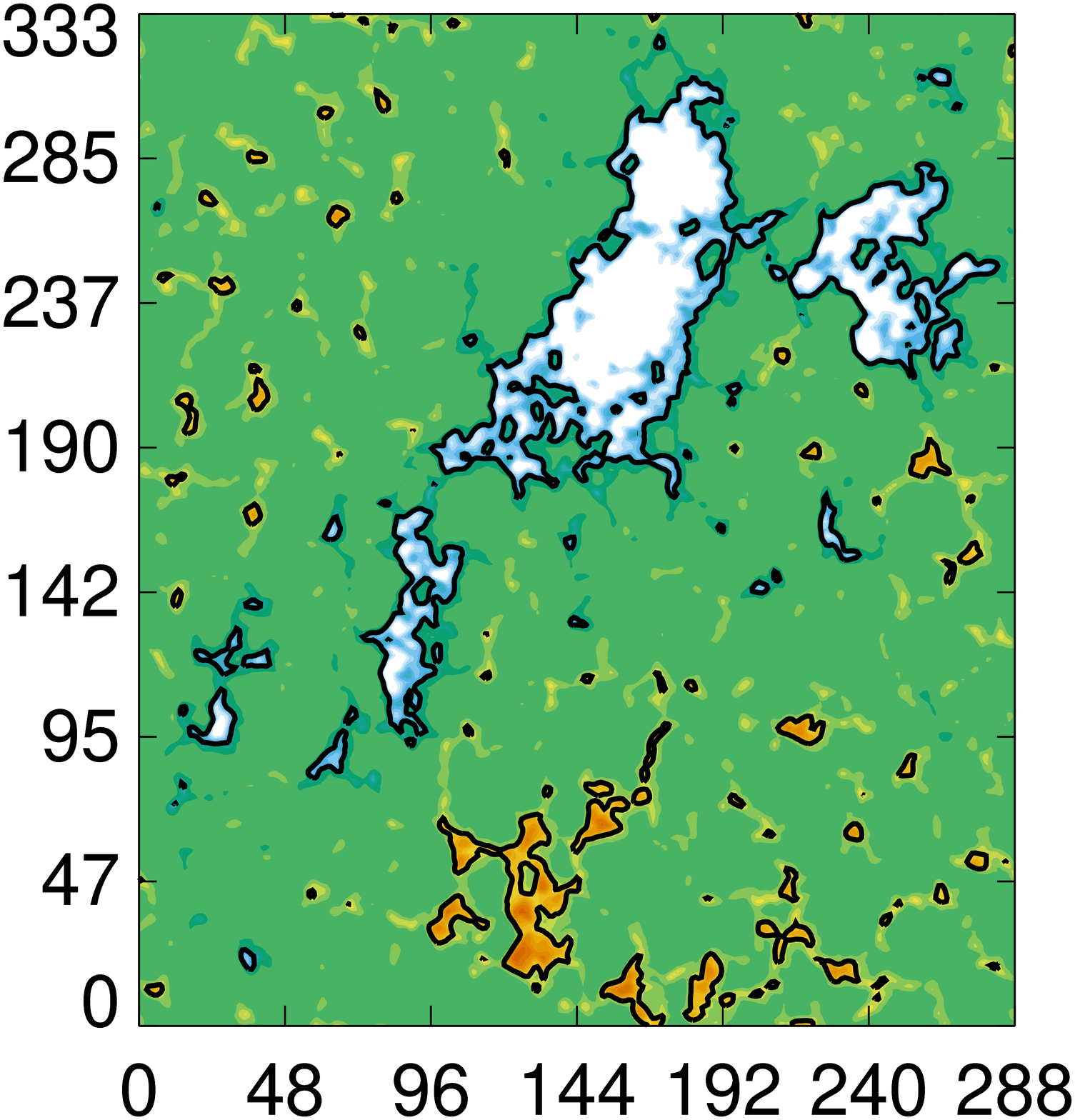}
 \includegraphics[width=\imsize]{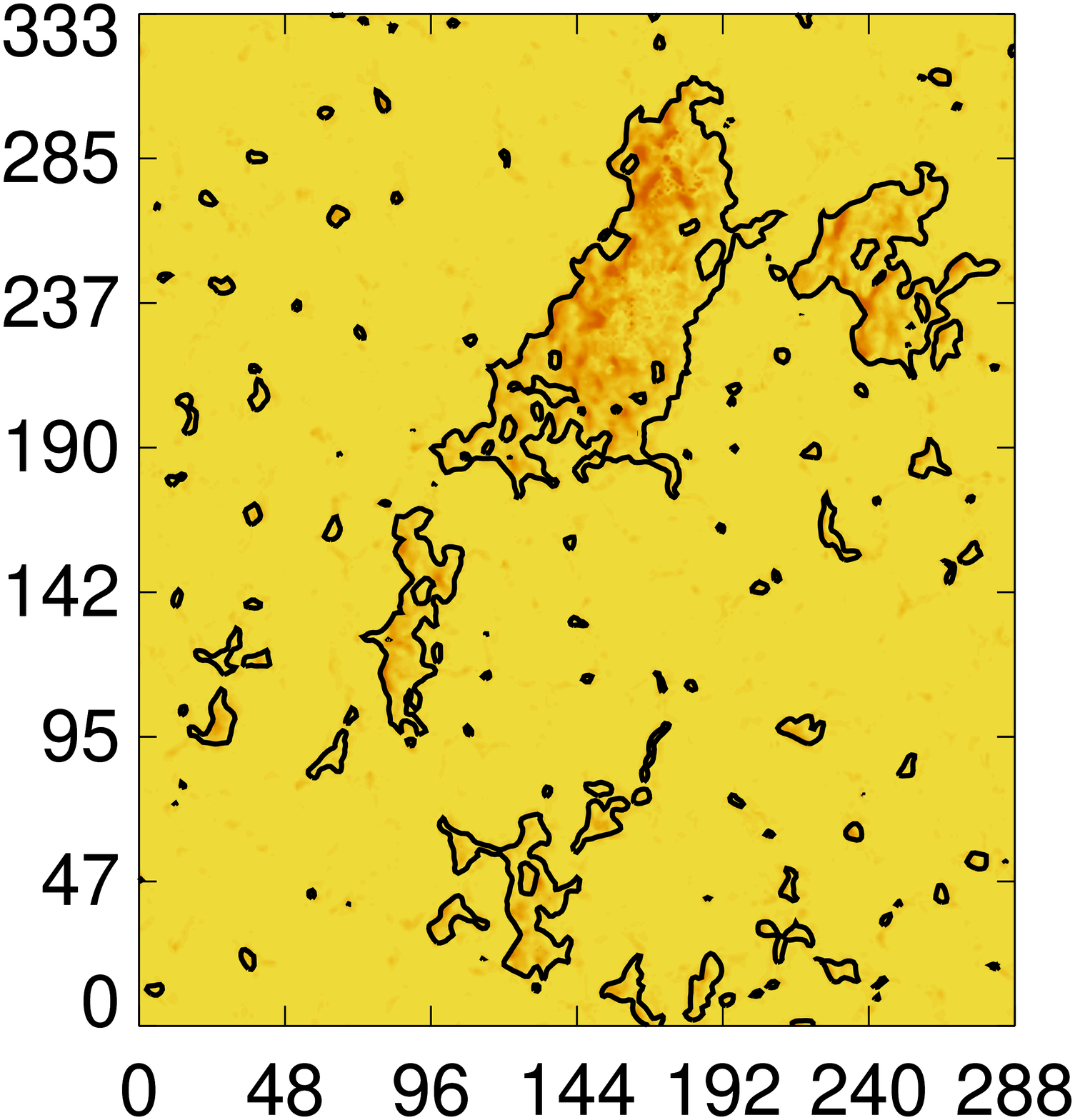}\vspace{-5mm}\\
 \raisebox{0.5\imsize}{$\gamma=40^\circ$}
 \includegraphics[width=\imsize]{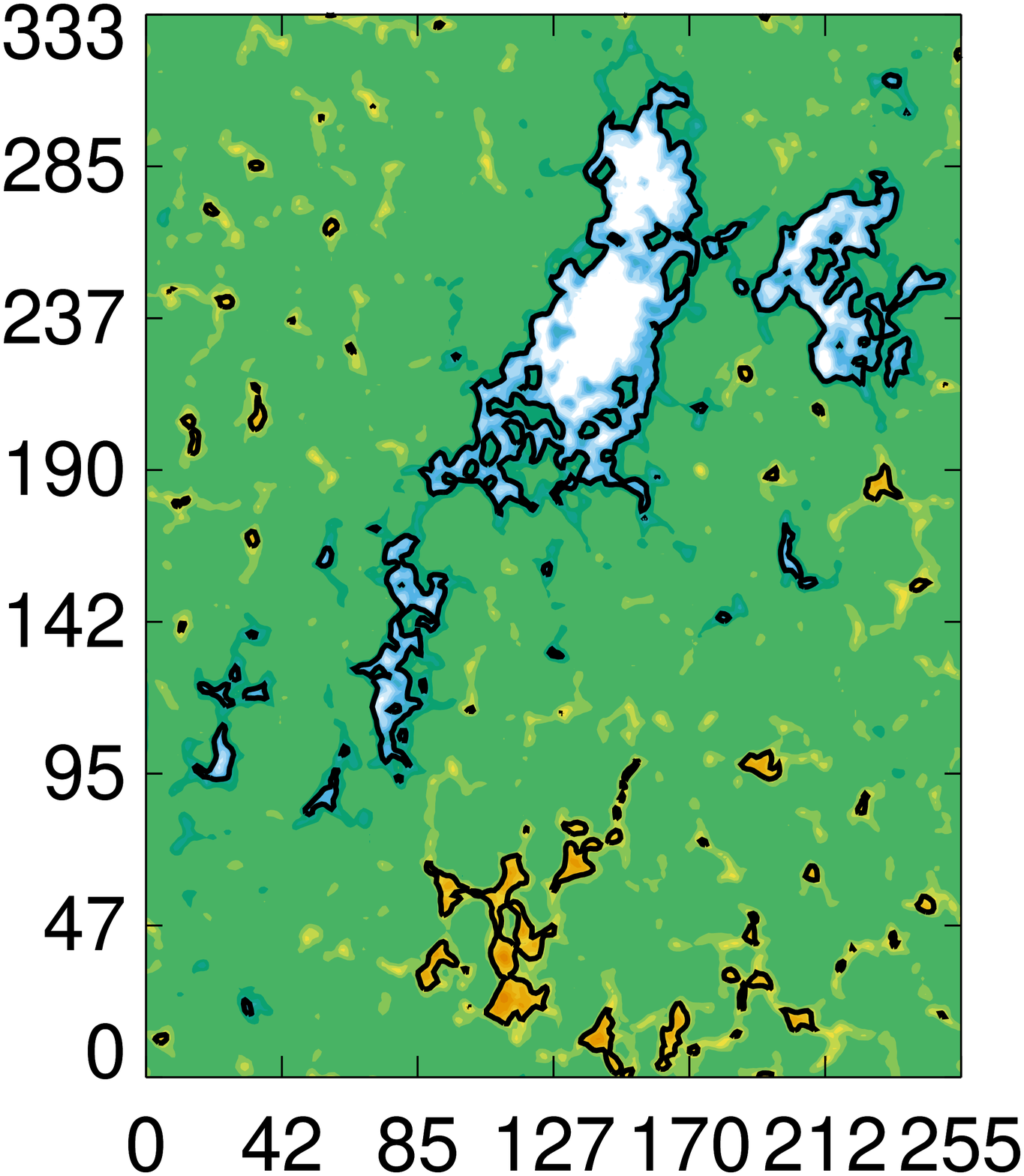}
 \includegraphics[width=\imsize]{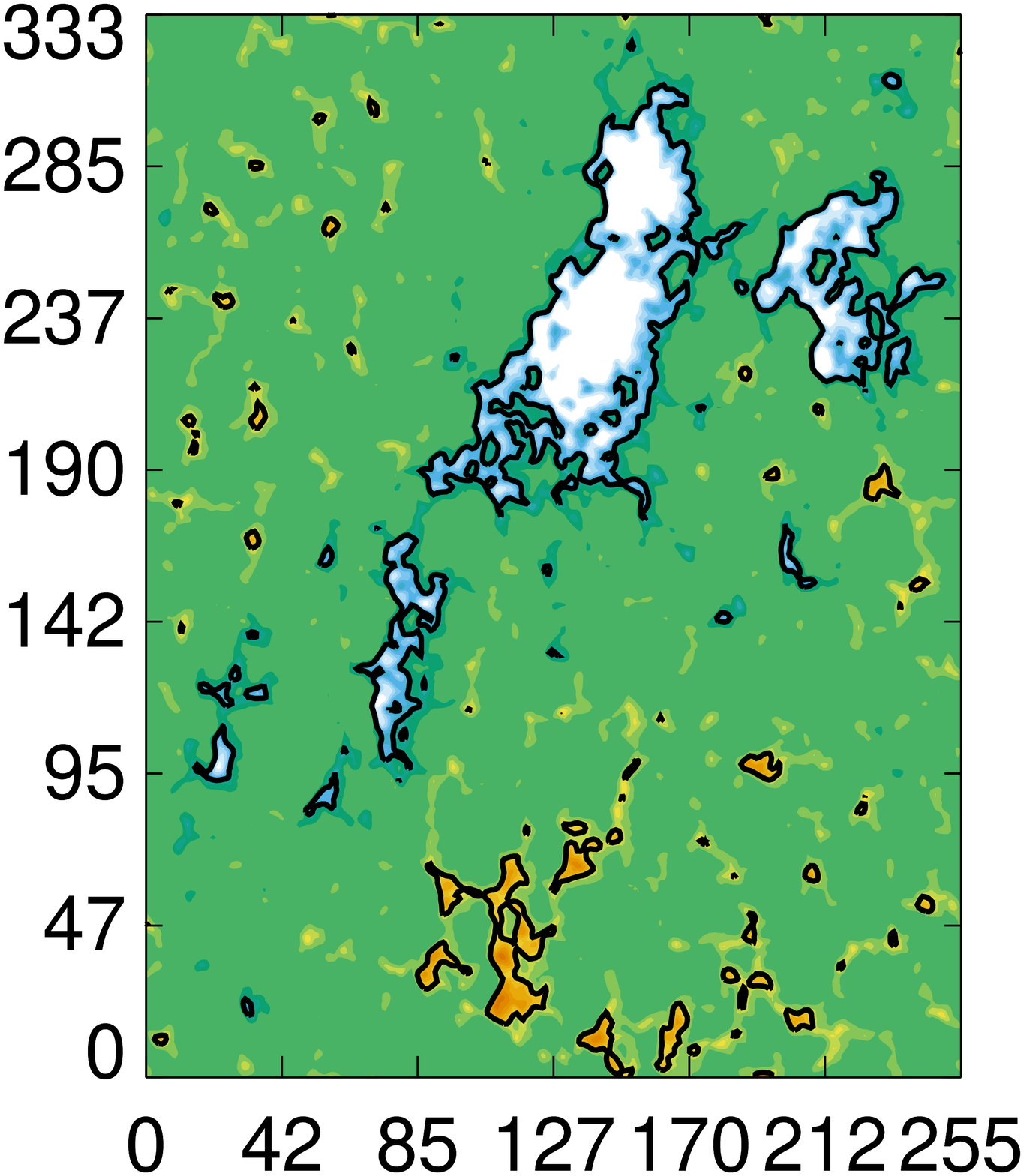}
 \includegraphics[width=\imsize]{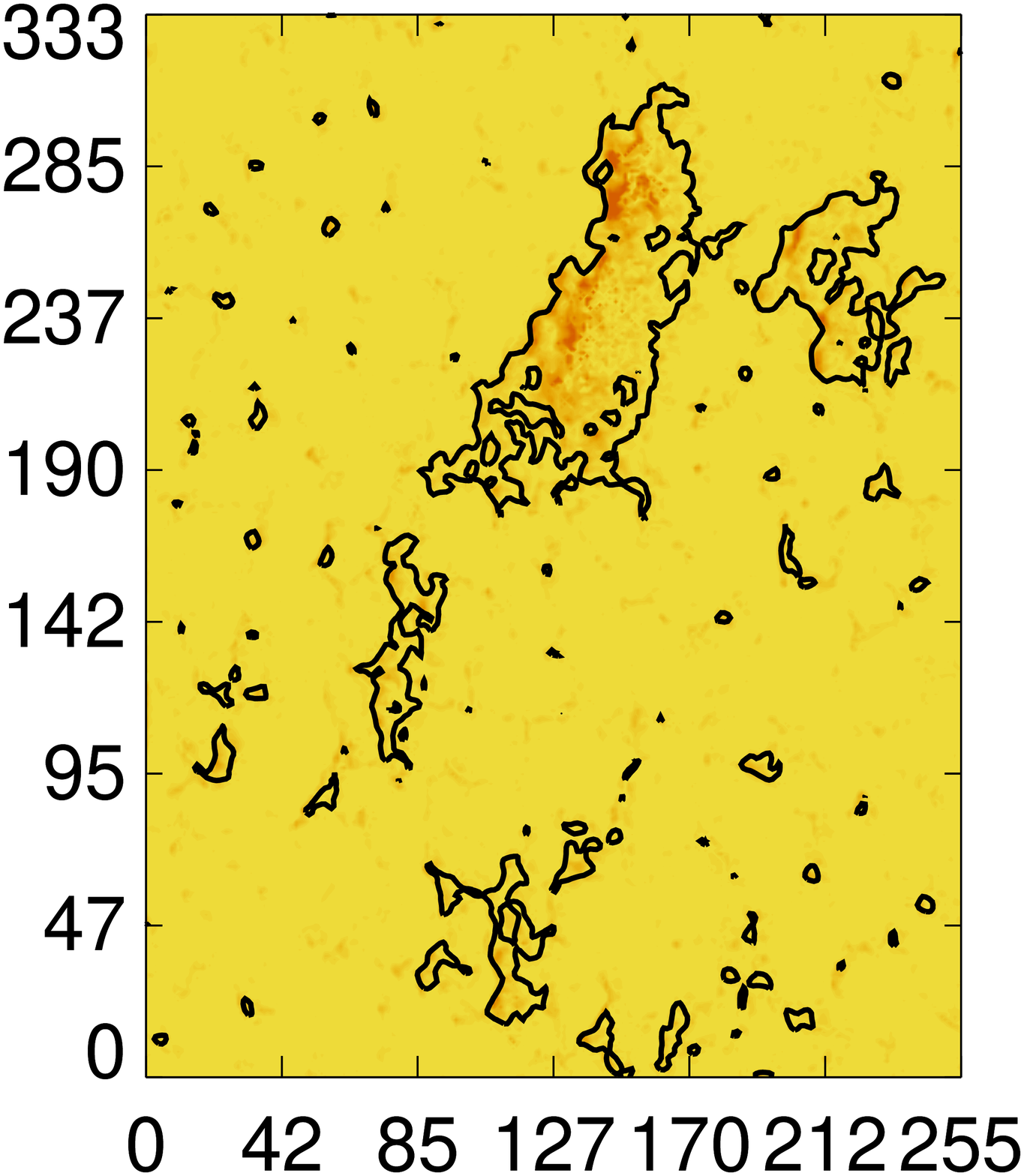}\vspace{-5mm}\\
 \raisebox{0.5\imsize}{$\gamma=50^\circ$}
 \includegraphics[width=\imsize]{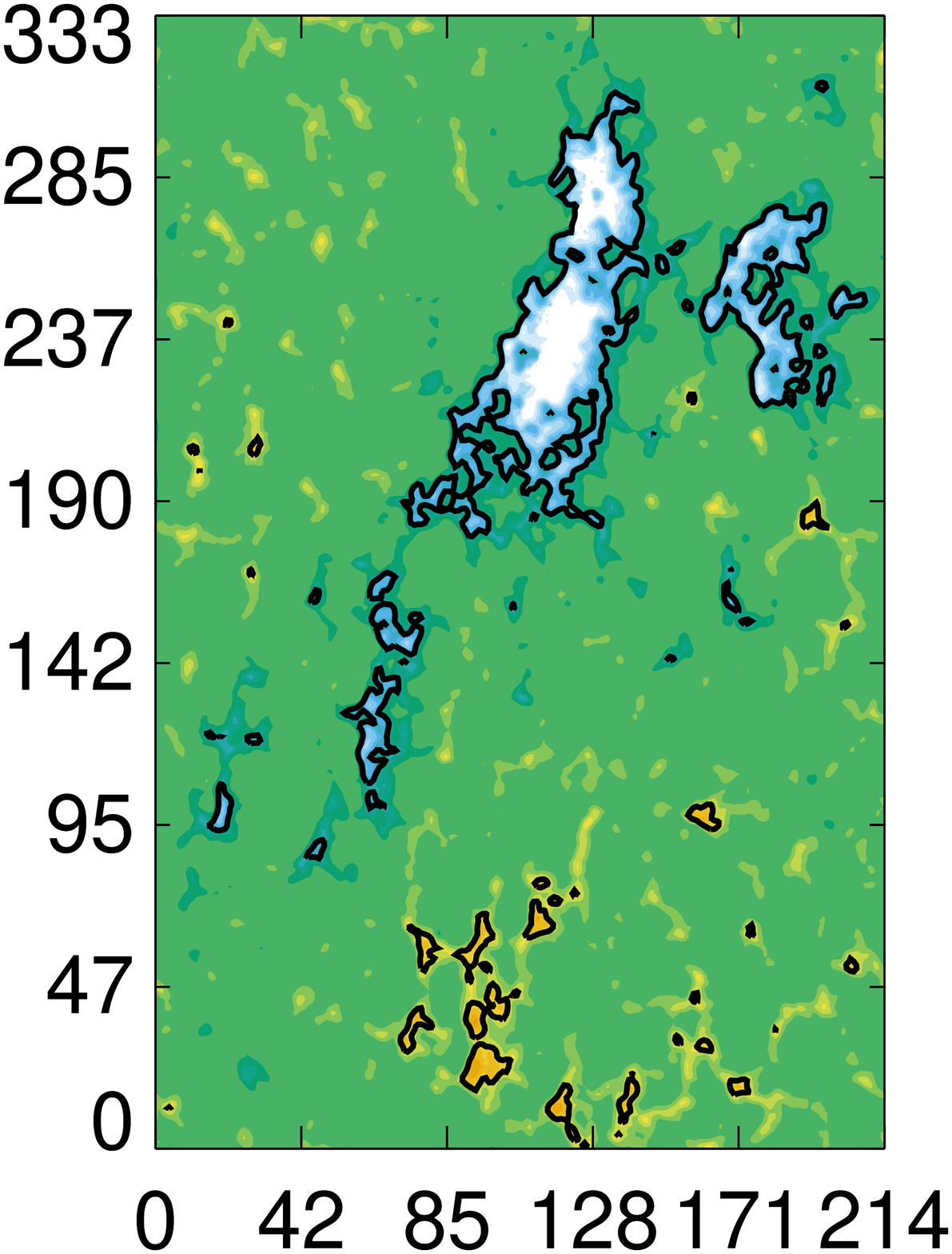}
 \includegraphics[width=\imsize]{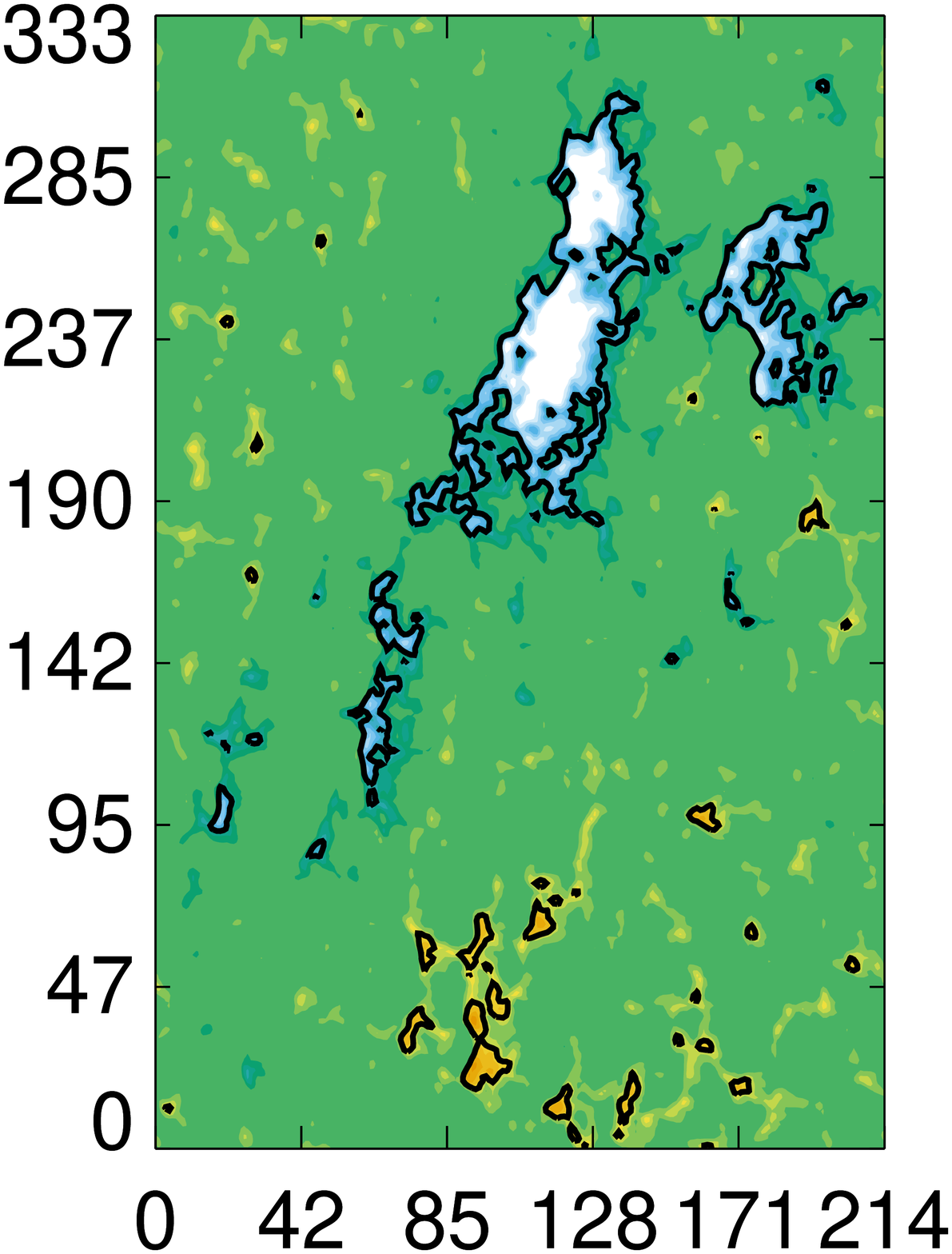}
 \includegraphics[width=\imsize]{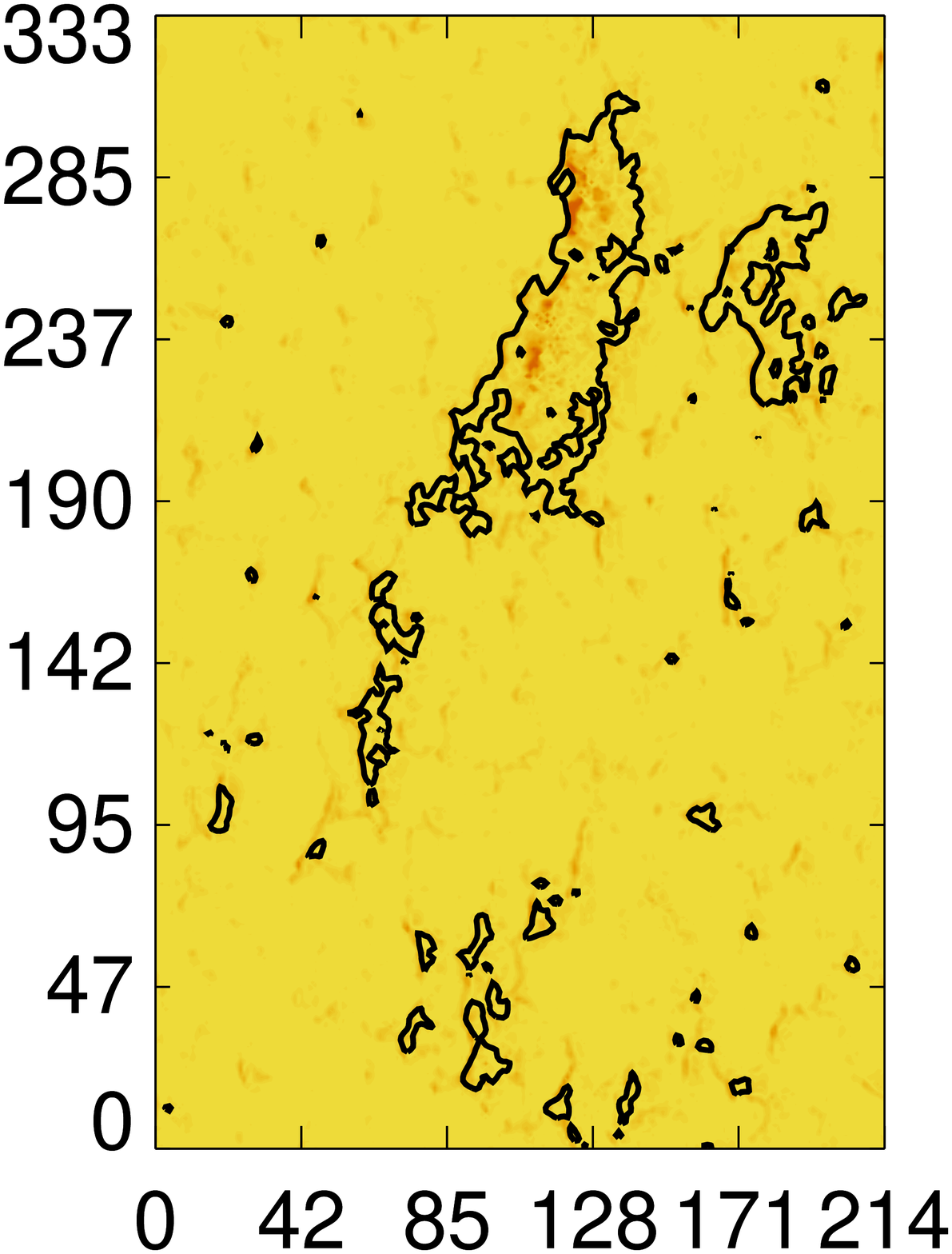}\vspace{-5mm}\\
 \raisebox{0.5\imsize}{$\gamma=60^\circ$}
 \includegraphics[width=\imsize]{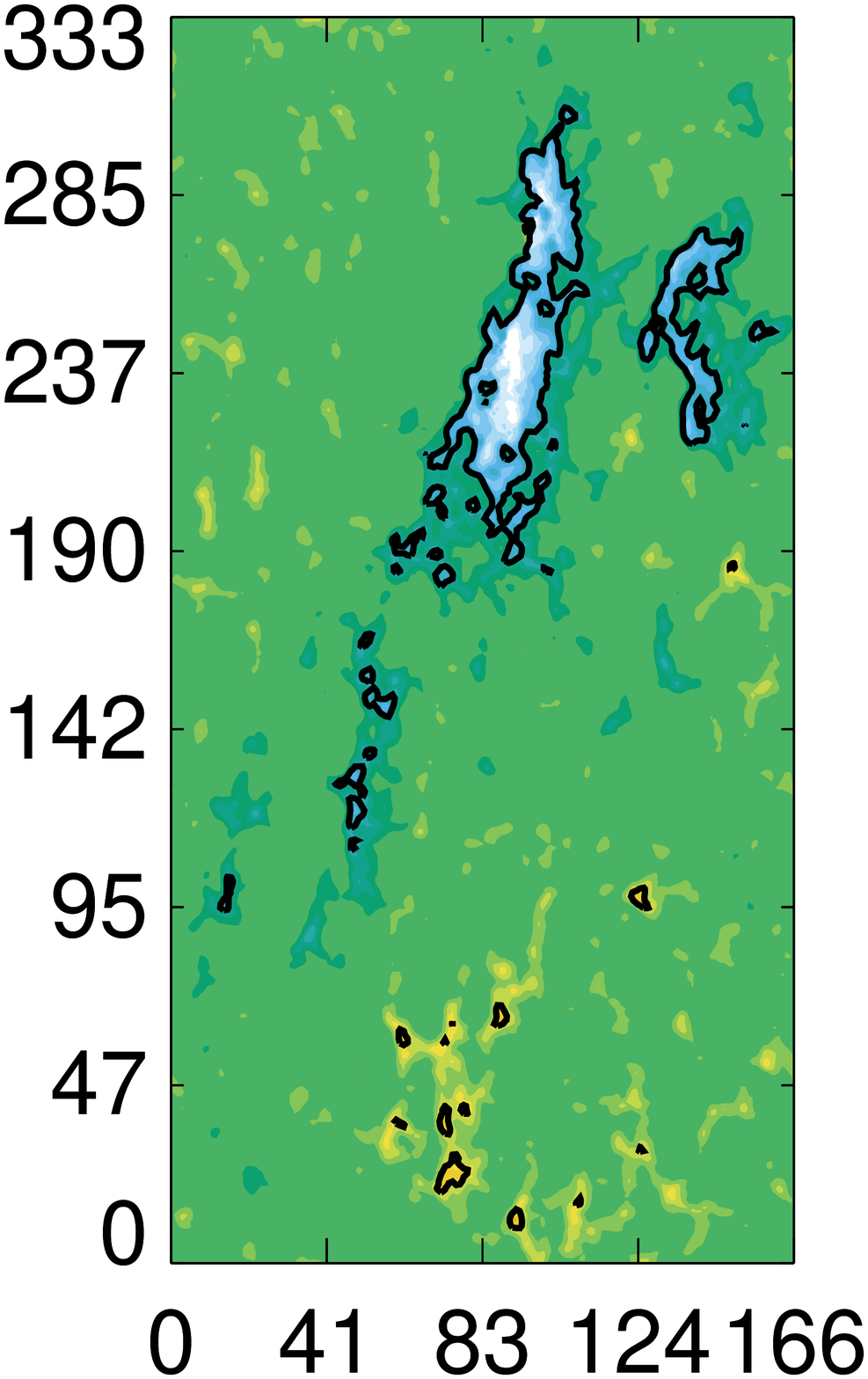}
 \includegraphics[width=\imsize]{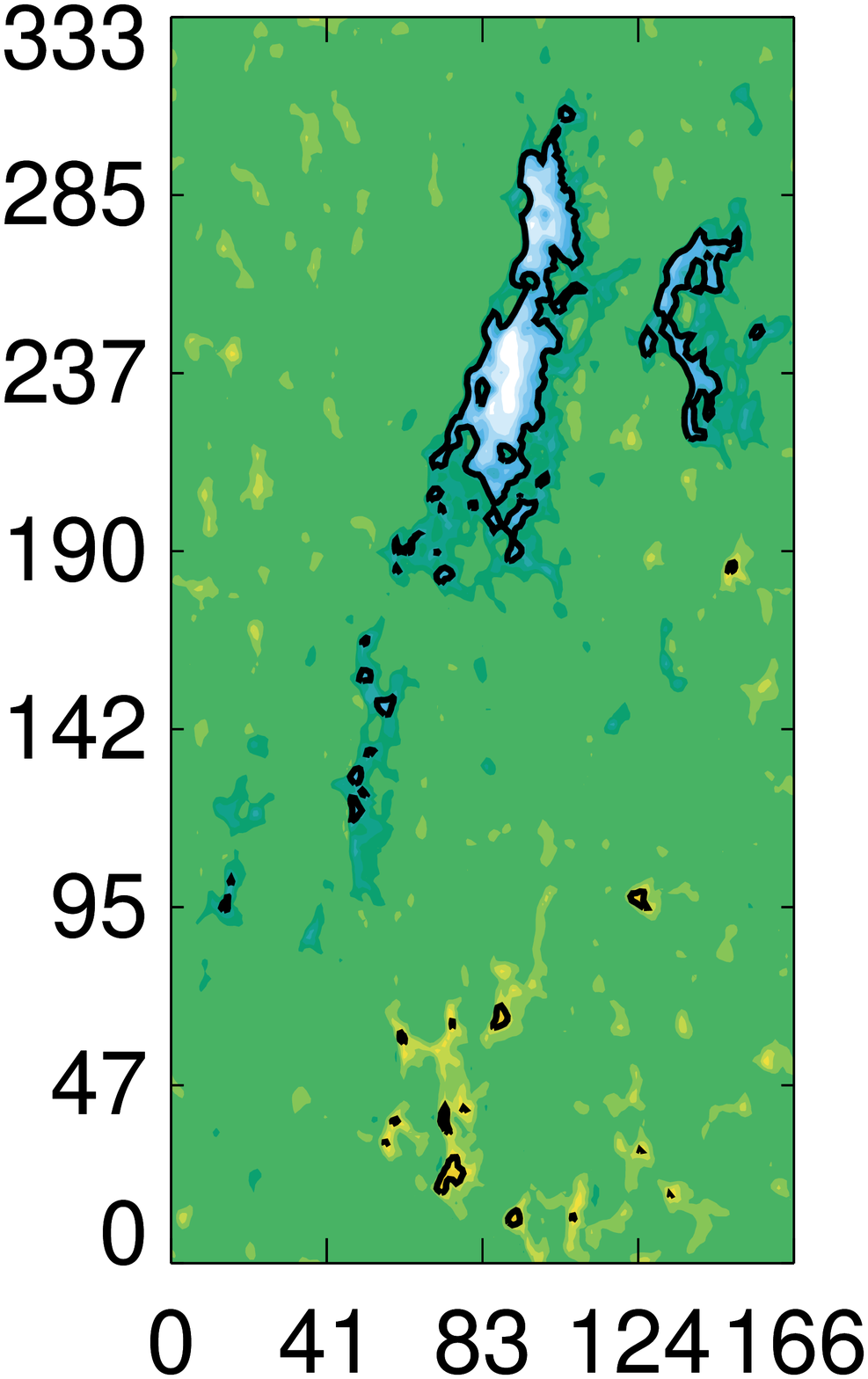}
 \includegraphics[width=\imsize]{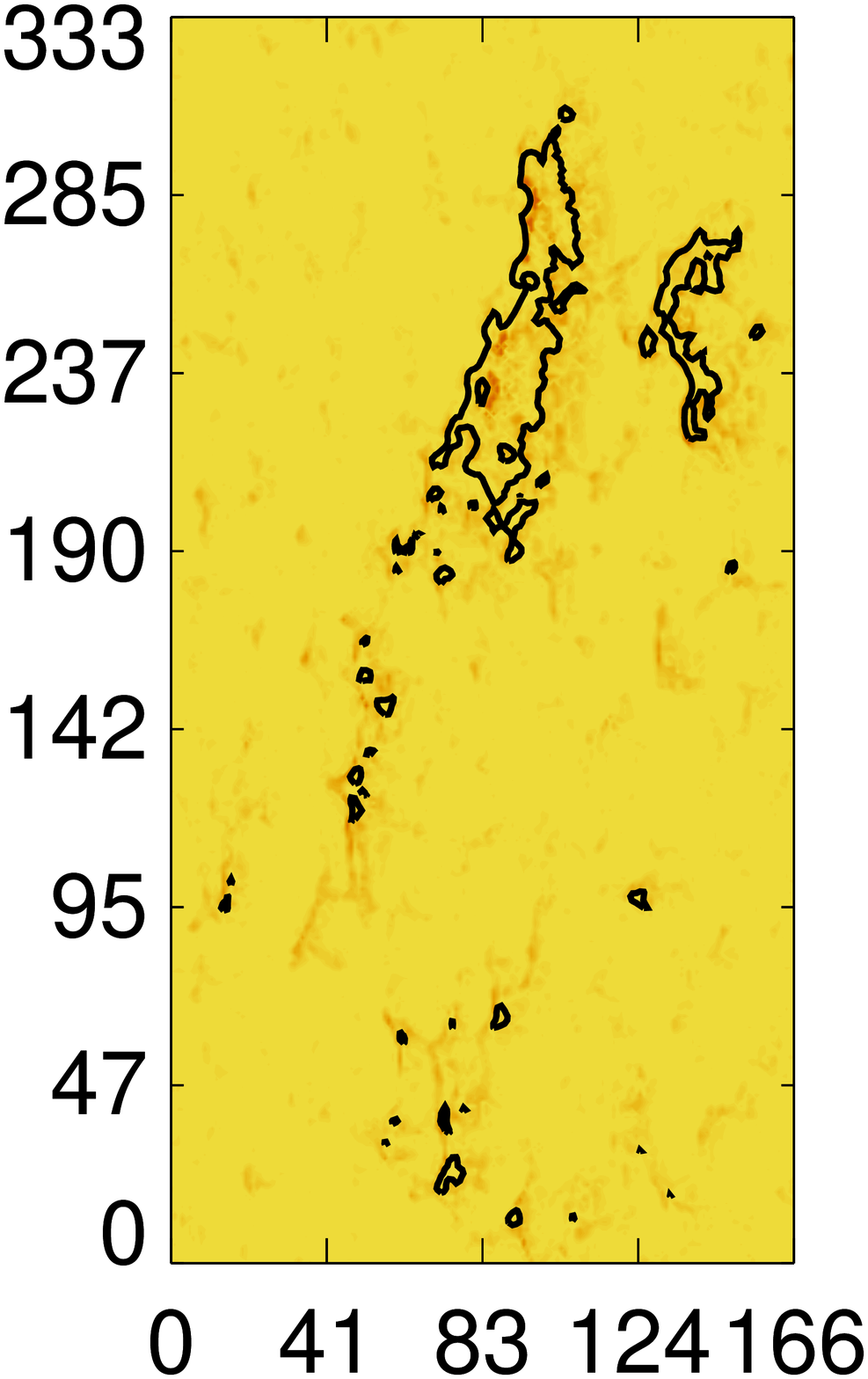}\vspace{-5mm}\\
 \raisebox{0.5\imsize}{$\gamma=70^\circ$}
 \includegraphics[width=\imsize]{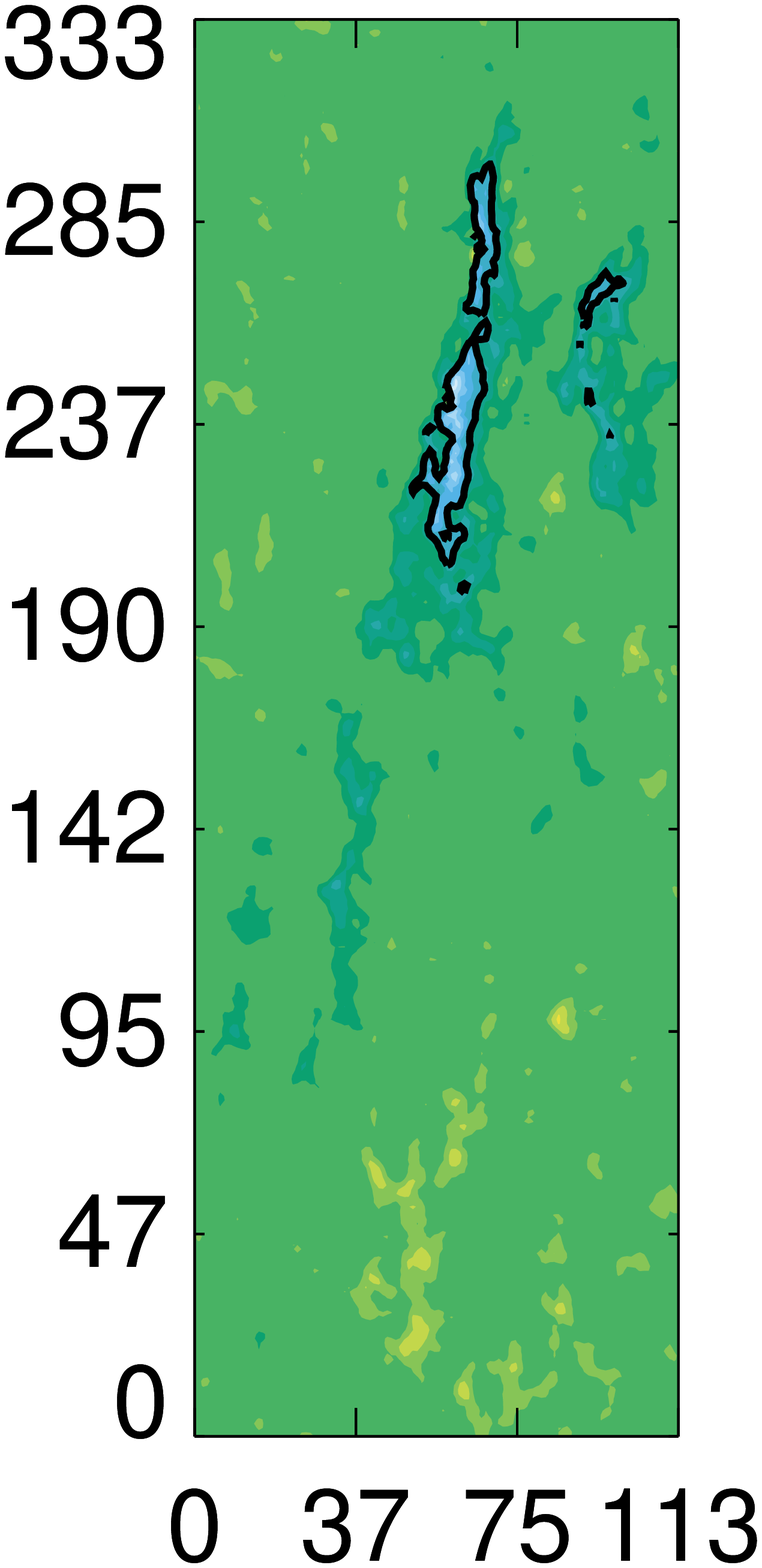}
 \includegraphics[width=\imsize]{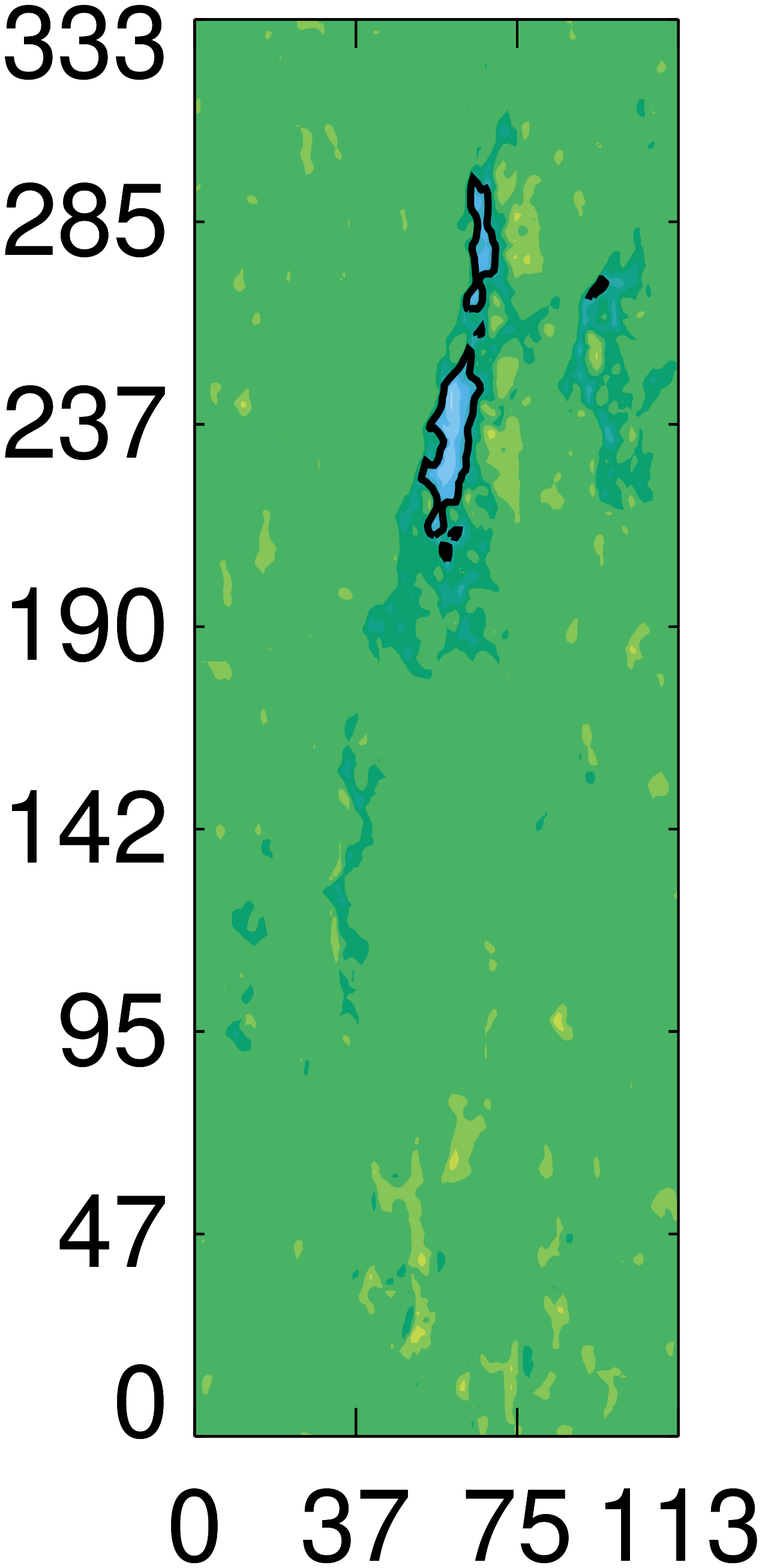}
 \includegraphics[width=\imsize]{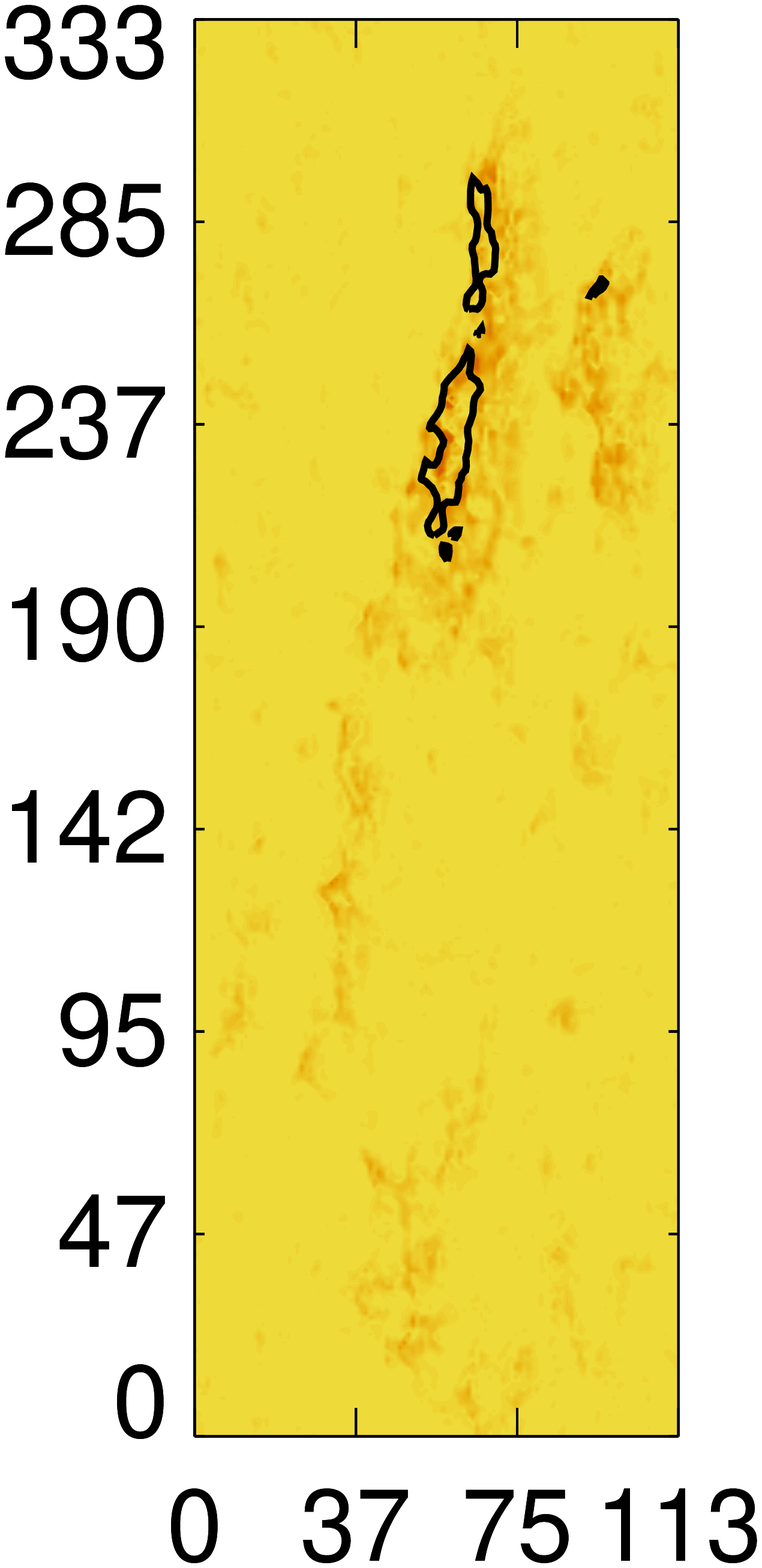}
 \caption{Isolines of the LoS-field component in the synthetic magnetogram (\textit{left column}) and the reconstructed magnetogram (\textit{central column}) saturated at $\pm3000$\,G, and the absolute value of their difference (\textit{right column}) saturated at 500\,G, for viewing angle, from top to bottom, $\gamma=[0,10,20,30,40,50,60,70]$ degrees.
     In all plots the \textit{blue contours} are the $\pm$500\,G-isoline, and axis units are in pixels.}
\label{f:spinor}
\end{figure}

The numerical model used in this test is the full 3D-{\Muram} simulation of a pore described in \sect{muram}.
We apply multiple steps of processing to this data set to produce a \HRT{}-like observation and a direct reconstruction from the contributing layers in the 3D-simulation for $\gamma\in [0, 10, 20, 30, 40, 50, 60, 70]$ degrees. 
The original resolution of the {\Muram} simulation is reduced by about 60\,\% 
resulting in a dimension of $324\times 324$\,pixels 
at $\gamma=0^\circ$.
Additionally, all three relevant resolutions (native of the model test [$\Dnat$], 
the {\HMI} \corrter{spatial sampling} [$\DHMI$], and the {\PHI} \corrter{spatial sampling} [$\DHMI$]) 
are taken to be all equal to 0.5\,arcsec, hence $\rD=1$. 
This implies that the test simulates the use of the HRT telescope when \SolO{} is at 1\,AU distance from the Sun. 
These choices, which are not optimal for the application of \SDM{} and not really representative of the typical case of application, were nevertheless necessary in order to limit the numerical efforts required by the reconstruction method described below. 
Because of the above limitations, the results in this section are not entirely  comparable with those in \sect{results_geo}, as we were forced to choose different parameters  from those in \tab{reference}. 
\corrbis{Hence, in order to limit the modifications to the model test case, no normalization of the field amplitude is adopted here.}

As a matter of fact, different types of magnetograms can be constructed from the \Muram{} simulations. 
We discuss these options in the following, and refer the reader to \tab{mgm} for a quick view of their essential properties and reference to relevant sections in the article.  

\subsubsection{Synthetic Magnetograms}\label{s:synth_mgm}
In the standard procedure for creating a synthetic \HRT{} magnetogram, the {\Muram} simulation data are used as input to the forward calculation mode of the SPINOR radiative-transfer inversion code \citep[see][]{Solanki1987,Frutiger2000,2000A&A...358.1109F} to compute synthetic Stokes spectra around the 
Fe\,{\sci}\,617.3\,nm 
line as observed by \HRT{}, \corr{see the blue box in \fig{flow_rec}}. 
The resulting spectra are then used as input to the SO/PHI Software siMulator \citep[\sophism{}; see][]{Blanco_Rodr_guez_2018} to produce {\HRT}-like observations including an inversion of the degraded spectra adopting the Milne--Eddington approximation as performed onboard SO/PHI. 
This operation results in a synthetic (ambiguous) magnetogram of the simulated region that contains fields on all scales down to the resolution limit and with strength up to 3500\,G. 
The LoS of these magnetograms at different viewing angles 
are shown in the left column of \fig{spinor}.

\SDM{} can then be applied to the synthetic \HRT{} magnetograms obtained in this way.
However, \corr{since the synthetic magnetograms are themselves ambiguous, such an application is not quite a test of the \SDM{}; since the real orientation of the transverse field is not known, the answer provided by \SDM{} cannot be validated.}
\corr{Instead, in}
order to assess the success of {\SDM} in resolving the ambiguity, one needs 
\corrbis{to construct a dataset best representing the expected correct result \citep{Leka2009}.
In order to obtain that, one needs}
to know how the transverse field that modulated the synthetic emission is truly oriented in each pixel of the image plane, for any angle $\gamma$.
To obtain such information requires more than simply simulating the {\HRT} observation \corr{of}
a 3D-simulation of the solar photosphere \corr{as done for the production of synthetic magnetograms}, as it implies tracing back the origin of the optical signal in the simulation, for any given $\gamma$, and reconstructing the orientation of the transverse field. 

\subsubsection{Reconstructed Magnetograms}\label{s:rec_mgm}

\begin{figure}
 \setlength{\imsize}{\columnwidth}
    \centering
    \includegraphics[width=\imsize]{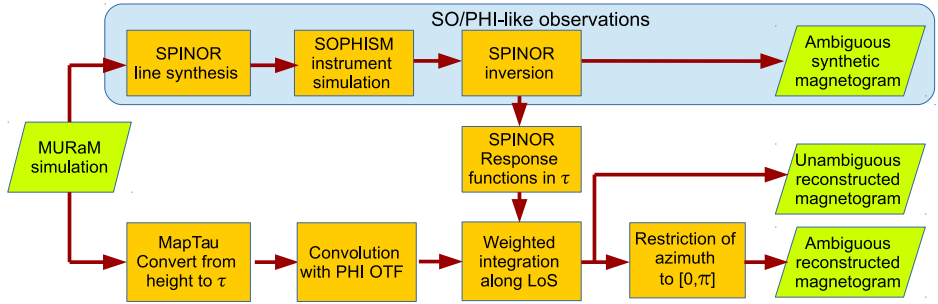}
    \caption{\corrbis{Flow-diagram of the construction of synthetic magnetograms (\textit{blue box} only; see \sect{synth_mgm}) and of the reconstructed magnetograms; see \sect{rec_mgm}. The procedure is applied on the image plane of each telescope for the given observing configuration specified by $\gamma$ and $\rD$. Input/output are visualized with \textit{green parallelograms}, operations with \textit{orange rectangles}.}}
    \label{f:flow_rec}
\end{figure}
Therefore, in addition to the synthetic \PHI{} observations, \corr{non-ambiguous}, 
vector magnetograms describing the solar scene  by the \Muram{} simulations as seen from different viewing angles $\gamma$ were produced\corr{, see \fig{flow_rec} for a flow diagram of the procedure described below}.

These data contain the reconstructed contribution of each cell along each ray through the {\Muram} cube. 
We compute these contributions from magnetic-response functions \citep{Landi1977,RuizCobo1994} as a function of the optical depth, which we use as a weighting function to extract and integrate over the relevant cells from the {\Muram} data. 
These magnetic-response functions are computed from the atmospheric model that SPINOR creates for an inversion of the {\HRT} like observations, \corr{as described in \sect{synth_mgm}}. 
Before they can be used to access the magnetic-field data in the {\Muram} cube, they must be transformed such that they depend on the geometrical height rather than on the optical depth.
This was achieved with the MapTau subroutine of SPINOR, which emulates the radiative transport through the {\Muram} simulation for each viewing angle [$\gamma$]. 
\corrbis{
From the temperature stratifications in the \Muram{} atmospheres, all other thermodynamic quantities such as the gas pressure, the electron pressure, or the density are derived assuming hydrostatic and local thermodynamic equilibrium.   
The optical depth is then computed using the frequency-dependent absorption coefficient for the continuum \citep[see, \eg][]{ToroIniesta2007}.
}
The output is a height coordinate along the given, inclined LoS, which is directly compatible with the optical-depth scale of the magnetic-response functions. 
\corrbis{
We rebin this dataset to match the resolution of the observation-based response functions.
In order to include the instrument degradation as simulated by SOPHISM, we convolve the rebinned dataset  
with the optical-transfer function for a 14\,cm telescope aperture  (this operation is marked as PHI OTF in \fig{flow_rec}).
The} subsequent weighted integration along the LoS for each pixel results in what we call a \emph{reconstructed} magnetogram, \ie{} a best-effort approximation to the ``true'' solar scene represented by the MHD simulation; see \fig{geo_inv_diff}b.
The reconstructed magnetograms at different values of $\gamma$ are used to test the \SDM{} in \sect{results_inv}, and they are shown (without any instrument degradation) in the central column of \fig{spinor} for different values of $\gamma$.
A case that also includes the effect of instrument degradation is shown in \fig{geo_inv_diff}b and discussed in the next section.
Since the reconstructed magnetograms retain the information about the orientation of the transverse field component, they can be used as a test for \SDM{} that includes the spectropolarimetric influence of the viewing angle; see \sect{results_inv}.

\subsubsection{Difference Between the Synthetic and the Reconstructed Magnetograms}\label{s:diff_syn_rec}
We notice that, in the reconstruction procedure described above, the weighted integration is done with respect to an average $\tau=1$ surface for the solar scene at the given $\gamma$. 
As a result, the formation of the reconstructed magnetograms are slightly shifted towards larger heights for larger $\gamma$ and they also do not necessarily coincide with the formation heights of the synthetic observations, even for the same viewing angle. 
The effect of this shift is shown in the right column of \fig{spinor}: the difference between the synthetic magnetogram (left column) and the reconstructed magnetogram (central column) for a same value of $\gamma$ can be locally very large, and it is on average of the order of \corrbis{20\,--\,35\,G} for the LoS-component and 70\,--\,140 G for the transverse one, depending on $\gamma$.

The natural discrepancy of reconstructed vs. synthetic magnetograms and the increase of the formation height with $\gamma$, which is identical to the center-to-limb variation, affect both the magnitude and orientation of the resulting field vector. 
Therefore, the reconstructed magnetogram never perfectly matches the synthetic magnetogram obtained from the direct application of the SPINOR inversion code. This is the price that we need to pay in order to retain the information about the orientation of the transverse component that is needed for testing \SDM{}.

\subsection{Difference Between the Reprojected-Reconstructed  and the Reconstructed magnetograms}\label{s:geo_vs_inv}
\begin{figure}
 \setlength{\imsize}{0.31\columnwidth}
 \labfigure{\imsize}{\locpath/proj_OTF2_diff_sat_L_40_sat_B_0_geo_Bx}{-0.1\imsize}{1.22\imsize}{a)}
 \labfigure{\imsize}{\locpath/proj_OTF2_diff_sat_L_40_sat_B_0_inv_Bx}{-0.1\imsize}{1.22\imsize}{b)}
 \labfigure{0.948\imsize}{\locpath/proj_OTF2_diff_sat_L_40_sat_B_0_diff_Bx}{-0.1\imsize}{1.22\imsize}{c)}
 \caption{
$\Bw$ component  on the {\HRT} image plane of \textbf{(a)} the reproj-reconstructed, and \textbf{(b)} the reconstructed magnetogram; \textbf{(c)} difference between the reproj-reconstructed and the reconstructed $\Bw$ components as quantified by $\epsilon(\Bw)$ defined in \eq{eps}, at $\gamma=40^\circ$ for $\rD=1$ and $\Dnat=\DHMI=0.5$\,arcsec. 
In all panels, the black contour represents the 500G isoline of $|\Bw|$, and axis units are in pixels.
     }
\label{f:geo_inv_diff}
\end{figure}

In the previous section we introduced the reconstructed magnetograms to be used as tests of \SDM{} and discussed how such magnetograms differ from the synthetic magnetograms (\fig{spinor}) for a given value of the viewing angle $\gamma$.
In this section we further discuss the geometrical effect of the viewing angle  on the reconstructed magnetograms.
This is an attempt to quantify the fact that the observed emission from a feature at different $\gamma$ will never originate from the exact same parcel of plasma. 
This particularly affects weak-magnetic-field structures, \corrbis{especially where such structures are found}  
to be predominantly horizontal 
\citep[see e.g.][and references therein]{Danilovic_2016}. 
\corrbis{In such a case, first} 
only a small change of the formation height can change the LoS-component from positive to negative if the field lines are even only slightly changing their orientation with height. 
Second, as the transverse Zeeman effect is a second-order effect, \ie{} the amplitudes of the corresponding Stokes parameters are proportional to $B^2$, the measurements of the transverse field strength are significantly more sensitive to noise effects in the observations than those of the LoS-components.

In order to test for the change in the depth of the $\tau$-surface due to the different viewing angles, we compare the reconstructed magnetograms at any given $\gamma$ with the magnetogram  obtained by applying the geometrical re-projection described in \sect{tests_geo} \corr{and \fig{flow_geo}} to the $\gamma=0$ reconstructed magnetogram.
We identify the latter as the (geometrically) reprojected-reconstructed magnetogram (``Reproj-reconstructed'' in \tab{mgm}).
This comparison between reconstructed and reprojected magnetograms  specifically addresses the question of what is the effect of different viewing angles on the reconstructed field, to the best of our abilities at simulating real observations that retain the information about the orientation of the transverse field.
Again, this is not a test for {\SDM} but, instead, it is a direct comparison where we do not apply any disambiguation, and indeed employs unambiguous reconstructed model fields. 
In fact, both magnetograms have the correct (reprojected or reconstructed) orientation of the transverse field, and we retain this information here since the above question is unrelated to the disambiguation problem.

Panels a and b in \fig{geo_inv_diff} shows the $\Bw$-component of the reproj-reconstructed (a) and reconstructed (b) magnetograms in the image plane of {\HRT} for $\gamma=40^\circ$. 
At a first glance, the two distributions looks similar enough, with the reconstructed field appearing smoother than the reprojected field.
This is confirmed by differences in the blue isolines, corresponding to $|\Bw|$=500\,G.

A more quantitative and local measure of the difference between the two magnetograms can be computed in each pixel as
\BE
\epsBw=\frac{\left |\Bw^{\rm reproj-reconstructed} - \Bw^{\rm reconstructed}\right|}
                  {\left |\Bw^{\rm reproj-reconstructed}\right| + \left|\Bw^{\rm reconstructed}\right|},
\label{eq:eps}
\EE
{\ie} by the absolute value of the average divided by the average of the absolute values.
In each pixel, $\epsBw=0$ means that $\Bw$ is identical in the reproj-recontructed and reconstructed magnetograms, whereas $\epsBw=1$ corresponds to the case where the field has the same amplitude in both magnetograms but is opposite in sign.
\Fig{geo_inv_diff}c shows that the differences between the two magnetograms are actually quite significant in low-field regions. 
The value of   $\epsBw$  is smaller than 0.5 in 31\,\% of the pixels in \fig{geo_inv_diff}c. 
Values close to zero are quite obviously found in high-field areas.
On the other hand, on small-field areas the distribution of  $\epsBw$ is rather granular, and is difficult to identify a pattern for  the location where $\epsBw$ is close to one.
Similar plots and conclusions are obtained for the other field components too (not shown).


Therefore, according to our computational model of the observation (\sect{inv_model}), the differences in the observed field as a function of the observing angle are significant.  
More generally, the discussion in this section and in \sect{synth_mgm} shows the complexity and intrinsic limitations of creating observation-like test magnetograms. 
Such a difficulty adds to the complexity of the \Muram{} numerical solution that is used as a basis for the test production, and that were extensively discussed throughout \sect{results_geo}. 
Hence, the application of \SDM{} to reconstructed magnetograms described in \sect{results_inv} offers an exceptionally challenging test that we regard as a first application.

\section{Test of the \SDM{} on Reconstructed Magnetograms}\label{s:results_inv}

\begin{figure}
 \centering
 \setlength{\imsize}{0.45\columnwidth}
 \labfigure{\imsize}{\locpath/proj_OTF2_separation_bt_0}{-0.08\imsize}{0.80\imsize}{a)}
 \labfigure{\imsize}{\locpath/proj_OTF2_separation_bt_0_kdJz}{-0.08\imsize}{0.80\imsize}{b)}\\
 \labfigure{\imsize}{\locpath/proj_OTF2_separation_bt_0_kdBt100}{-0.08\imsize}{0.80\imsize}{c)}
 \labfigure{\imsize}{\locpath/proj_OTF2_separation_bt_0_kdBt500}{-0.08\imsize}{0.80\imsize}{d)}
 \caption{\corrbis{The metrics 
         \textbf{(a)} $\Marea$; \textbf{(b)} $\MJz$; \textbf{(c)} $\MBtone$; \textbf{(d)} $\MBtfive$, 
         as a function of the separation angle [$\gamma$], for the reconstructed magnetogram. 
          {\direct} and {\reverse} application are coded in \textit{orange} and \textit{blue} colors, respectively.
          }
}
\label{f:geo_inv_sep}
\end{figure}
%

In this section we apply \SDM{} to the \corrbis{ambiguous} reconstructed magnetograms obtained as described in \sect{rec_mgm} for $\gamma=[10,20,30,40,50,60,70]$ degrees, \ie{} using only the magnetograms shown in the central column in \fig{spinor}.
This test is similar to the geometrical test of \sect{tests_geo} except that we used the reconstructed magnetograms obtained in \sect{rec_mgm} for its application. 
In particular, we first apply Point~iii of \sect{mgm} to the reconstructed magnetograms at $\gamma=0$ and at finite values of $\gamma$ to introduce the ambiguity; second, we apply {\SDM}  as described in \sect{appl_geo} to the (now) ambiguous reconstructed magnetograms, for $\rD=1$ and different separation angles [$\gamma$]; third, we then use the knowledge of the real orientation of the transverse component in the reconstructed magnetograms to assess the correctness of the disambiguation using the success \corr{metrics $\Marea$, $\MJz$, and $\MBt$ for $\Tau=100$\,G and $\Tau=500$\,G defined in \sect{success}, as a function of $\gamma$ (see \Fig{geo_inv_sep})}.

\Fig{geo_inv_sep}a shows that
the accuracy of the disambiguation \corr{as quantified by $\Marea$} is almost monotonically decreasing from  79\,\% to 65\,\% in the \direct{} case, while it is more flat between 78\,\% and 83\,\% in the \reverse{} case, with a peak at $\gamma=20^\circ$.
\corr{
The other three metrics ($\MJz$, $\MBtone$, and  $\MBtfive$, shown in \fig{geo_inv_sep}(b), (c), and (d), respectively) have basically opposite trends for the \direct{} and \reverse{} cases, as a function of $\gamma$.
For instance, the total vertical current metric [$\MJz$, Panel b] increases in the \reverse{} application from \corrbis{57\,\% to 71\,\%} for increasing $\gamma$, whereas it decreases to about 50\,\% in the \direct{} application.
A similar anti-symmetric dependence on $\gamma$ is true for the transverse-field metrics $\MBtone$ and $\MBtfive$
In this case it is worth noticing that for separation angles larger than about 30$^\circ$, the fraction of the transverse field above 100\,G that is correctly disambiguated is above 90\,\% in the \direct{} case, and this fraction becomes almost 100\,\% if the threshold [$\Tau$] is raised to 500\,G.
}
We recall that, in the case examined in this section, the spatial sampling ratio is unity [$\rD=1$], hence, from the geometrical point of view, there is a pure foreshortening effect but the 
\corrter{spatial sampling} of the two detectors is the same. 
Therefore, the decrease of \corr{the success metrics} 
with increasing $\gamma$ in the \direct{} case is likely the consequence of probing increasingly different layers in the modelled atmosphere. 
Since, as noticed already above, the \Muram{} field is mostly vertical, such an effect is less pronounced when using the top-view from \HMI{} to resolve the ambiguity of \HRT{}, \ie in the  \reverse{} case.


When considered together, the most notable qualitative change between the geometrical case in \fig{sep_res}e and the \SDM{} application to reconstructed magnetograms of \fig{geo_inv_sep}a,b is the absence of a clear maximum at intermediate values of $\gamma$.
Moreover, the introduction of a threshold in the computation of $\Marea$, while sensibly improving the rate of successful disambiguation in the \reverse{} case, it is counter-productive in the \direct{} case. 

\begin{figure}
 \centering
 \setlength{\imsize}{0.45\columnwidth}
 \includegraphics[width=\imsize]{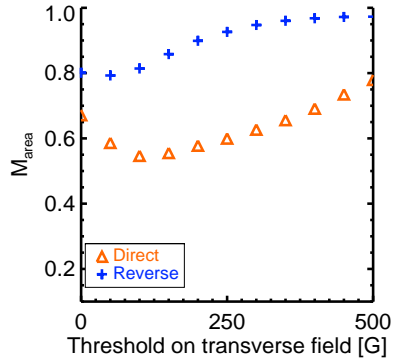}
 \caption{\corr{
         Dependence of the metric $\Marea$ as a function of the the threshold [$\Tau$] on the amplitude of the transverse component (see also \sect{thresh_results}) for the reconstructed magnetogram, at $\gamma=40^\circ$.
          {\direct} and {\reverse} application are coded in orange and blue colors, respectively.
          }
}
\label{f:geo_thresh}
\end{figure}
To complete the above picture, \fig{geo_thresh} shows the variation of $\Marea$ as a function of the threshold on included pixels, for  the fixed value $\gamma=40^\circ$, \corr{computed as described in} \sect{thresh_results}.
Again, we find a more complex qualitative dependence  than in the geometrical case in \fig{thr}c; as the computation of $\Marea$ is restricted to pixels with increasingly larger values of the transverse component, the \corr{area metric} for the \reverse{} case is increasing, up to a very high value of $\Marea =97\,\%$ for a threshold of 500\,G. 
On the other hand, the same threshold has largely the opposite effect in the \direct{} case, where a value of almost 400\,G is required in order to increase on the $\Marea$ obtained without any threshold.     
The lower success rate in the \direct{} case is identified as a consequence of the intrinsic properties of stereoscopic measurements which do not allow to identically probe small-scale and low magnetic solar regions when seen from different vantage points.

In conclusion, the spectropolarimetric effect of different viewing angles has significant consequences on the application of (possibly any) disambiguation method.
In consideration of the analysis in \sects{inv_model}{geo_vs_inv}, this is hardly surprising, given that {\SDM} assumes that the same field is observed from different viewpoints. 
Since observing from different directions produces significantly different fields (see \sect{geo_vs_inv}), any geometrically based method of disambiguation such as {\SDM} translates such differences into inaccuracies. 
On the other hand, the application in the \reverse{} case (\ie{} basically using \HMI{} to remove the ambiguity of \HRT{}) has, even in such a complex case, an accuracy in excess of 80\,\%, which can rise to above 90\,\% if a 200\,G-threshold on the transverse component is introduced. 
Considering the unfavorable (and unrealistic) parameters that we were forced to adopt for the \SDM{} test presented in this section, namely the simulated observation of a small pore surrounded by large area of quiet-Sun with  the limited resolution provided by \SolO{} at 1\,AU ($\rD=1$), we find the above results very encouraging for the application of the \SDM{} even when the effect of optical path is included. 
A more extended exploration of parameter space, as well application to higher-resolution simulation of active-region field, are needed, and they planned for future publications.

\section{Conclusions}\label{s:conclusions}
We present in this article the analytical formulae, \eqs{s_l}{s_w}, that enable observational resolution of the ambiguity of the transverse component in the image planes of two telescopes observing the same region on the Sun from two different vantage points. 
These formulae are implemented in the stereoscopic disambiguation method ({\SDM}) for application to any two observatories.
{\SDM} is designed to be applied to any chance of stereoscopic observations, as currently given by {\HMI} and {\PHI}, but also to other existing observatories as well as forthcoming missions ({\eg} \textit{Lagrange}).
In particular, we notice that the {\SDM} can in principle be applied even if one of the two detectors only measures the LoS-component of the magnetic field (\ie{} using the latter as detector B in \eq{s_l} and discarding \eq{s_w}).

{\SDM} is then thoroughly benchmarked using geometrical tests and tests employing reconstructed magnetograms.
In the first type of test, an identical 2D vector magnetogram is viewed from different angles \corr{and distances, \ie{} 
\corrter{spatial sampling}.}
We consider three types of test magnetograms going from very smooth field ({\TD}), to field with both large and small scales ({\Feng}), to a simulation of a pore surrounded by a large quiet-Sun area ({\Muram}). 

The result of the geometrical tests can be summarized as follows: 
\begin{itemize}
\item  in an idealized situation such as the \TD{} where the field is smooth and well resolved on both detectors, the application of {\SDM} is able to remove the ambiguity with 100\,\% accuracy, at all separation angles and considered 
\corrter{spatial sampling};
\item for an active-region field (\Feng{}), the disambiguation accuracy is also 100\,\%, with small decreases at expected locations (\eg{} when the spacecrafts are co-aligned or close to quadrature), or for the lower resolution of the \FDT{} telescope;
\item the accuracy is modulated by both the separation angle between detectors as well as by the {\HRT} effective resolution, {\ie} by the distance of the {\SolO} spacecraft from the Sun; 
\item the disambiguation accuracy is found to be mostly sensitive to the small scales present in the test magnetogram, going from perfect disambiguation (\eg{} for the smooth \TD{}-test magnetogram) to  almost 50\,\% accuracy, equivalent to a random resolution of the ambiguity, in the least favorable case of observing fields with very small scales (such as the \Muram{}) with the lower-resolution \FDT{} telescope when \SolO{} is at 1\,AU; 
\item furthermore, the success rate is found to increase sensibly even in the most challenging {\Muram} case if pixels with small values of the transverse component are not included in the disambiguation procedure; 
\end{itemize}

\begin{figure}
 \centering
 \setlength{\imsize}{0.9\textwidth}
 \includegraphics[width=\imsize]{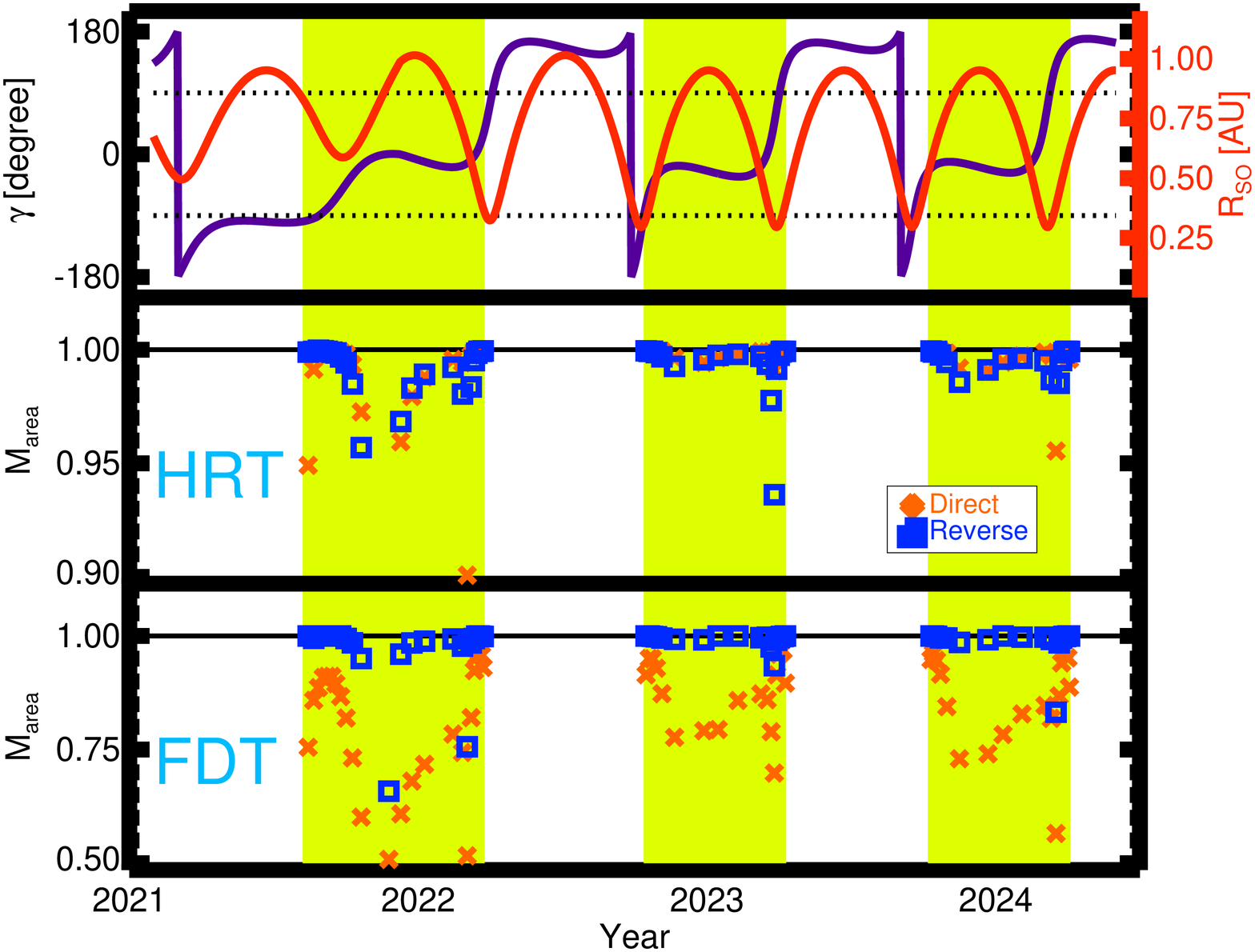}
 \caption{Expected  $\Marea$ during the years 2021\,--\,2024 as \SolO{} travels on its orbit. 
          The \textit{top panel} shows the orbit of the spacecraft in terms of the separation angle with \HMI{} [$\gamma$] and the distance from the Sun ($R_{SO}$, related to $\rD$ by \eq{rD}); the \textit{middle panel} shows the expected $\Marea$ from application of \SDM{} using the \HRT{} telescope, while the \textit{bottom} one is for the \FDT{} telescope (notice the \textit{different vertical axis range} for $\Marea$ in the \textit{middle- and bottom-panels}).
\textit{Orange crosses} (respectively, \textit{blue squares}) show the results of the \direct{} (respectively, \reverse{}) application of the \SDM{}.
          The \textit{green} areas correspond to times where \SDO{} and \SolO{} are on the same side of the Sun, \ie{} when stereoscopic disambiguation is possible. 
          The model field  used for this test is the \Feng{} case; see \sect{feng}.  
  }
 \label{f:eta_orbit}
\end{figure}
In addition to testing the method, our parametric study allows meaningful application to the \SolO{} observations.
The practical case of the expected accuracy of {\SDM} during  the years 2022\,--\,2024 can be obtained by using the orbit information. 
\Fig{eta_orbit} shows the expected accuracy for the \Feng{} case, 
as a function of time from 2021 to 2024. 
This type of study can be used to place remote-sensing windows at times that are favorable to stereoscopic disambiguation.
The systematic exploration of the best placement of observing remote-sensing windows that includes such effects is straightforward but extensive, and we reserve this study for a dedicated  work.   
Here we limit ourselves to notice the anti-correlation (clearly visible for the \FDT{} telescope) between distance, \ie{} 
\corrter{spatial sampling}, and accuracy, and that this effect is stronger than the separation angle effect.

The second type of test that we perform employs reconstructed magnetogram, \ie{} synthetic spectropolarimetric observations at different angles obtained using a 3D-{\Muram} simulation of the upper photospheric layers; see \sect{rec_mgm}.
By testing the \SDM{} with reconstructed magnetograms we can draw the following conclusions:
\begin{itemize}
\item  viewing the same area from different angles has a significant influence on the observed values. 
       Since stereoscopic methods are based on combining two different viewpoints, this implies that any stereoscopic disambiguation method faces serious challenges, in particular in low-field areas, regardless of the method details;
\item as a consequence, the application of {\SDM} to reconstructed magnetograms translates the differences in the reconstructed magnetograms as seen from different angles into inaccuracies of the disambiguation. 
Such inaccuracies are expected to be small in the case of strong-field areas such as active regions, but proper testing of such supposition requires suitable numerical simulations.
However, even in the challenging case of reconstructed observations of a pore surrounded by quiet Sun, the {\SDM} accuracy can attain significant improvements (above 80\,\%) with respect to a random choice of orientation when applied in the \reverse{} mode, \ie{} when \HMI{} is used to remove the ambiguity of \HRT{}. 
\end{itemize}
We regard these results as preliminary. 
On the grounds of the geometrical tests summarized above, there is basically a 10\,\% difference between geometrical and reconstructed tests in the disambiguation accuracy of the \Muram{} magnetogram. 
As in the geometrical test case the success rate in strong magnetic-field regions of these simulations is easily 10-15\,\% better than in the quiet Sun, we would expect a success rate in excess of 90\,\% if applied to magnetograms thresholded for only sunspot areas (see, \eg{} \fig{thr}c). 
We have chosen this type of \Muram{} simulation to test \SDM{} under conditions as close as possible to a realistic case. 
It has to be emphasized that during periods of high solar activity typical magnetograms obtained by \HMI{} contain only a very small area of regions with 
\corrter{pixel-averaged signal above 150\,Mx\,cm$^{-2}$} \citep[cf.][]{Liu2017a}. 
In the weak-field regions the success rate of \SDM{} is highly dependent on the ratio of spatial resolutions of the input data since a reliable disambiguation needs to take into account the magnetic fine scale of the solar photosphere. 
Despite the intrinsic limitations of observing particularly the exact fine structure of weak transverse field regions from different viewpoints (see discussion in \sect{tests_inv}) a success rate of above 80\,\% (\fig{geo_inv_sep}a) can be achieved with \HRT{} observations with a 
\corrter{spatial sampling} comparable or better than that of \HMI{}.

In conclusion, the stereoscopic disambiguation method [\SDM{}] is in principle an exact method of removing the ambiguity, which is proven to yield 100\,\% accuracy when applied to \corrbis{idealized} conditions.
The accuracy of {\SDM} is found to depend mostly on the amount of unresolved structure at a given 
\corrter{spatial sampling} (\ie{} distance from the Sun), but it can otherwise remove the ambiguity with excellent accuracy.  
Even for the quiet-Sun case (\Muram{}), the rate of success of \SDM{} is found to be above 80\,\%, which is significantly better than the random orientation currently adopted in many situations.  
In this respect, the \SDM{} is expected to improve significantly disambiguation with respect to the current state-of-the-art using one viewpoint only, and it can be used as a reliable benchmark for other single-viewpoint methods not only in strong-field areas but also on quiet-Sun areas.

\corr{The study presented here is preliminary in many ways.
First, a systematic study of the influence of noise and unresolved scales on the accuracy of \SDM{} along the lines of \cite{Leka2009} is required to assess such effects in a realistic way.
On the other hand, a} direct extension of the present work will be to test \SDM{} employing reconstructed magnetograms of an extended sunspot, such as modelled by \cite{Siu_2018}.

\corr{Second, and more specific to the stereoscopic nature of the \SDM{}, a difference in the calibration of the two employed telescopes may potentially be an issue in real applications.
Our preliminary investigations indicate that a difference in the field strength measured by the two telescopes up to 10\,\% results in a decrease of accuracy in \SDM{} disambiguation of about 2\,\% at most.
In other words, a moderate difference in the calibration functions of the observing instruments is not expected to greatly affect the accuracy of the \SDM{}.
However, a proper estimation of the effect of calibration differences on the \SDM{} is best done once the actual calibration functions of \HMI{} and \PHI{} become available. 
}

\corr{Third, }
the application to real \SolO{} observations will require the implementation of {\SDM} in the {\SolO} pipeline, and the inclusion of maps of estimated  errors based on the parameter optimization presented here.
\corr{In this respect, one}
notable missing test is the dependence of \SDM{} on pointing errors. 
Since the pointing information is essentially a labeling of pixels with coordinates that is an integral part of the data processing of each telescope, we postpone this test to the implementation of \SDM{} in the \SolO{} data pipeline.

Finally, a comparison with traditional, single-view disambiguation methods would be desirable in order to quantitatively assess how \SDM{} can be used to benchmark more traditional, single-viewpoint disambiguation methods. 


\begin{acks}
\corr{We wish to thank the anonymous reviewer whose constructive comments helped to improve the clarity and focus of the article.}
This work was developed within the community-led effort by European Space Agency's Modelling and Data Analysis Working Group (MADAWG).
 G.\,Valori acknowledges the support from the European Union’s Horizon 2020 research and innovation programme  under  grant  agreement  No  824135  and  of  the STFC grant number ST/T000317/1.
 D.\,Stansby is supported by STFC grant ST/S000240/1.
 E.\,Pariat acknowledges financial support from the Programme National Soleil Terre (PNST) of the CNRS/INSU and from the French National Space Agency CNES.
G.\,Valori, P.\,L\"oschl, and J.\,Hirzberger acknowledge funding by the Bundesministerium f{\"u}r Wirtschaft und Technologie through Deutsches Zentrum f{\"u}r Luft- und Raumfahrt e.V. (DLR) Grants No. 50 OT 1001/1201/1901 as well as 50 OT 0801/1003/1203/1703, and by the President of the Max Planck Society (MPG). 
P.\,L\"oschl also acknowledges support from the International Max Planck Research School for Solar System Science at the University of G{\"o}ttingen.
 The PENCIL-AR simulation was supported by the computational resource from Yellowstone (ark:/85065/d7wd3xhc) provided by NCAR's Computational and Information Systems Laboratory, sponsored by the National Science Foundation. 
\end{acks}

\begin{dataavailability}
The datasets generated and analyzed during the current study are available from the corresponding author on reasonable request.
\end{dataavailability}

\begin{ethics}
\begin{conflict}
The authors declare that they have no conflicts of interest.
\end{conflict}
\end{ethics}


 \bibliographystyle{spr-mp-sola}
\end{article}
\end{document}